\documentclass{ws-ijmpd}
\usepackage[super,compress]{cite}
\usepackage[table,xcdraw]{xcolor}
\usepackage{graphics}
\usepackage{url}

\newcommand{\beq}{\begin{equation}}
\newcommand{\eeq}{\end{equation}}
\newcommand{\ga}{\lower.7ex\hbox{$\;\stackrel{\textstyle>}{\sim}\;$}}
\newcommand{\la}{\lower.7ex\hbox{$\;\stackrel{\textstyle<}{\sim}\;$}}

\usepackage{tikz}
\usetikzlibrary{shapes.geometric}
\usetikzlibrary{arrows}
\usepackage{amsmath}

\DeclareMathOperator{\csch}{csch}
\newcommand{\vphi}{\varphi}

\begin{document}

\markboth{Ellis, Garc{\' i}a, Nagata, Nanopoulos, Olive and Verner}
{Building Models of Inflation in No-Scale Supergravity}

%
\catchline{}{}{}{}{}
%

\title{Building Models of Inflation in No-Scale Supergravity}

\author{John Ellis}

\address{Theoretical Particle Physics and Cosmology Group, Department of
  Physics, King's~College~London, London WC2R 2LS, United Kingdom;\\
Theoretical Physics Department, CERN, CH-1211 Geneva 23,
  Switzerland;\\
  National Institute of Chemical Physics and Biophysics, R{\" a}vala 10, 10143 Tallinn, Estonia\\
john.ellis@cern.ch}

\author{Marcos~A.~G.~Garc{\' i}a}

\address{Instituto de F\'isica Te\'orica (IFT) UAM-CSIC, Campus de Cantoblanco, 28049, Madrid, Spain \\
marcosa.garcia@uam.es}

\author{Natsumi Nagata}

\address{Department of Physics, University of Tokyo, Bunkyo-ku, Tokyo
 113--0033, Japan \\
 natsumi@hep-th.phys.s.u-tokyo.ac.jp}

\author{Dimitri~V.~Nanopoulos}

\address{George P. and Cynthia W. Mitchell Institute for Fundamental
 Physics and Astronomy, Texas A\&M University, College Station, TX
 77843, USA;\\
Astroparticle Physics Group, Houston Advanced Research Center (HARC),
 \\ Mitchell Campus, Woodlands, TX 77381, USA;\\ 
Academy of Athens, Division of Natural Sciences,
Athens 10679, Greece\\
dimitri@physics.tamu.edu}

\author{Keith~A.~Olive}

\address{William I. Fine Theoretical Physics Institute, School of
 Physics and Astronomy, \\
 University of Minnesota, Minneapolis, MN 55455,
 USA\\
 olive@umn.edu}

\author{Sarunas~Verner}

\address{William I. Fine Theoretical Physics Institute, School of
 Physics and Astronomy, \\
 University of Minnesota, Minneapolis, MN 55455,
 USA\\
 nedzi002@umn.edu}

\maketitle

\newpage
\begin{abstract}
\begin{center}
{\bf Abstract} \\
\end{center}
After reviewing the motivations for cosmological inflation formulated in the formalism of supersymmetry,
we argue that the appropriate framework is that of no-scale supergravity. We then show how to
construct within this framework inflationary models whose predictions for the tilt in the spectrum of
scalar perturbations, $n_s$, and the ratio, $r$, of tensor and scalar perturbations
coincide with those of the $R + R^2$ model of inflation proposed by Starobinsky. A more detailed study of
no-scale supergravity reveals a structure that is closely related to that of $R^2$ modifications of the 
minimal Einstein-Hilbert action for general relativity, opening avenues for constructing no-scale
de Sitter and anti-de Sitter models by combining pairs of Minkowski models, as well as generalizations of
the original no-scale Starobinsky models of inflation. We then discuss the phenomenology of no-scale
models of inflation, including inflaton decay and reheating, and then the construction of explicit scenarios
based on SU(5), SO(10) and string-motivated flipped SU(5)$\times$U(1) GUT models. The latter provides a possible
model of almost everything below the Planck scale, including neutrino masses and oscillations, the 
cosmological baryon asymmetry and cold dark matter, as well as $n_s$ and $r$.
\begin{center}
    KCL-PH-TH/2020-44, CERN-TH-2020-136,
    ACT-6-20, MI-TH-2024, UMN-TH-3926/20, 
    FTPI-MINN-20/29, IFT-UAM/CSIC-20-126
\end{center}

\end{abstract}




\section{Introduction}	

Inflation was initially proposed as a possible simultaneous solution to several
fundamental problems in cosmology \cite{Guth}. 
These included the horizon problem, i.e., the
fact that the Universe is nearly homogeneous and isotropic on scales that are
much larger than would ever have been within the range of causal contact in the conventional
Big Bang cosmology framework, and the flatness problem, i.e., the fact the average density of 
the visible Universe today is very close to the critical density corresponding to a flat geometry
with negligible curvature. Moreover, inflation also had the added bonus of suppressing by
large factors the densities of unobserved massive relics from the first stages of the
Big Bang, e.g., magnetic monopoles.

Most early formulations of inflationary models~\cite{reviews} were based on one or more effectively 
elementary scalar fields, though a proposal by Starobinsky~\cite{Staro} was based on an extension of the
Einstein-Hilbert action - which is linear in the Ricci tensor $R$ - to include an 
additional $R^2$ term. However, it was noticed later that this $R + R^2$ model is
equivalent via a conformal transformation to the Einstein-Hilbert action complemented
by a scalar field with a very specific form of effective potential~\cite{WhittStelle}. In this and many other
scalar field models one finds that the change in the effective scalar inflaton field
is ${\cal O}(M_P)$\cite{primordial}, where $M_{P} = 2.4 \times 10^{18}$~GeV is the reduced Planck mass, suggesting that some assumption about gravity would be needed for their
formulation.

A major step forward in the phenomenological interpretation of inflationary models
came with the realization that the scalar inflaton field would be subject to
quantum fluctuations~\cite{MC,pert}. At the time, observational upper limits on possible
perturbations in the cosmological microwave background (CMB) already imposed strong
constraints on these fluctuations, which in turn implied that the effective inflaton
potential must contain a small parameter, e.g., the inflaton mass and/or a potential
coupling $\lambda$.

This realization posed a cosmological hierarchy problem: what was the dynamics
permitting the inflaton mass to be $\ll M_P$ - a situation resembling that of the
Higgs mass and the electroweak hierarchy problem - and/or the coupling $\lambda$ to be $\ll 1$.
Three of us (JE, DN and KO) promptly suggested that this cosmological hierarchy
problem of inflation {\em cries out} for stabilization by supersymmetry.~\cite{Cries,primordial,fluct}.
Many supersymmetric models of inflation have been proposed subsequently.

The cosmological perturbations predicted by models of inflation share some
generic features. The quantum fluctuations in the scalar inflaton field should be 
predominantly Gaussian in all models whose inflaton field rolls slowly down its 
effective potential, their spectrum would not in general be scale-invariant but exhibit
a small tilt, $n_s \ne 1$, and the gravitational background would also be subject to small
tensorial quantum fluctuations, with a ratio $r \ll 1$ relative to the larger scalar
perturbations. Present measurements of the CMB confirm the presence of a tilt~\cite{planck18},
$n_s = 0.965 \pm 0.004$, and set stringent upper limits on the tensor-to-scalar ratio~\cite{rlimit},
$r < 0.06$. These measurements conflict with the predictions of single-scalar inflationary
models based on simple monomial potentials, but are completely consistent with the 
Starobinsky $R + R^2$ model, which predicts $r \simeq 0.003$.

The question then arises, how to write down supersymmetric models of inflation that
are compatible with the available CMB data, with predictions similar to those
of the Starobinsky model? More specifically, how may this be achieved within the
framework of supergravity?

Everyone can agree that general relativity is an essential complication in cosmology, 
and we have already commented that the magnitude of the variation in the inflaton field 
confirms the need to include gravity in any complete model of inflation. Concretely,
supergravity~\cite{cremmer,susy} is the extension of supersymmetry to include gravity, so it is the
appropriate framework for any cosmological scenario involving supersymmetry.

However, there is a significant theoretical hurdle to overcome\cite{eta}. Generic supergravity models
with matter fields have effective potentials that do not obey the flatness conditions needed
by slow-roll models of inflation, and have anti-de Sitter (AdS) vacua with field energies 
$-{\cal O}(m_{3/2}^2 M_P^2)$, where $m_{3/2}$ is the gravitino mass. Moreover, a second desideratum for any particle physics model,
and in particular any model of inflation, is that it not be incompatible with ultraviolet
completion in some string model. So, are there any supergravity models that yield effective
potentials with flat directions and no AdS `holes', which emerge naturally as infrared
limits of string models?

The affirmative answer to both questions is provided by no-scale supergravity models~\cite{no-scale,EKN1,EKN,LN},
so named because their effective potentials feature flat directions with no specific
dynamical scale selected at the classical (tree) level, which have been derived from
string models as their effective low-energy theories\cite{Witten}. 

In this paper we review the construction of models of inflation based on no-scale
supergravity, paying particular attention to their predictions for inflationary
observables and the extent to which they mimic those of the Starobinsky $R + R^2$
model. As we discuss, there is a natural close relation between the no-scale
supergravity and Starobinsky models~\cite{eno6}. We also discuss the evolution of the Universe after
no-scale supergravity inflation, paying particular attention to the reheating
following inflaton decay and the production of gravitinos and supersymmetric dark matter.
We discuss these issues, together with baryogenesis, neutrino masses and nucleon decay,
in three grand unified theories (GUTs) incorporated into the no-scale supergravity 
framework, namely minimal SU(5), SO(10) and flipped SU(5)$\times$U(1). The latter is of particular
interest, as SU(5)$\times$U(1) is the only GUT gauge group whose symmetry can be broken 
down to the Standard Model (SM) via Higgs fields in the representations available in the
weakly-coupled limits of heterotic string models~\cite{AEHN}. The inflationary SU(5)$\times$U(1) 
no-scale supergravity model is therefore a prototype theory of (almost) everything
below the Planck scale.

The review is structured as follows.
We begin in Section 2 with a brief motivation for supersymmetry,
a review of the essential elements of supergravity with a focus on
no-scale supergravity, as well as aspects of the Polonyi model~\cite{pol} for supersymmetry breaking. 
Simple models of inflation in minimal and no-scale supergravity 
are discussed in Section 3 and the computation of inflationary observables are described in Section 4. A pioneering no-scale Starobinsky~\cite{Staro} inflation model~\cite{eno6} is introduced in Section 5. At this point, we delve deeper into the structure of no-scale supergravity in Section 6, and the connections between
no-scale supergravity and higher-order gravity theories are reviewed in Section 7. We discuss generalized no-scale models, Minkowski and de Sitter solutions in Section 8 and generalized inflationary models in Section 9. The stabilization of fields that do not drive inflation is discussed in Section~10. Phenomenological aspects of no-scale inflationary models including supersymmetry breaking and reheating are covered in Section 11. In Section 12, we review some UV completions of the 
inflationary models based on SU(5), SO(10) and flipped SU(5)$\times$U(1)
grand unified theories.  Our conclusions are summarized in Section 13.

\section{Supergravity Primer}
\label{sec:sugrp}

\subsection{Why Supersymmetry and Supergravity?}

Supersymmetry is thought to be an essential feature of string theory. However, the absence
of supersymmetric particles at the TeV scale~\cite{nosusy} implies that it must be broken somewhere between
there and the Planck scale. It should certainly be incorporated in any model of inflation
if the supersymmetry-breaking scale is not larger than the scale of inflation, which is
typically ${\cal O}(10^{13})$~GeV. In fact, there are specific features of supersymmetric
field theories that led some of us (JE, DN and KO)~\cite{Cries} to propose that it should play an essential
role in inflation, and therefore should be broken only at some scale $< {\cal O}(10^{13})$~GeV.

Supersymmetric field theories have unique renormalization properties: the masses of scalar
particles have no quadratic divergences in any order of perturbation theory, and their trilinear
and quartic couplings have only multiplicative logarithmic divergences. It was the absence of quadratic
divergences that led to the suggestion that supersymmetry at the TeV scale could stabilize the
electroweak scale and its hierarchy with the Planck scale\cite{Maiani:1979cx}, and the absence of non-multiplicative
renormalization of trilinear and quartic couplings helped stabilize the hierarchy of mass scales
in GUTs.

In view of the difficulties with the completion of the first-order phase transition in the
first scalar field theory of inflation~\cite{Guth:1982pn},
radiatively-driven GUT-based models of inflation were suggested~\cite{new}.
However, in these models the magnitude of the effective potential was determined by the GUT gauge
coupling, and was too large to be compatible with the CMB data\cite{pert}. The solution proposed in Ref.~\refcite{Cries}
was to postulate that inflation was driven by a gauge singlet with a quartic scalar coupling that was kept naturally 
small by the fact that its renormalization was purely multiplicative in a supersymmetric theory.
Moreover, the absence of quadratic divergences keeps the inflaton mass naturally small.

The field theories discussed in this early work featured global supersymmetry, but there were several
reasons to extend them to theories with local supersymmetry, namely supergravity theories\cite{nost}. One
motivation was the principle that Nature abhors global symmetries\cite{hpp}, but embraces local symmetries
such as the gauge symmetries of the Standard Model, and supergravity is the unique local (gauge)
extension of global supersymmetry. Another motivation, already mentioned in the Introduction, is
the fact that any treatment of cosmology should incorporate gravity, and supergravity is the unique
framework combining it with supersymmetry. Finally, as a corollary of this observation, since the
low-energy effective theory obtained from string theory necessarily contains gravity, and should also
include supersymmetry so as to render inflation natural, as discussed above, it must be a supergravity
theory.

\subsection{Structure of Supergravity Models}

We now review briefly the structure of the bosonic part of a generic
supergravity Lagrangian to second order in the derivatives of the
physical scalar fields, leaving until later aspects concerning auxiliary fields and fermions. 

This supergravity bosonic Lagrangian for an $\mathcal{N} = 1$ supersymmetric theory
may be written in terms of a
Hermitian function of complex chiral scalar fields $\phi^i$, 
called the K{\" a}hler
potential, $K (\phi^i, \phi_j^*)$, that characterizes the geometry of the theory, and a holomorphic function of these fields, called the superpotential, $W(\phi^i)$, responsible for interactions among these fields and their fermionic partners. These
may be combined into the function 
$G \equiv K + \ln |W|^2$.~\footnote{We use natural units with $M_P = 1$ in this Section.}
The kinetic terms and scalar potential in the bosonic Lagrangian
can then be written in the form 
\beq
{\mathcal L} = - \frac12 R + K^j_i \partial_\mu \phi^i \partial^\mu \phi^*_j - V - \frac14 {\rm Re} (f_{\alpha\beta}) F^\alpha_{\mu\nu} F^{\beta \mu\nu} -\frac{1}{4} {\rm Im} (f_{\alpha\beta}) F^\alpha_{\mu\nu} \tilde{F}^{\beta\mu\nu} \, ,
\label{Lkin3J}
\eeq
where the first term is the minimal Einstein-Hilbert
term of general relativity and in the second term $K^j_i \equiv \partial^2 K/\partial \phi^i \partial \phi_j^*$.
The effective scalar potential,
\begin{equation}
\label{effpot}
V \; = \; e^{G} \left[G_i \left({G^{-1}}\right)^i_j G^j -3 \right],
\end{equation}
where $G_i \equiv \partial{G}/{\partial \phi^i}$, $G^j \equiv \partial{G}/{\partial\phi_j^*}$, and $\left(G^{-1}\right)^i_j$ is the inverse of the matrix of second
derivatives of $G$. In addition, 
there are also $D$-term contributions for gauge non-singlet
chiral fields. Finally, $f_{\alpha\beta}$ is the gauge kinetic function, which is in general a function of the chiral fields, $\phi^i$.~\footnote{For a review of local supersymmetry, see Ref.~\refcite{susy}.}
Minimal supergravity (mSUGRA) is characterized by a K{\" a}hler potential
of the form 
\beq
K = \phi^i \phi_i^* \, ,
\label{Kmin}
\eeq 
in which case the effective potential (\ref{effpot}) can be written in the form
\beq
V(\phi^i,\phi_j^*)  = e^{\phi^i {\phi^*_i}}  \left[ |W_i + \phi_i^* W
|^2 - 3|W|^2
\right] , 
\label{sgpotJ}
\eeq
where $W_i \equiv \partial{W}/{\partial\phi^i}$. 

The effective potential (\ref{sgpotJ}) illustrates some of
the pitfalls of supergravity cosmology. The first is that,
unlike the effective scalar potential in global supersymmetry, where $V = |W_i|^2$, 
the minimal supergravity potential is {\it not} positive
semi-definitive. Indeed, the negative term $\propto |W|^2$ in
(\ref{sgpotJ}) generates in general AdS `holes' with depth
$- {\cal O}(m_{3/2}^2 M_P^2)$, 
inducing a cosmological instability.
More generally, (\ref{sgpotJ}) does not have flat directions
in field space except under special conditions such as those we discuss
below. This is an issue for constructing models of
inflation because, as we discuss in more detail below, a 
period of inflation that is long enough to solve the horizon
and flatness problems should satisfy slow-roll conditions
that require the scalar potential to have a(n
almost) flat direction. This can be problematic
in minimal supergravity (\ref{Kmin}), as the effective scalar potential 
(\ref{sgpotJ}) is proportional to $e^K$,
which is $(1+ \phi \phi^* + ...)$ for small field values.
Scalars therefore typically pick up masses proportional
to $V \sim H^2$, where $H$ is the Hubble parameter~\cite{eta,drt},
violating the slow-roll conditions. This is known as the $\eta$-problem.

\subsection{No-Scale Supergravity}

These difficulties are avoided in no-scale supergravity models,
of which the simplest is the single-field example with\cite{no-scale,EKN1}
\begin{equation}
        K \; = \; -3 \ln (T + T^*) \, ,
\label{CFKN}
\end{equation}
where $T$ may be identified with the volume modulus in a string compactification.
This leads to an Einstein-Hilbert Lagrangian for gravity accompanied by the following kinetic term
for the modulus field, derived from Eq.~(\ref{Lkin3J}):
 \begin{eqnarray}
{\mathcal L}_{kin}  & = &  \frac{3}{(T+T^*)^2} \partial^\mu T \partial_\mu T^*  \nonumber \\
& = &  \frac{1}{12} (\partial_\mu K)^2 + \frac34 e^{2K/3} |\partial_\mu (T- T^*)|^2  \, ,
\label{LkinnsJ}
\end{eqnarray}
where we note that, up to a factor of $\sqrt{6}$, $K$ has a canonical kinetic term. 
In the absence of a superpotential for the modulus, the effective scalar potential vanishes:
\begin{equation}
    V \; = \; 0 \, ,
    \label{V0}
\end{equation}
which satisfies trivially the flatness condition, in particular the absence of negative-energy AdS solutions.

In the minimal no-scale model (\ref{CFKN}) the single complex field $T$ parametrizes a non-compact
SU(1,1)/U(1) coset space. It can be generalized by including matter fields $\phi^i$
that parametrize, together with $T$, an SU(N,1)/SU(N)$\times$U(1) coset space, 
defined by the K{\" a}hler potential~\cite{EKN}~\footnote{There are other generalizations
based on other non-compact coset spaces, which also appear in some string models.} 
\beq
\label{v0}
K \; = \; -3 \ln (T + T^* - |\phi_i|^2/3) \, .
\eeq
In this case we find the following scalar-field Lagrangian
\begin{eqnarray}
{\mathcal L} & = &   \frac{1}{12} (\partial_\mu K)^2 + e^{K/3} |\partial_\mu \phi^i|^2 \nonumber \\
&& +\frac34 e^{2K/3} |\partial_\mu (T- T^*) + \frac13 (\phi_i^{*}\partial_{\mu}\phi^i-\phi^i\partial_{\mu}\phi^{*}_i)|^2  - V \, ,
\label{LmanyJ}
\end{eqnarray}
where the effective scalar potential can be written as 
 \beq
 V \; = \; e^{\frac{2}{3}K} {\hat V} \; = \; \frac{\hat V}{\left((T + {T^*})- \frac{1}{3} |\phi^i |^2 \right)^2} \, ,
 \label{VJ}
 \eeq
 with
  \beq
{\hat V} \; \equiv \; \ \left|  W_{i} \right|^2  +\frac{1}{3} (T+T^*) |W_T|^2 +
\frac{1}{3} \left(W_T (\phi_i^* W^{*i} - 3 W^*) + {\rm h.c.}  \right) \, .
\label{effVJ}
\end{equation}
We see that when $W_T = 0$ the potential takes a form related to that in global supersymmetry,
though with a proportionality factor of $e^{2K/3}$, as seen in Eq.~\eqref{VJ}, where $K$ is
the canonically-redefined modulus. Hence large mass terms are not generated~\cite{deln}, and
the $\eta$-problem is avoided~\cite{GMO}. 

\subsection{Supersymmetry Breaking}
\label{sec:susyb}

In globally-supersymmetric models, supersymmetry is broken when $\langle F \rangle = W_\phi \ne 0$ for some field $\phi$.
However, because the potential is simply equal to $|W_\phi|^2$, broken supersymmetry always leads to a non-zero
vacuum energy density of the same order as the supersymmetry breaking. In contrast, 
in locally-supersymmetric models, supersymmetry is broken when 
an $F$-term given by
\beq
F_i = -m_{3/2} (G^{-1})^j_i G_j
\label{fterm0}
\eeq
picks up a vacuum expectation value (VEV). In minimal $N=1$ supergravity, this corresponds to
$\langle F_i \rangle = \langle e^{K/2} D_{\phi^i} W \rangle  = \langle  e^{K/2} (W_i + K_i W) \rangle \ne 0$.
In this case, because of the form of the potential in Eq.~(\ref{sgpotJ}), it is possible to break supersymmetry 
and at the same time cancel the vacuum energy density, with a gravitino mass given by $m_{3/2} = e^{\langle G \rangle/2}$.

\begin{figure}[ht]
\centerline{\psfig{file=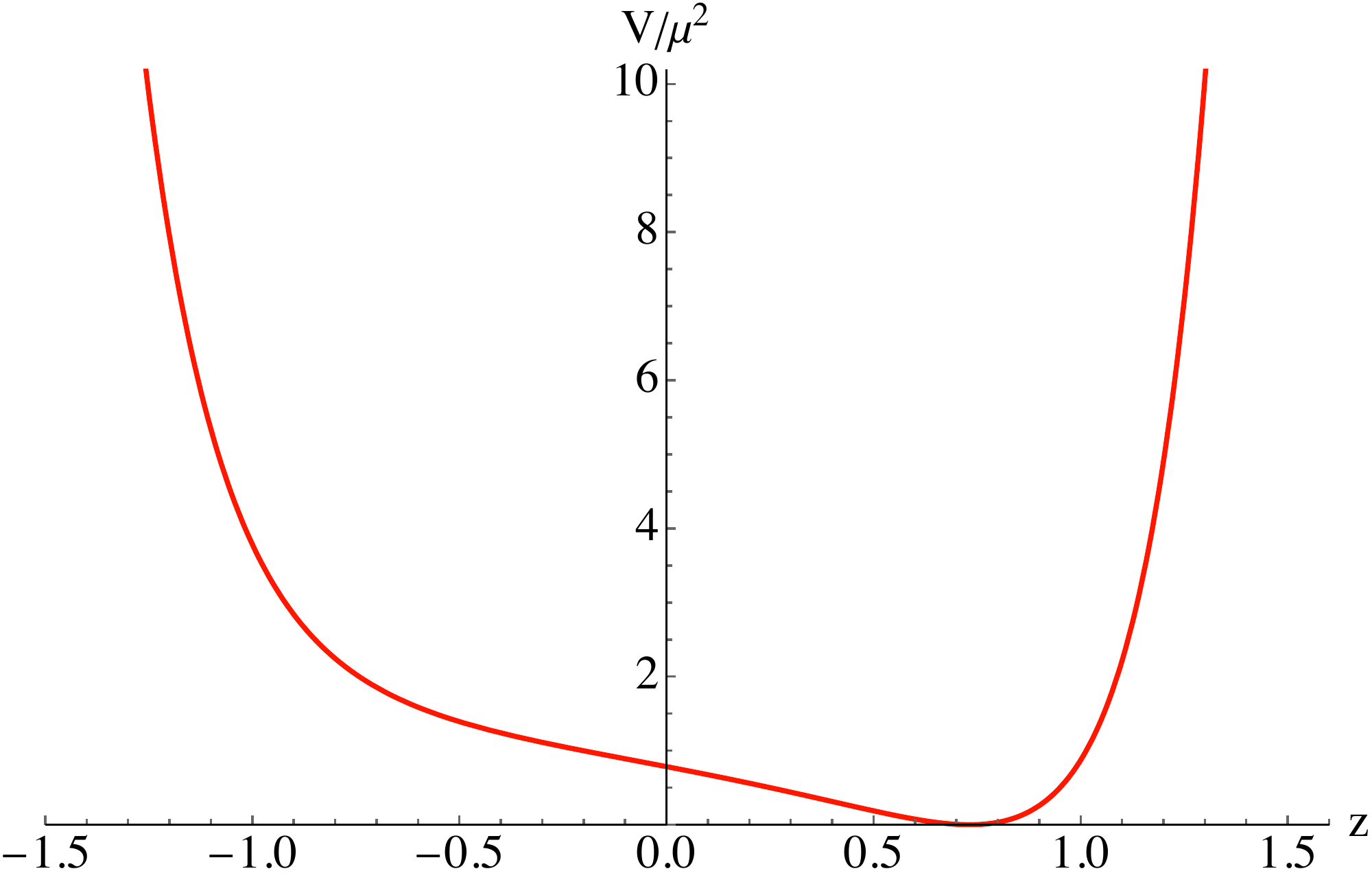,width=7.7cm}}
\vspace*{8pt}
\caption{\it The effective potential in the simplest Polonyi model of supersymmetry breaking
in supergravity. \label{f1}}
\end{figure}

The simplest example of a model that breaks supersymmetry and allows $V = 0$ is the Polonyi model \cite{pol}.
The model is based on adding a single chiral 
superfield that breaks supersymmetry spontaneously through the super-Higgs mechanism~\cite{cremmer,superh},
which has two
physical scalar fields whose fermionic partners are eaten by the gravitino. 
In the simplest version of the model, the superpotential
is separable in  the Polonyi field, $z$, and the matter fields, $\phi^i$:
\beq 
W(z,\phi^i) = f(z) + g(\phi^i) \, ,
\label{sep}
\eeq
with the particular choice
\beq
f(z) = \mu(z + \zeta) \, ,
\label{polonyi}
\eeq
where $\zeta$ is a constant. 
Ignoring for the moment the matter fields,
the potential for $z$ is
\beq
V(z,z^*) = e^{zz^*} \mu^2 \left[ |1 + z^* (z + \zeta)|^2 - 3|(z+\zeta)|^2 \right] \, ,
\label{VZ}
\eeq
where we have assumed a minimal supergravity framework, i.e., $K = zz^*$.
Minimizing the potential and insisting that $V(\langle z \rangle) = 0$
(which requires fine-tuning), 
we find $\langle z \rangle = \sqrt{3} - 1$ and $\zeta = 2 - \sqrt{3}$. 
The Polonyi potential is shown in Fig.~\ref{f1}. 

In this example, the masses of the two real scalars,
denoted by $A$ and $B$, are 
\beq
m_A^2 = 2\sqrt{3} m_{3/2}^2 \qquad m_B^2 = 2(2-\sqrt{3}) m_{3/2}^2 \, ,
\label{eq:maandmb}
\eeq
where the gravitino mass
\beq
m_{3/2} = e^{\langle G \rangle/2} = e^{2-\sqrt{3}} \mu \, .
\eeq
These satisfy the mass relation $m_A^2 + m_B^2  = 4 m_{3/2}^2$,
which is a consequence of the supertrace formula in supergravity~\cite{cremmer}. 

Including now the matter fields, one can calculate their soft supersymmetry-breaking mass terms~\cite{bfs}
by evaluating the potential at $\langle z \rangle$ and 
dropping terms in the potential that are of dimension higher than four, 
as these would scale with inverse powers of the Planck mass. The scalar potential then becomes
\beq
V =
m_{3/2} e^{(2 - \sqrt{3})} \left(\phi^i \frac{\partial g}{\partial \phi^i} - \sqrt{3} g
+{\rm h.c.} \right)  + m_{3/2}^2 \phi^i \phi_i^* \, .
\label{cpot}
\eeq
Rescaling the superpotential by a factor $e^{\sqrt{3}-2}$, and noting that $\sum \phi \partial g / \partial
\phi = 3 g$ for trilinear terms and $\sum \phi \partial g / \partial
\phi = 2 g$ for bilinear terms, we can read off the soft masses
\beq
 m_0 = m_{3/2} \ , \qquad\,\ B_0= (2-\sqrt{3}) m_{3/2}\ , \qquad  \,\  A_0= (3-\sqrt{3}) m_{3/2} \, ,
 \label{softpol}
\eeq
where $m_0$ is a universal soft scalar mass, $A_0$ is universal 
soft trilinear term, and $B_0$ is a universal soft bilinear term.

This simple paradigm for supersymmetry breaking has important consequences for the minimal supersymmetric extension of the Standard Model (MSSM). The soft masses in Eq.~(\ref{softpol}) represent universal boundary conditions for all scalar masses, $A$-terms and $B$-terms. In the constrained MSSM (CMSSM)~\cite{cmssm,elos,eelnos,eemno,eeloz,ehow++},
all scalars are assumed to be universal at some high energy scale 
often taken to be the GUT scale, $A$-terms are left free but universal, and the $B$-term (there is only one in the MSSM)
is obtained from the minimization of the Higgs potential
when $\tan \beta$ (the ratio of the two Higgs VEVs) is taken as a free parameter. 

A supersymmetry-breaking gaugino mass requires a non-trivial gauge kinetic function for a canonically-normalized gauge field,
\beq
m_{1/2} = \left| \frac12 e^{G/2} \frac{{f^*_i}}{{\rm Re} f} (G^{-1})^i_j G^j \right| \, ,
\label{m12}
\eeq
where we have assumed a universal gauge kinetic function, $f_{\alpha\beta} = f \delta_{\alpha \beta}$.

In addition to the gaugino mass, $m_{1/2}$, 
$m_0$, $A_0$, and $\tan \beta$ make up the four continuous free input parameters of the CMSSM. The boundary conditions in Eq.~(\ref{softpol}) are more restrictive, as $m_0$, $A_0$, and $B_0$
are all determined by the gravitino mass. Indeed, the relation $B_0 = A_0 - m_0$, is a common feature of many models based on supergravity \cite{vcmssm}. 

However, this simple paradigm for supersymmetry breaking is not without problems. 
In particular, the potential shown in Fig.~\ref{f1} has a serious cosmological 
problem~\cite{polprob} of excess entropy production. 
Since we expect $\mu$ to be of order the weak scale whereas the VEV
of $z$ is of order the Planck scale, the potential is very flat. This means 
that if $z$ is displaced from its minimum after inflation (and we would expect an $\mathcal{O}(M_P)$
displacement), the subsequent evolution of $z$
would lead to huge entropy generation. 
The problem appears when the $z$ field
begins oscillating about its minimum, which occurs when the Hubble parameter 
drops to $H\sim m_z \sim \mu$, {where $m_z$ corresponds to $m_A$ or $m_B$ in Eq.~\eqref{eq:maandmb}. } At this time,
the Universe becomes matter-dominated by Polonyi oscillations until they decay when 
$H \sim \Gamma_z \sim m_z^3/M_P^2$. This leads to late reheating and an 
entropy increase by a factor $M_P/\mu \sim 10^{16}$. 
Furthermore, the late decay almost inevitably leads to an overproduction of 
cold dark matter in the form of the
lightest supersymmetric particle (LSP)~\cite{myy}.
We note, however, that the Polonyi problem can be alleviated by a mechanism of 
strong stabilization~\cite{ego}, as discussed in Section \ref{sec:stable} below.

The gravitino poses other cosmological problems for supergravity models of cosmology, stemming from the
abundance of gravitinos produced after inflation\cite{weinberg,elinn,nos,ehnos,kl,ekn,Juszkiewicz:gg,mmy,Kawasaki:1994af,Moroi:1995fs,enor,Giudice:1999am,bbb,cefo,kmy,stef,Pradler:2006qh,ps2,rs,kkmy,egnop,Garcia:2018wtq}. If the gravitino is the LSP,
the relic gravitino abundance could exceed the permitted density of cold dark matter, depending on its
mass. On the other hand, if the gravitino is not the LSP, the fact that the gravitino 
couplings to other particles are Planck-suppressed implies that its lifetime for
decays into other particles may be quite long: $\tau \sim M_P^2/m_{3/2}^3$. 
Also in this case the mass and abundance of the gravitino are constrained,
by experimental limits on late-decaying particles, particularly from big bang nucleosynthesis (BBN) \cite{bbn,Kawasaki:1994af,cefo,ceflos,ps,SFT,kkmy,Kawasaki:2017bqm}.

It is easy to break supersymmetry in no-scale supergravity. Even in the minimal SU(1,1)/U(1) case,
simply taking a constant superpotential, $W = \mu$,
leads to a non-zero gravitino mass:~\footnote{We point out that simply adding a constant
superpotential in minimal supergravity does not break supersymmetry. Rather,
minimization of the potential in this case leads to a supersymmetry-preserving AdS vacuum.}
\beq
m_{3/2} = \frac{\mu}{(T+T^*)^{3/2}} \, ,
\eeq
whereas the scalar potential vanishes (as in Eq.~(\ref{V0})). Hence the magnitude of
the gravitino mass is undetermined so long as the modulus $T$
remains unfixed. On the other hand, in this case there is no supersymmetry breaking in the matter sector:
\beq
 m_0 = 0 \ , \qquad\,\ B_0= 0\ , \qquad  \,\  A_0= 0 \, .
 \label{softns}
\eeq
We discuss later other mechanisms for breaking supersymmetry in the matter sectors of
no-scale models, and how the Polonyi and gravitino problems may be avoided. 

\section{Introduction to Inflation in Supergravity}

\subsection{Basic Principles of Models of Inflation}

Before focusing on supersymmetric models of inflation,
we first review briefly the dynamics of single-field models of inflation.

The contribution to the energy-momentum tensor of the scalar field $\phi$ is
\beq
	T_{\mu\nu}  
	    = \partial_{\mu} \phi \partial_{\nu} \phi 
- \frac12 g_{\mu\nu}  \partial_{\rho} \phi
\partial^{\rho} \phi + g_{\mu\nu}V(\phi)	\, .
\eeq
Assuming that it may be treated as a perfect fluid, we can express as follows the 
energy density $\rho$ and pressure $p$ due to the scalar $\phi$:
\begin{eqnarray}
	\rho & = & \frac12 {\dot{\phi}}^2  + \frac12 R^{-2}(t)
(\nabla\phi)^2  + V(\phi) \, ,		\\
	p & = & \frac12 {\dot{\phi}}^2  - \frac16 R^{-2}(t)
(\nabla\phi)^2  - V(\phi)	\, ,
\label{prho}
\end{eqnarray}
where $R(t)$
is the cosmological scale factor.
Ignoring the spatial-gradient terms (which is appropriate as the Universe becomes almost homogeneous 
as it expands),
the equation of motion of $\phi$ obtained from the conservation of $T_{\mu\nu}$ is
\beq
{\ddot \phi} + 3 H {\dot \phi} + \frac{\partial V}{\partial \phi} = 0 \, ,
\label{sevol}
\eeq
where $H = {\dot R}/R$ is the Hubble parameter.
If we 
approximate $\partial V/\partial \phi \simeq (\partial^2 V/\partial \phi^2) \phi = m^2(\phi) \phi$, the solution to the equation of motion when $|m^2| \ll H^2$ is $\phi \sim \exp(|m^2|t/3H)$, and the field
evolves very slowly over a time period $H \tau \sim 3 H^2/|m^2|$, during which the 
universe expands (near-)exponentially. Enough inflation
is obtained if the potential is sufficiently flat, i.e., $|m^2|$ is small enough. 

\subsection{Simple Supergravity Models of Inflation}
\label{simple}

Restricting our attention to theories of inflation in the context of 
supersymmetry, for the reasons discussed earlier we focus on supergravity models.
These include some Planck-scale effects which may be important for inflation and, as already discussed,
make possible the breaking of supersymmetry while (almost) cancelling the cosmological constant. 

\begin{figure}[ht]
\centerline{\psfig{file=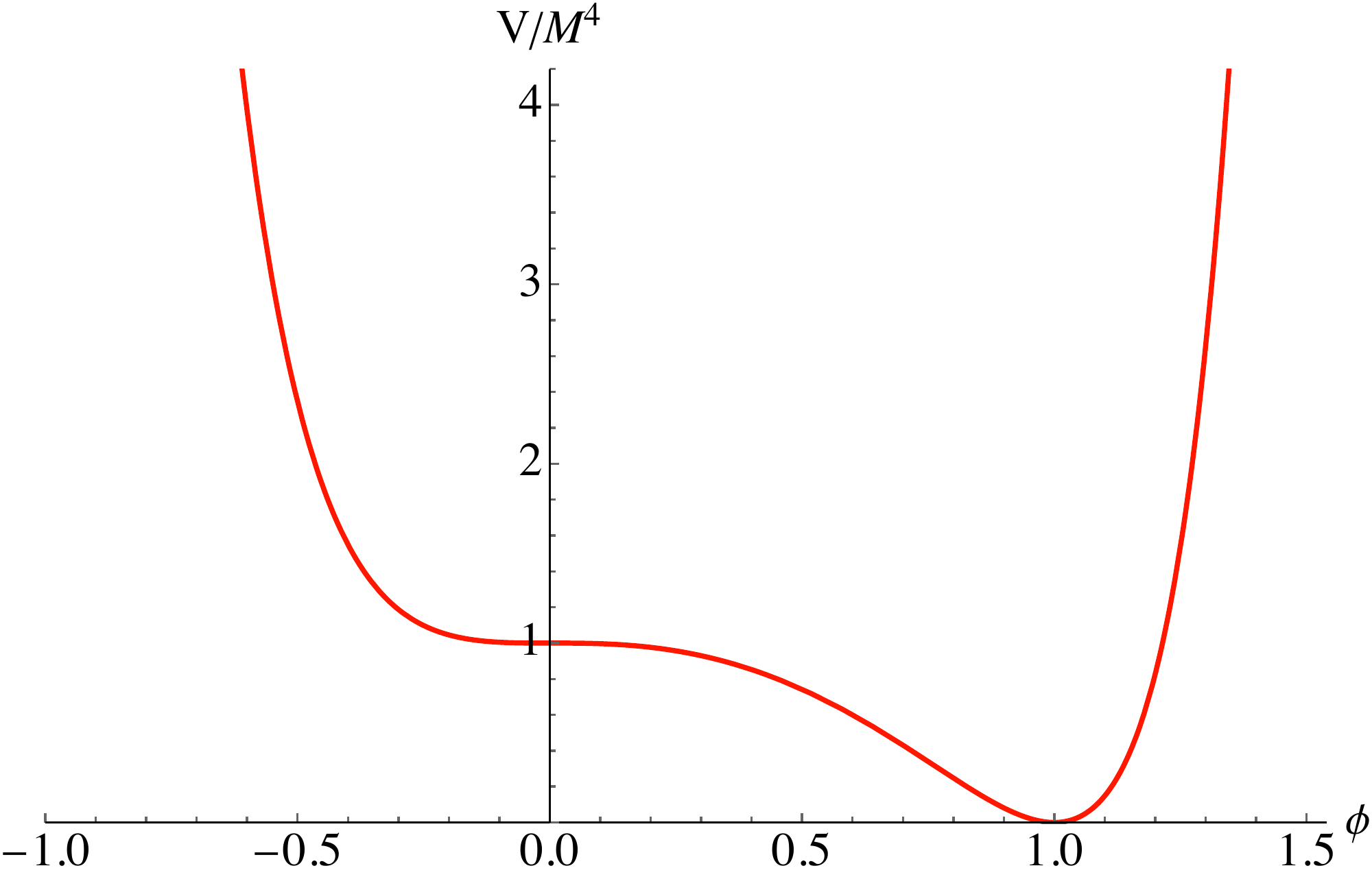,width=7.7cm}}
\vspace*{8pt}
\caption{\it The inflaton potential in a simple model~\cite{hrr} based on minimal supergravity. \label{fig:hrr}}
\end{figure}

We begin with the simplest such model, which is based on a single chiral superfield, 
$\phi$, the inflaton \cite{nos},
in minimal supergravity with $K = \phi \phi^*$. 
One can consider a general polynomial form for the superpotential \cite{nost},
the simplest being~\cite{hrr}
\beq
W = M^2 (1-\phi)^2 \, ,
\label{hrr}
\eeq
which leads to 
\begin{eqnarray}
V &  = & M^4 e^{|\phi|^2} \left[ 1+ |\phi|^2 - (\phi^2 + {\phi^*}^2) - 2  |\phi|^2 (\phi + {\phi^*}) +5  |\phi|^4 \right. \nonumber \\
& & + \left.  |\phi|^2 (\phi^2 + {\phi^*}^2) - 2  |\phi|^4 (\phi + {\phi^*}) + |\phi|^6 \right] \, .
\end{eqnarray}
This can be expanded about the origin in the real direction $\phi = \phi^*$ to give 
\beq
V \simeq M^4 \left(1 - 4 \phi^3 + \frac{13}{2} \phi^4 + \dots \right) \, ,
\eeq
which is shown in Fig. \ref{fig:hrr}. 
This is an example of new inflation~\cite{new} driven by the cubic term,
and the mass scale $M \sim 10^{-4}$ is determined by normalization of the CMB fluctuation spectrum \cite{fluct}. 
We note that, although the theory defined by (\ref{hrr})
is constructed to avoid the $\eta$-problem, a generic inflationary model is in general plagued by the problem of
large masses. This simple model is an example of ``accidental" inflation~\cite{accident}, as the ratio of the constant 
and linear terms in (\ref{hrr}) must be equal to 1 to very high accuracy in order to avoid the $\eta$-problem.

It is possible to construct many more examples
of inflationary models by adding an auxiliary chiral field, $S$~\cite{kyy,rube,klor}. 
Consider, for example
\beq
K = S S^* -\frac12 (\phi - \phi^*)^2 \, ,
\eeq
which can still be viewed as minimal because $K^i_j =  \delta^i_j$.
Then, for the simple choice
\beq
W = S f(\phi)
\eeq
one finds\beq
V = \left| f(\phi) \right|^2 \, ,
\eeq
for $S = {\rm Im}\, \phi = 0$,
and one can easily generate any scalar potential that is a perfect square. 

Given the relative ease of constructing inflationary models in minimal supergravity,
it is natural to ask about the possibilities in the context of no-scale supergravity \cite{GL,KQ,EENOS,otherns}.
One interesting example is given by~\cite{EENOS}
\beq
W = M^2 (\phi - \phi^4/4) \, ,
\label{eenos}
\eeq
which gives
\beq
\hat{V} = M^4 |1 - \phi^3|^2 \, ,
\eeq
a potential that is very similar to that shown in Fig.~\ref{fig:hrr}.

\section{Inflationary Observables}

\begin{figure}[!ht]
\centerline{\psfig{file=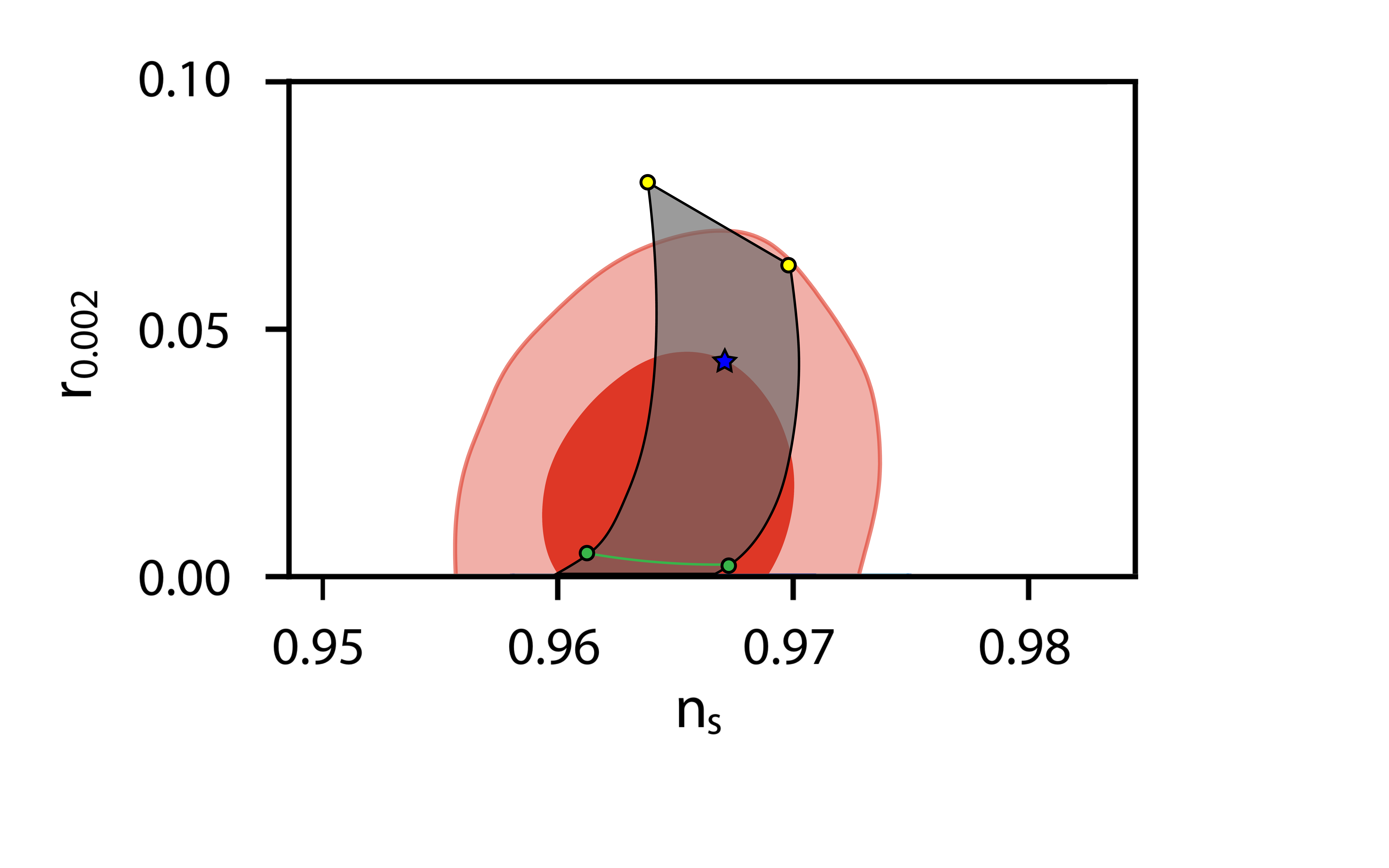,width=10cm}}
\vspace*{-0.4in}
\caption{\it Plot of the CMB observables $n_s$ and $r$.  
The red  shadings correspond to the 68\% and 95\%
confidence level regions from Planck data combined with BICEP2/Keck results~\cite{planck18}.
The pairs of dots are the predictions of the $\alpha$-Starobinsky potential~(\ref{alstaro}) discussed later for $N_* = 50$ (left) and 60 (right). 
The upper (lower) pair of yellow (green) dots are the predictions when $\alpha = 100$ 
($\alpha  = 1$, corresponding to the Starobinsky model~(\ref{staropot})), while the lower end of the swath represents the 
cosmological observables in the limit $\alpha \rightarrow 0$. 
The 68\% upper bound $r_{0.002} \lesssim 0.041$, indicated by the blue star, is attained
for $\alpha \sim  27$ when $n_s \sim 0.967$, for a nominal choice of $N_* \simeq 55$.  \label{fig:plancklimits}}
\end{figure}

Quantum fluctuations during inflation generate scalar density and tensor metric perturbations
that leave imprints on the CMB and large-scale structure \cite{MC,pert}. 
The overall scale of the potential is related to the amplitude, $A_s$, of the power spectrum of scalar
perturbations in the CMB,~\cite{planck18}
and other CMB observables include the tilt in the spectrum of scalar perturbations, 
$n_s$, and the tensor-to-scalar ratio, $r$. It is convenient to
characterize these in terms of the slow-roll parameters $\epsilon$ and $\eta$~\cite{reviews}. 
For a given single-field scalar potential, these are given by
\begin{equation} 
\epsilon \; \equiv \; \frac{1}{2} M_{P}^2 \left( \frac{V'}{V} \right)^2 ; \; \;  \eta \; \equiv \; M_{P}^2 \left( \frac{V''}{V} \right)   \, ,
\label{epsilon}
\end{equation}
where, here and subsequently, the prime denotes a derivative with respect to the inflaton field $\phi$.
The tilt in the spectrum of scalar perturbations, $n_s$, and the tensor-to-scalar ratio, $r$
are the principal CMB observables~\cite{planck18,rlimit}, and can be expressed as follows in terms of the slow-roll parameters at the pivot scale $k_*=0.05 \,{\rm Mpc}^{-1}$:
\begin{eqnarray}
{\rm Amplitude~of~scalar~perturbations}~A_s:\; 
A_s \;& = &\; \frac{V_*}{24 \pi^2 \epsilon_* M_{P}^4 } \simeq 2.1 \times 10^{-9} \, , \label{As} \\ \notag
\hspace{-15mm} {\rm Scalar~spectral~tilt}~n_s:\;  n_s \; & \simeq &\; 1 - 6 \epsilon_* + 2 \eta_*\\
& = &\; 0.965 \pm 0.004 \; (68\%~{\rm CL}) \, ,
\label{ns} \\
\hspace{-5mm} {\rm Tensor\mbox{-}to\mbox{-}scalar~ratio}~r:\;  r \; &
\simeq & \; 16 \epsilon_* < 0.061 \; (95\%~{\rm CL}) \, . \label{r}
\label{observables}
\end{eqnarray}
The region of the $(n_s, r)$ plane allowed by the Planck data is shown in Fig.~\ref{fig:plancklimits}.
In addition to the above expressions, we note that 
the number of $e$-folds, $N_i$, of inflation between the initial and final values of the inflaton field $\phi_{i, \, \rm{end}}$
is given by the formula
\begin{equation}
N_i \;\equiv\; \ln\left(\frac{R_{\rm{end}}}{R_i}\right) \; = \; \int_{t_i}^{t_{\rm{end}}} H dt \; \simeq \;  - \int^{\phi_{\rm{end}}}_{\phi_i} \frac{1}{\sqrt{2 \epsilon}} \frac{d \phi}{M_P} \, .
\label{efolds}
\end{equation}
The number of $e$-folds between when the scale $k_*$ exits the horizon and the end of inflation is denoted by $N_*$. Typical values of $N_*$ are in the range $\sim$ 50--60, dependent on the mechanism ending inflation~\cite{LiddleLeach,MRcmb,planck18}.

We can now test the simple models of inflation discussed in the previous Section. 
For the potential determined by Eq.~(\ref{hrr}),
the amplitude of density fluctuations (\ref{As}) implies that $M \sim 10^{-4} M_{P}$,
as expected. In this case the slow-roll parameter $\epsilon \simeq 3.6 \times 10^{-10}$ is very small, yielding a value of the tensor-to-scalar
ratio that is allowed, but unobservably small. However, the scalar tilt 
in this model is $n_s \simeq 1+2 \eta \simeq 0.928$, which is strongly excluded by the Planck data~\cite{planck18}.
Similarly,  the  no-scale potential defined by (\ref{eenos}) gives $\epsilon \simeq 1.5 \times 10^{-9}$,
with similar values of $M$ and $\eta$, resulting again in $n_s \simeq 0.928$, excluding the model. 
This value of $n_s$ does not even lie with the range shown in Fig.~\ref{fig:plancklimits}.

\begin{figure}[!ht]
\centerline{\psfig{file=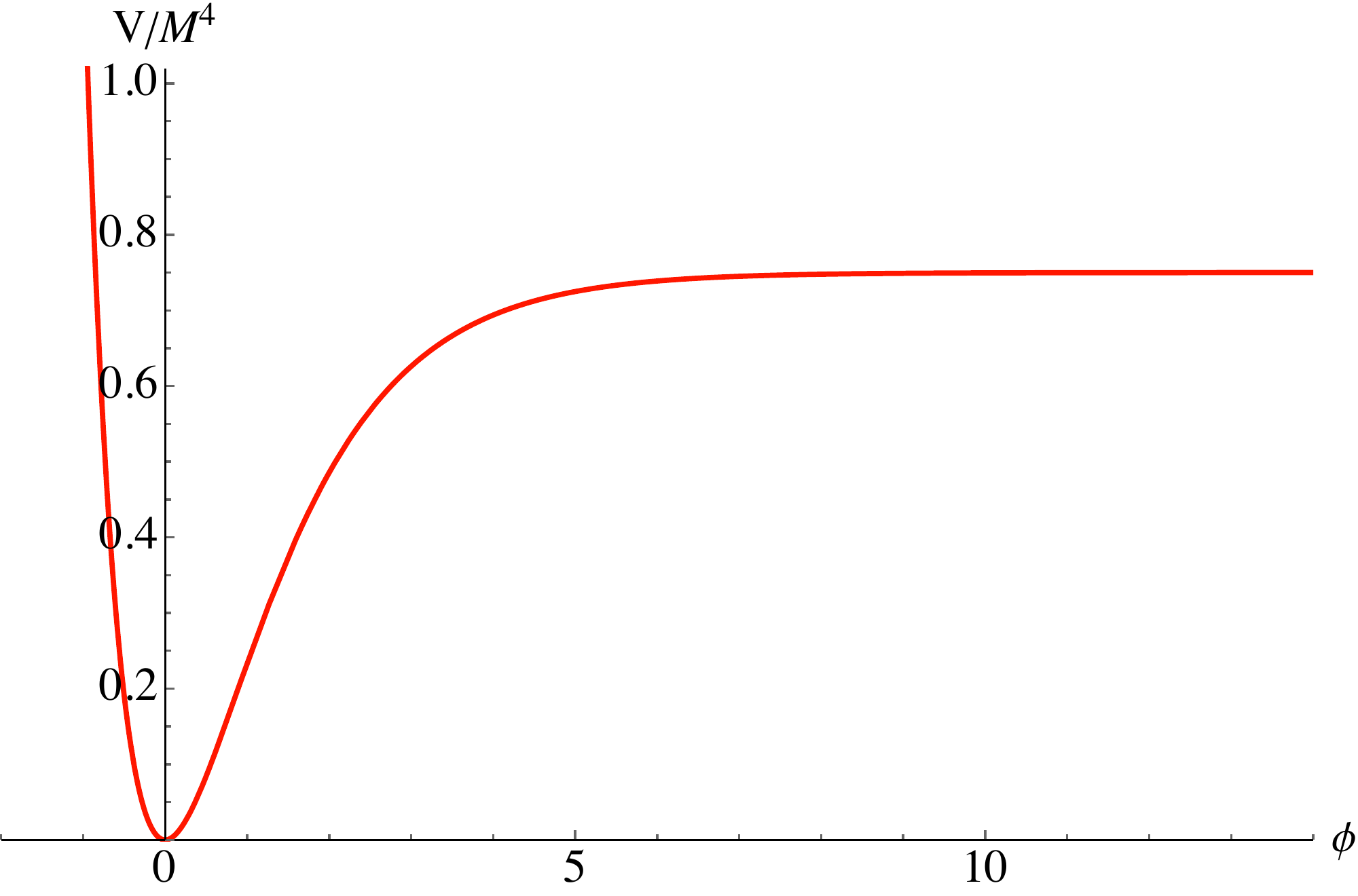,width=7.7cm}}
\caption{\it The effective scalar potential in the Starobinsky model of inflation~\cite{Staro}. \label{fig:staro}}
\end{figure}

In contrast to the models discussed above, one of the first models 
of inflation, namely
the Starobinsky model~\cite{Staro},
yields a value of $n_s$ that is in excellent agreement with observation,
and a value of $r$ that is testable in the next generation of CMB experiments.
As originally written, the model was based on an
$R + R^2$ theory of gravity. However, a suitable conformal transformation~\cite{WhittStelle}
brings the theory into the form of a theory with an Einstein-Hilbert action for gravity,
and a canonical scalar field with a scalar potential of the form 
\beq
V = 3 M^2 e^{-\sqrt{2/3} \phi} \sinh^2 (\phi/\sqrt{6}) = \frac34 M^2 \left(1 - e^{-\sqrt{2/3} \phi} \right)^2 \, ,
\label{staropot}
\eeq
as depicted in Fig.~\ref{fig:staro}.

This model was the first to predict a slightly red scalar perturbation spectrum ($n_s < 1$) \cite{MC}.
It is easy to determine in analytic form the slow-roll parameters for this potential:~\cite{eno6}
\begin{eqnarray}
A_s &  = &  \frac{3 M^2}{8\pi^2} \sinh^4 (\phi/\sqrt{6}) \, , \\
\epsilon & = & \frac13 \csch^2 (\phi/\sqrt{6}) e^{-\sqrt{2/3}\phi} \, , \\
\eta & = &  \frac13 \csch^2 (\phi/\sqrt{6}) \left( 2 e^{-\sqrt{2/3}\phi} -1\right) \, .
\end{eqnarray} 
For $N_*=55$, we find $M = 1.25 \times 10^{-5} M_{P}$, $n_s = 0.965$, and $r = 0.0035$.
The line between the two green dots in Fig.~{\ref{fig:plancklimits} corresponds to the predictions
of the Starobinsky model for $N_* = 50$ to 60.  We make the connection between this model and 
no-scale supergravity in the next Section. 

\section{A No-Scale Starobinsky model}
\label{wzeno6}

As already mentioned, the Starobinsky inflationary model based on the potential in Eq.~(\ref{staropot})
had its origins in higher-derivative gravity.
It can, however, be seen to arise rather 
simply and directly within the no-scale supergravity framework.~\cite{eno6} 

Our starting point is the no-scale supergravity scalar potential given in Eqs.~(\ref{VJ}) and (\ref{effVJ}). We see immediately that it is not possible to construct a Starobinsky-like model using only a single field, $T$.
In that case, the scale invariance of the Starobinsky potential at large field values would require a constant potential at large $T$, which is possible only if
the superpotential scales as $W \sim T^{3/2}$~\cite{eno7}. In that case, however, the leading term in ${\hat V}$ would be negative, $\hat{V} \sim -\frac32 T^2$. Therefore a minimal model requires two fields, which we take as $T$ and the inflaton, $\phi$. 

For now, we assume that the modulus is fixed by some unspecified mechanism with $\langle T \rangle = 1/2$ for illustration (the value of $\langle T \rangle$ is unimportant and its stabilization is discussed in Section \ref{sec:stable} below). 
Further, we postulate the following Wess-Zumino form for the superpotential~\cite{eno6}
\begin{equation}
f(\phi) = \frac{M}{2}\phi^2 - \frac{\lambda}{3}\phi^3  \, ,
\label{wi}
\end{equation}
 which is a function of $\phi$ alone. In this case, $W_T = 0$, and we see from Eq.~(\ref{effVJ}) that $\hat{V} = |W_\phi|^2$.
 The resulting potential is 
 \beq
V(\phi) = M^2 \frac{|\phi|^2 |1-\lambda \phi/M|^2}{(1-|\phi|^2/3)^2}. 
\eeq
We can rewrite the kinetic terms in Eq.~(\ref{LmanyJ}) as 
\begin{eqnarray}
\hspace{-5mm}
{\cal L}_{KE} \; =  \; \left( \partial_\mu \phi^*, \partial_\mu T^* \right) \left(\frac{3}{(T + T^* - |\phi|^2/3)^2} \right)
 \left( \begin{array}{cc} \frac{(T + T^*)}{3} & - \phi/3 \\ - \phi^*/3 & 1 \end{array} \right)
\left( \begin{array} {c} \partial^\mu \phi \\ \partial^\mu T \end{array} \right) \, ,
\label{no-scaleL}
\end{eqnarray}
indicating that neither $T$ nor $\phi$ is normalized canonically. 
When $T$ is fixed, we can define the canonically-normalized field
$\chi$:
\beq
\chi \; \equiv \; \sqrt{3} \tanh^{-1} \left( \frac{\phi}{\sqrt{3}} \right) \, .
\eeq
Decomposing $\chi$ into its real and imaginary parts:
$\chi = (x + iy)/\sqrt{2}$, we find that the potential is minimized for $y=0$, and that
the potential in the real direction
takes the same form as the Starobinsky potential in Eq.~(\ref{staropot}) when $\lambda = M/\sqrt{3}$.

To get a feel for how ``accidental" this result is, the potential is plotted for several values of $\lambda/M \simeq 1/\sqrt{3}$
in Fig.~\ref{fig:eno6-1}. Requiring $N_*=50$ to 60 $e$-folds specifies the value of the field $x$ at the beginning of inflation. For example, the nominal choice $N_*=55$ corresponds to
$x=5.35$ and, as one can see in Fig.~\ref{fig:eno6-1}, inflation is still possible for $\sqrt{3} \lambda/M$ slightly less than 1.
However, deviations from 1 by more than a few parts in $10^{-4}$ would not provide a suitable inflationary potential, as seen in
Fig.~\ref{fig:eno6-2}.

\begin{figure}[ht]
\vspace*{-2.5in}
\centerline{\psfig{file=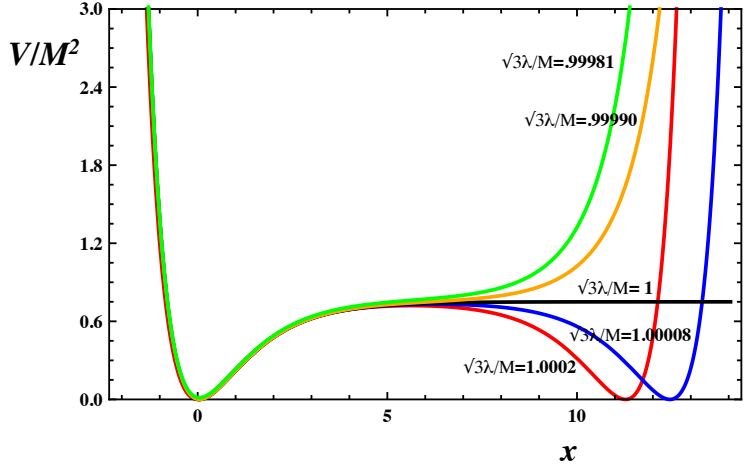,width=15cm}}
\vspace*{-2.5in}
\caption{\it Starobinsky-like inflationary potential in the no-scale supergravity model with superpotential (\ref{wi}) for choices of $\lambda \sim M/\sqrt{3}$, as indicated. \label{fig:eno6-1}}
\end{figure}

Fig.~\ref{fig:eno6-2} displays the predictions for $(n_s, r)$ in this model
for five choices of the coupling $\lambda$ that yield $n_s \in [0.93,1.00]$ and $N_* \in [50, 60]$. 
The last 50 to 60 $e$-folds of inflation arise as $x$ rolls to zero from $\sim 5.1$--5.8, with
the exact value depending on $\lambda$ and $N_*$.
We see again that the values of $\lambda$ are constrained to be close to the critical value $M/\sqrt{3}$,
for which we find
extremely good agreement with the Planck determination of $n_s$.  
The values of $r$ are well below the current experimental limit\cite{rlimit}, varying over the range 0.0012 -- 0.0084,
in the models displayed, with $r \simeq 0.003$ for $\lambda = M/\sqrt{3}$.

\begin{figure}[ht]
\vspace*{-2.5in}
\centerline{\psfig{file=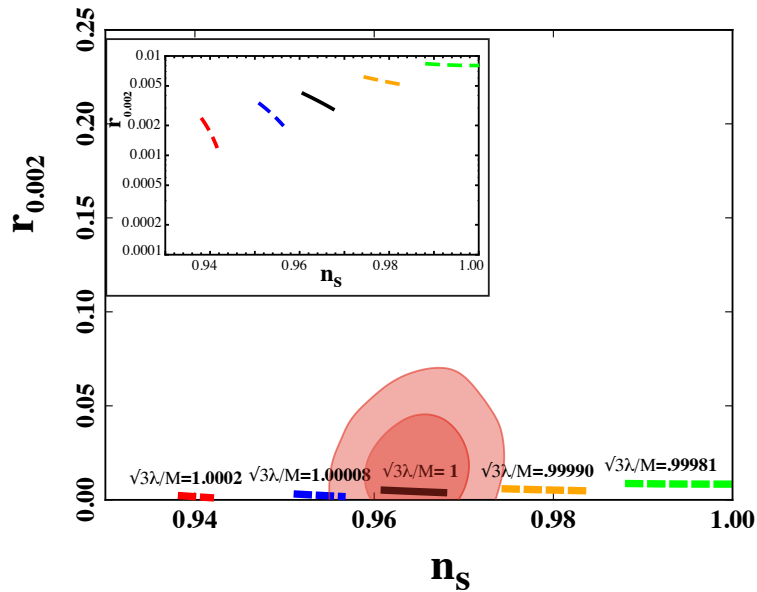,width=15cm}}
\vspace*{-2.0in}
\caption{\it Predictions 
for the tilt $n_s$ in the spectrum of scalar perturbations and for the
tensor-to-scalar ratio $r$, compared with the 68 and 95\%
CL regions from Planck data combined with BICEP2/Keck 
results~\cite{planck18}. 
In the main panel the lines are labelled by the values of $\sqrt{3} \lambda/M$ (in Planck units) assumed
in each case. In the inset, the same cases are shown on a log scale to display more clearly the values of $r$.
\label{fig:eno6-2}}
\end{figure}

We turn in later Sections to some more formal aspects of no-scale supergravity and its relationship to higher-derivative gravity
and the Starobinsky model, to some generalizations of the superpotential (\ref{wi}) and to aspects of the
phenomenology of no-scale models of inflation.
However, we comment first how modifying the no-scale K{\" a}hler potential 
can affect the observables, causing deviations from the Starobinsky predictions.~\cite{eno7} 

The Starobinsky potential can be expressed in
the simple form
\begin{equation}
V \; = \; A \left( 1 - e^{-Bx} \right)^2 \, ,
\label{StaroAB}
\end{equation}
with $B = \sqrt{2/3}$, and we note that
the inflationary predictions are
derived in the large-field regime where the leading non-constant term is $\propto e^{-Bx}$.
We considered in Ref.~\refcite{eno7} phenomenological generalizations of (\ref{StaroAB}) in which
\begin{equation}
V \; = \; A \left( 1 - \delta e^{-Bx} + {\cal O}(e^{-2Bx}) \right) \, ,
\label{StaroAlambdaB}
\end{equation}
with $\delta$ and $B$ treated as free parameters that  may deviate from the
Starobinsky values $\delta = 2$ and $B = \sqrt{2/3}$. At leading order when $e^{-Bx}$ is small, one finds
\begin{eqnarray}
n_s & = & 1 - 2 B^2 \delta e^{-Bx} + \cdots \, , \nonumber \\
r & = & 8 B^2 \delta^2 e^{-2Bx} + \cdots \, , \nonumber \\
N_* & = & \frac{1}{B^2 \delta} e^{+ Bx} + \cdots \, ,
\label{predictions}
\end{eqnarray}
and we have
\begin{equation}
n_s \; = \; 1 - \frac{2}{N_*} + \cdots \; ,\ \ r \; = \; \frac{8}{B^2 N_*^2} + \cdots \, .
\label{relations}
\end{equation}
Requiring $N_* = 50$ to 60 yields the characteristic predictions $n_s = 0.960$ to 0.967, independent of $B$,
and the Starobinsky choice $B = \sqrt{2/3}$ yields $ r = 12/N_*^2 = 0.0048$ to 0.0033. Since the
experimental upper limit is $r < 0.06$,~\cite{planck18} it is clear that substantial deviations from the
Starobinsky value of $B$ are currently allowed.
We note finally that the predictions (\ref{relations}) are independent of $\delta$. 

Different values of $B$ may be obtained by considering generalized K{\" a}hler potentials~\cite{eno7}
that include an inflationary sector:
\begin{equation}
\label{Kinf}
    K \; \ni \; K_{\rm inf} \; = \; - 3 \, \alpha \ln (T + T^*)  \, ,
\end{equation}
where $\alpha \ne 1$ in general. In such a case
\begin{equation}
    B \; = \; \sqrt{\frac{2}{3\alpha}} \, ,
    \label{avatarB}
\end{equation}
leading to the prediction
\begin{equation}
\label{rstaro}
    r \; \simeq \; \frac{12 \alpha}{N_*^2} .
\end{equation}
For example, a simple string compactification contains
3 complex moduli $T^i$ whose product is the volume modulus $T$ considered above. Inflation
might be driven by just one or a pair of the $T^i$, in which case $\alpha = 1/3$ or $2/3$.
In these examples $r$ is decreased below the Starobinsky prediction, rendering its
experimental detection more challenging. However, it is interesting that, within
the class of string-inspired no-scale models discussed here, measuring $r$ might
cast light on the dynamics of string compactification.~\cite{eno7}

One may also consider models with larger values of $\alpha$, as illustrated in Fig.~\ref{fig:plancklimits},
where we see that values of $\alpha \lesssim 27$ are allowed by the current upper limit on $r$.
However, we note that, for large values of $\alpha \gg 1$, sub-asymptotic corrections to the leading-order
predictions (\ref{predictions}, \ref{relations}) become important, causing the curvature in
the predicted bands in Fig.~\ref{fig:plancklimits}.

We discuss such models with $\alpha \ne 1$ in more detail in subsequent Sections.

\section{On the Structure of No-Scale Supergravity}
\label{structure}

Motivated by this phenomenological connection to the Starobinsky $R + R^2$
model of inflation, we now go beyond the brief introduction to the minimal and no-scale supergravities
given in Section~\ref{sec:sugrp}.
We explore some more theoretical aspects of the 
construction of the supergravity
Lagrangian, highlighting particular aspects that will 
help in connecting no-scale supergravity with $R^2$ and $R + R^2$ gravity.

As is well known in globally-supersymmetric models, the supersymmetry algebra 
includes auxiliary fields that can be removed using their equations of motion.
The Lagrangian can be written as~\cite{cremmer}
\beq
{\mathcal L} = {\mathcal L}_{\rm aux} + {\mathcal L}^\prime \, ,
\eeq
with
\begin{eqnarray}
 {\mathcal L}_{\rm aux}  & = &  \frac{9}{\Phi} \left| \frac12 g^* + \cdots \right|^2  \frac{3}{\Phi J_{\phi\phi^*} } \left| \frac12 g^* \left( \frac{g^*_{\phi^*}}{g^*} - J_{\phi^*} \right) + \cdots \right|^2  \nonumber \\
 && - \frac{1}{4\Phi} \left[ ( \Phi_{\phi^*} \partial_\mu \phi^* - \Phi_\phi \partial_\mu \phi) + \cdots \right]^2 \, ,
\end{eqnarray}
and 
\beq
{\mathcal L}^\prime  =  - \Phi_{\phi \phi^*} |\partial_\mu \phi|^2 - \frac16 \Phi R \ + \cdots .
\label{Lprime}
\eeq
In writing this greatly abbreviated version of the Lagrangian, 
we have dropped all terms involving fermions (including the gravitino)
and vectors, keeping only the purely scalar part of the Lagrangian coupled to the gravitational curvature $R$.~\footnote{Note that sign differences here relative to those in Ref.~\refcite{cremmer,eno9} are due to our differences in the conventions for $R$ and the metric signature. Here, we are using  the $(+ - - -)$ metric.}
Here, the fields $\phi^i$ correspond to scalars though, for clarity, we suppress the index $i$, 
and $J = 3 \ln (\Phi/3)$. As we will see, the function $J$ can be identified (up to a sign)
as the K\"ahler potential: $J = -K$, while $g$ is related to the superpotential. 

We note that the Lagrangian written this way does not correspond to minimal Einstein gravity, as it contains
a coupling of $R$ to the function $\Phi$. We will return to this form later when we consider 
higher-order theories of gravity. 
We can perform a conformal transformation of the metric: $g_{\mu\nu} \to e^{2\Omega} g_{\mu\nu}$, with 
$e^{2\Omega} =  \frac13 \Phi$, corresponding to $J = 6 \Omega$.  
Thus we can bring the curvature term into its Einstein-Hilbert form
by making a conformal transformation proportional to the K\"ahler potential, leading to 
\beq
- \frac16 \Phi R \to -\frac12 R + \frac34 (\partial_\mu (\ln \Phi))^2 \, .
\eeq
We can now combine ${\mathcal L}^\prime$ with the final term in ${\mathcal L}_{\rm aux} $ to write
the kinetic terms in the Lagrangian as
\begin{eqnarray}
\hspace{-5mm}
{\mathcal L}_{\rm kin}  & = & - \frac12 R  +  \frac34 (\partial_\mu (\ln \Phi))^2  - \Phi_{\phi \phi^*} |\partial_\mu \phi|^2 \left(\frac{3}{\Phi} \right) \nonumber \\ 
&& -  \frac{1}{4\Phi} ( \Phi_{\phi^*} \partial_\mu \phi^* - \Phi_\phi \partial_\mu \phi)^2  \left(\frac{3}{\Phi} \right) \, ,
\label{Lkin1}
\end{eqnarray}
where we have included the appropriate conformal factor in the last two terms. 
We can then rewrite (\ref{Lkin1}) as
\beq
{\mathcal L}_{\rm kin} = - \frac12 R - \left( \frac{3}{\Phi} \Phi_{\phi \phi^*} - \frac{3}{\Phi^2} \Phi_\phi \Phi_{\phi^*} \right) |\partial_\mu \phi|^2 =  - \frac12 R - J^j_i \partial_\mu \phi^i \partial^\mu \phi^*_j \, .
\label{Lkin2}
 \eeq
The identification of the K\"ahler potential $K = -J$ allows us to write the 
kinetic terms in the conventional manner seen in Eq.~(\ref{Lkin3J}), 
recalling that the K\"ahler potential is (up to a constant) the conformal factor: $K = - 6\Omega$.

Next, we consider the remaining highlighted terms in ${\mathcal L}_{\rm aux}$, which we
identify with the scalar potential:
\beq
V = \frac{9}{4 \Phi} |g|^2  \left(\frac{9}{\Phi^2} \right)  + \frac{3}{4\Phi} \frac{|g|^2}{J_{\phi\phi^*}} \left| \frac{g_\phi}{g} - J_\phi \right|^2 \left(\frac{9}{\Phi^2} \right) \, ,
\eeq
where we have included once again the conformal factor needed to write the potential in the Einstein frame. 
We define the function 
\begin{equation}
G \; \equiv \;  -J + \ln \left| \frac{g}{2} \right|^2 = K + \ln~|W|^2 \, ,
\end{equation}
where we have associated the function $g$ with the superpotential: $g = 2 W$. Then, after some simple algebra we
arrive at the expression given in Eq.(\ref{effpot}) for the effective scalar potential.

To obtain minimal supergravity,  we must choose
\beq
\Phi = 3 e^{-\phi \phi^*/3} \, ,
\eeq
which corresponds to the K\"ahler potential given in Eq.~(\ref{Kmin}) and the potential (\ref{sgpotJ}).
There are of course simpler choices for the function $\Phi$. Of particular interest,
will be the class of functions for which there is no kinetic term for scalars in Eq.~(\ref{Lkin1}) 
and hence $\Phi_{\phi \phi^*} = 0$, as appropriate in $R^2$ theories of gravity as we discuss below.
One choice of $\phi$ with this property is
\beq
\Phi = 3 ( \phi + \phi^*) \, ,
\label{confPhi}
\eeq
corresponding to $K = -3 \ln (\phi + \phi^*)$ as in Eq.~(\ref{CFKN}), after 
the identification $\phi \to T$. The scalar potential vanishes as in Eq.~(\ref{V0}),
which is characteristic of no-scale supergravity \cite{no-scale,EKN1,EKN,LN}.

\section{No-Scale Supergravity, $R+R^2$ Inflation, and de Sitter Solutions}
\label{r+r2}

The parallels between no-scale supergravity and $R+ R^2$
models of inflation suggest that there must be a deeper
connection between them, and there are indeed many 
such connections with both $R^2$ and $R+R^2$ 
gravity models~\cite{DLT,eno9}, as we now discuss.
We begin by recalling some basic features of 
$R^2$ gravity and its transformation to Einstein gravity.
We then repeat the procedure for $R + R^2$ gravity,
which leads to the Starobinsky model discussed above.
We also consider the addition to these theories of scalar matter and a scalar potential defined in the conformal frame.

The action for a pure, scale-invariant $R^2$ theory of gravity 
can be written as
\begin{equation}
{\cal A} \; = \;  \frac{1}{2}\int d^4x \sqrt{-g} \alpha R^2 \, ,
\label{pureR2}
\end{equation}
where $\alpha$ is a dimensionless constant. 
This action may be rewritten in the following form, 
by inserting a Lagrange multiplier field $\Phi$:
\begin{equation}
{\cal A} \; = \; - \frac{1}{2}\int d^4x \sqrt{-g} \left( 2 \alpha \Phi R + \alpha\Phi^2 \right) \, .
\label{phiR2}
\end{equation}
We recall at this point the form of the Ricci
curvature term in the original supergravity Lagrangian 
in Eq.~(\ref{Lprime}) with $\alpha = 1/6$.  The action (\ref{phiR2}) may be rewritten in the Einstein-Hilbert form, if
one rescales the metric by a conformal factor $\Omega$, as follows:
\begin{equation}
{\tilde g}_{\mu \nu} \; = \; e^{2\Omega} g_{\mu \nu}  \; = \; {2 \alpha \Phi} g_{\mu \nu} \, .
\label{rescaledmetricR2}
\end{equation}
After the conformal transformation, the curvature
can be rewritten as 
\beq
  R = e^{2\Omega}\left( {\tilde R} - 6 \partial^\mu \Omega \partial_\mu \Omega + 6 \Box \Omega \right)  
\eeq
and, after eliminating the total divergence, 
we see that the action in the Einstein frame is: 
\begin{equation}
{\cal A} \; = \; - \frac{1}{2}\int d^4x \sqrt{- {\tilde g}} \left( {\tilde R} - 6  \partial^\mu \Omega \partial_\mu \Omega + \frac{1}{4 \alpha} \right) \, ,
\label{R2Einstein0}
\end{equation}
or equivalently:
\beq
{\cal A} \; = \; - \frac{1}{2}\int d^4x \sqrt{- {\tilde g}} \left( {\tilde R} - \frac{3}{2} \frac{\partial ^\mu \Phi \partial_\mu \Phi}{\Phi^2} + \frac{1}{4 \alpha} \right) \, .
\label{R2Einstein1}
\end{equation}
After a field redefinition: $\phi \equiv \sqrt{6} \Omega = \sqrt{\frac32}\ln 2 \alpha \Phi$,
the action now takes the simple form
\begin{equation}
{\cal A} \; = \; - \frac{1}{2}\int d^4x \sqrt{- {\tilde g}} \left(  {\tilde R} - \partial ^\mu \phi \partial_\mu \phi + \frac{1}{4 \alpha} \right) \, .
\label{R2Einstein2}
\end{equation}
This is a well-known result~\cite{old,KLT,Alvarez-Gaume:2015rwa}. Pure $R^2$ gravity is equivalent to the conventional Einstein-Hilbert
theory with a massless scalar field $\phi$ and a cosmological constant $\Lambda = 1/8 \alpha$.  Thus, 
the dimensionless parameter $\alpha$ in (\ref{pureR2}) specifies the magnitude of $\Lambda$ in
Planck units.

We can extend this discussion to include
the Starobinsky model of inflation~\cite{Staro}, which was
formulated by
adding the conventional linear Einstein-Hilbert term to the pure $R^2$ action (\ref{pureR2}):
\begin{equation}
{\cal A} \; = \;  \int d^4x \sqrt{-g} \left(- \frac{1}{2} R + \frac{\alpha}{2} R^2 \right) \, .
\label{EH}
\end{equation}
We introduce once again a Lagrange multiplier:  
$ - \alpha R^2 \to 2 \alpha \Phi R + \alpha\Phi^2$ 
and perform a conformal transformation~\cite{WhittStelle,Kalara:1990ar}:
\begin{equation}
{\tilde g}_{\mu \nu}  \; = \; e^{2\Omega} g_{\mu \nu} \; = \;  \left(1 + 2 {\alpha} \Phi \right) g_{\mu \nu} \, ,
\end{equation}
finding
\begin{equation}
{\cal A} \; = \; - \frac{1}{2}  \int d^4x \sqrt{- {\tilde g}} \left[ {\tilde R} - \frac{6 {\alpha}^2}{(1 + 2 {\alpha}  \Phi)^2} \left(\partial^\mu \Phi \partial_\mu \Phi -  \frac{\Phi^2}{6 {\alpha}} \right) \right] \, .
\label{almostStaro}
\end{equation}
Setting $\phi \equiv \sqrt{3/2} \ln \left(1 +  2 { \alpha} \Phi \right)$, 
(\ref{EH}) may be written in the Einstein frame as:
\begin{equation}
{\cal A} \; = \; - \frac{1}{2} \int d^4x \sqrt{- {\tilde g}} \left[ {\tilde R} - \partial^\mu \phi \partial_\mu \phi + \frac{1}{4 {\alpha}} 
\left(1 - e^{-\sqrt{\frac{2}{3}} \phi }
 \right)^2 \right] \, .
\label{FullStaro}
\end{equation}
One recovers the inflationary 
potential (\ref{staropot}) with ${\alpha} = 1/6 M^2$.
The scale invariance of the pure $R^2$ theory (\ref{pureR2})
is broken by the Einstein-Hilbert term 
in (\ref{EH}), and leads to an effective potential
(\ref{staropot}) with a constant, scale-invariant asymptotic
limit that is approached exponentially at a rate controlled by the Planck scale.

We now introduce $N-1$
additional complex fields $\phi^i$ with 
conformal couplings to $R$:
\begin{equation}
{\cal A} \; = \;  - \frac{1}{2} \int d^4x \sqrt{-g} \left[ \delta R - {\alpha} R^2 - 2 \left(\partial^\mu \phi^i \partial_\mu \phi^*_i  + \frac{1}{6} |\phi^i|^2 R \right) \right] \, ,
\label{manyphi}
\end{equation}
where $\delta = 0$ corresponds to the $R^2$ theory 
and $\delta = 1$ corresponds to the Starobinsky model. 
Following the same procedure, we transform to the Einstein frame, 
using a modified conformal factor $\Omega$:
\begin{equation}
{\tilde g}_{\mu \nu}  \; = \; e^{2\Omega} g_{\mu \nu} \; = \; \left(\delta + 2 {\alpha} \Phi - \frac{1}{3} \sum_{i = 1}^{N-1} |\phi^i |^2 \right) g_{\mu \nu}  \, ,
\label{newtilde}
\end{equation}
leading to the following generalization of (\ref{R2Einstein0}):
\begin{eqnarray}
\label{R2Einstein3}
\hspace{-1cm}
{\cal A} \; & = & \;  - \frac{1}{2} \int d^4x \sqrt{-{\tilde g}} \left[ {\tilde R}  - 6 \partial^\mu \Omega \partial_\mu \Omega \right. \nonumber \\ 
&&
\left. \; - \; \sum_{i=1}^{N-1} \frac{ 2  \partial^\mu \phi^i \partial_\mu \phi^*_i }{\left(\delta + 2 {\alpha}  \Phi - \frac{1}{3} 
 |\phi^i |^2 \right)} + \frac{ {\alpha}  \Phi^2}{\left(\delta + 2 { \alpha}  \Phi - \frac{1}{3}  |\phi^i |^2 \right)^2}  \right] \, ,
\end{eqnarray}
which may be compared with the effective action of SU(N,1)/SU(N)$\times$U(1) no-scale supergravity.

First, we notice that the coefficients of the curvature terms
in Eqs.~(\ref{Lprime}) and (\ref{phiR2}) 
match for $\alpha = 1/6$. 
In the case of the SU(1,1)/U(1) no-scale supergravity model, 
$2 \alpha \Phi \to \Phi/3 \to (T+T^*)$, 
resulting in an equivalence in all kinetic terms,
though bearing in mind that all the scalar
fields in the supergravity theory are complex, and the association is only possible along the real direction.
The second term in Eq.~(\ref{phiR2}) is effectively a 
potential term, which we see in the Einstein frame is a constant, $1/8\alpha$. 
In the supergravity context, we are required 
therefore to add a superpotential term such as 
\beq
W = T^3 -\frac{1}{12\alpha} \, ,
\label{T3}
\eeq
which, using Eqs.~(\ref{VJ}) and (\ref{effVJ}),
generates a scalar potential of the form
\beq
V(T, T^*) = \frac{1}{4 \alpha} \frac{T^2 + {T^*}^2}{(T + {T^*})^2}
\label{VT}
\eeq
that reduces to a constant ($1/8\alpha$) along the real direction.
We return below to the question of generating de Sitter 
solutions in no-scale supergravity. 

In the case of the $R+R^2$ theory, the correspondence 
becomes $1+ 2 \alpha \Phi \to \Phi/3 \to (T+T^*)$
and, once again, a superpotential must be added to 
generate the Starobinsky potential, as
will be discussed in detail below.

Finally, when matter fields are conformally coupled to curvature as in Eq. (\ref{manyphi}), 
we must compare Eq.~(\ref{R2Einstein3}) and Eqs.~(\ref{LmanyJ})-(\ref{effVJ}),
recalling that the comparison 
can only be made along the real direction in the 
supergravity Lagrangian. Terms related to the 
imaginary parts of 
fields can also be accounted for by incorporating additional 
terms in the gravitational action~\cite{eno9}.
The kinetic terms are easily seen to be identical,
with the replacements $-6 \Omega \to K$ and 
$(\delta + 2 \alpha \Phi ) \to (T + T^*)$.
With these identifications, the conformal factors in 
front of the matter kinetic terms are equal, $e^{K/3}$,
as is the conformal factor in front of the potential, $e^{2K/3}$.

\section{Generalized No-Scale Models, Minkowski Pairs and (Anti-)de Sitter Solutions}
\label{pairs}

The original no-scale framework describes a 
non-compact coset field space with 
constant K{\"a}hler curvature, $R = 2/3$~(we use the convention that $R > 0$ for a hyperbolic manifold and $R < 0$ for a spherical manifold).
As mentioned in Section~5, this can be generalized by considering a K\"ahler 
potential of the form
\begin{equation}\label{kah2}
K \; = \; - \, 3 \, \alpha \, \ln (T + T^* ) \, ,
\end{equation}
which also parametrizes a non-compact SU(1, 1)/U(1)} 
coset manifold, but with a constant curvature 
$R = \frac{2}{3 \alpha}$ that is positive if we 
assume that $\alpha > 0$. This unique structure was 
first discussed in 1984 in Ref.~{\refcite{EKN1}}, and similar models 
have been studied more recently,~\cite{alpha1,klr,RS,alpha3,enno,enov3,ennov,Aldabergenov:2019aag}
where they were termed $\alpha$-attractors. 

The de Sitter solution given by Eq.~(\ref{T3})
is a special case of a more general class of 
superpotentials that were also first derived in Ref.~\refcite{EKN1}:
\begin{eqnarray}
1) \qquad W & = & \lambda \qquad {\rm with} \qquad \alpha = 1 \, , \label{ekn1} \\
2) \qquad W & = & \lambda \, T^{3 \alpha/2} \, , \label{ekn2} \\
3) \qquad  W & = & \lambda \, T^{3\alpha/2} (T^{3 \sqrt{\alpha}/2} - T^{-3 \sqrt{\alpha}/2})\, ,  \label{ekn3}
\end{eqnarray}
where we note that (\ref{ekn2}) corresponds to an AdS space,
since $V < 0$. 

These may be further generalized by considering first a 
general form for the superpotential that
produces a Minkowski solution \cite{enno}
\begin{equation}\label{mink1}
W_M = \lambda \cdot T^{n_{\pm}} \, ,
\end{equation}
where $n_{\pm}$ is given by
\begin{equation}\label{npm}
n_{\pm} = \frac{3}{2} \left( \alpha \pm \sqrt{\alpha} \right) \, .
\end{equation}
The effective potential vanishes, $V = 0$,
along the real $T$ direction. Generalizing (\ref{ekn3}),
de Sitter solutions can also be expressed
in a compact form,~\cite{enno,enov3,ennov}
\begin{equation}\label{ds22}
W_{dS} =  \lambda_1 \, T^{n_{-}} - \lambda_2 \, T^{n_{+}}  \, ,
\end{equation}
where $n_{\pm}$ is again given by~(\ref{npm}). 
Along the real $T$ direction the effective scalar potential~(\ref{effpot}) is:
\begin{equation}
V = 3 \cdot 2^{2 - 3 \alpha} \cdot \lambda_1 \, \lambda_2 , 
\end{equation}
giving a de Sitter solution when $\lambda_1 \lambda_2 > 0$, 
and AdS when $\lambda_1 \lambda_2 < 0$.
It is fascinating that a de Sitter vacuum construction~(\ref{ds22}) 
is obtained by combining two 
Minkowski vacuum solutions~(\ref{mink1}).

This construction based on Minkowski pairs can be generalized 
to cases with $N > 1$ moduli. We first choose
\begin{equation}
K = -3 \sum_{i = 1}^N \, \alpha_i \ln \left(\mathcal{V}_i \right) \, ,
\label{kahmult}
\end{equation}
where $\mathcal{V}_i = T_i + T^*_i$. Next, 
we restrict to real values of all the fields, 
so that $T_i = T^*_i$,\footnote{This condition can be achieved dynamically by introducing quartic terms in the K\"ahler potential that stabilize the field in the imaginary direction, as we discuss in Section~\ref{sec:stable}.} which leads to:
\begin{equation}
\mathcal{V}_i \longrightarrow \xi_i, \quad \text{for}~i = 1,2,...,N \, ,
\end{equation}
with $\xi_i = 2 T_i$.
The general $N$-field Minkowski vacuum solutions are given by the following superpotential choice
\begin{equation}
W_{M} = \lambda \cdot \prod_{i =1}^N \xi_i^{n_{i}}.
\label{mink3}
\end{equation}
Combining Eq.~(\ref{effpot}) with the superpotential~(\ref{mink3}), 
we find
\begin{equation}
V = \lambda^2 \cdot \prod_{i = 1}^N \xi_{i}^{2 n_{i} - 3 \alpha_i} \cdot \left( \sum_{i = 1}^N \frac{(2n_i - 3 \alpha_i)^2}{3 \alpha_i} - 3 \, \right)\, ,
\label{mink4}
\end{equation}
and it can be seen from Eq.~(\ref{mink4}) that to find the Minkowski vacuum solutions with $V = 0$, we must satisfy the condition
\begin{equation}
\sum_{i = 1}^N  \frac{(2n_i - 3 \alpha_i)^2}{3 \alpha_i} = 3 \, .
\label{cond1}
\end{equation}
We can parametrize the constraint~(\ref{cond1}) as a radial unit $N$-vector $\vec{r} = (r_1, \, r_2, \, \dots, \, r_N)$, where
\begin{equation}
r_i \equiv \frac{2n_i - 3 \alpha_i}{3 \sqrt{\alpha_i}}, \quad~{\text{for}}~i=1, 2, ..., N \, .
\label{cond2}
\end{equation}
Combining equations~(\ref{cond1}) and~(\ref{cond2}),
we find the unit vector condition
\begin{equation}
\sum_{i = 1}^N r_i^2 = 1 \, .
\label{cond3}
\end{equation}
Eq.~(\ref{cond2}) shows that the $N$-field  Minkowski  vacuum solutions are parametrized
by coordinates compactified on the surface of an $(N - 1)$-sphere.

\begin{figure}[ht]
\centerline{\psfig{file=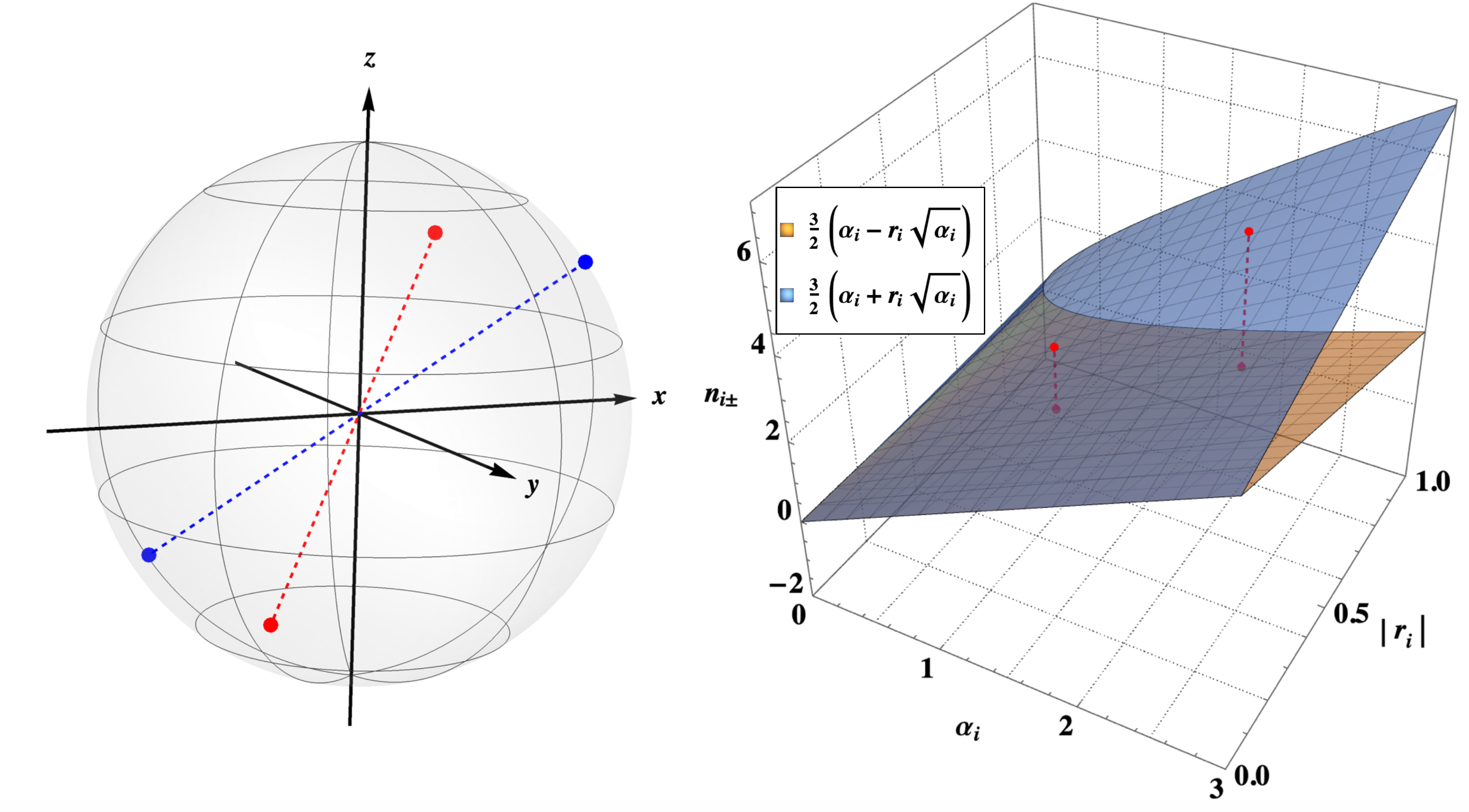,width=13cm}}
\caption{\it {\it Left:}  Illustration of Minkowski pairs 
on the surface of a sphere. Two distinct Minkowski pairs lie in different octants of the sphere, where the red dots correspond to a Minkowski-pair solution 
$r = (1/\sqrt{3}, - 1/\sqrt{3}, 1/\sqrt{3})$ (4th octant) and $\bar{r} = (-1/\sqrt{3},  1/\sqrt{3}, -1/\sqrt{3})$ (6th octant), 
and the blue dots correspond to a Minkowski-pair solution 
$r =(1/\sqrt{3}, 1/\sqrt{3}, 1/\sqrt{3})$ (1st octant) and $\bar{r} = (-1/\sqrt{3}, -1/\sqrt{3}, -1/\sqrt{3})$ (7th octant).
{\it Right:} Illustration of the Minkowski-pair solution for possible choices for powers $n_{i}$ (blue sheet) and $\bar{n}_{i}$ (yellow sheet). 
The Minkowski-pair solutions are shown by red dots, and their coordinates, $(\alpha_i,r_i,n_i)$, are given by $\left(1, \frac{1}{2}, \frac{3}{4} \right)$ with $\left( 1, \frac{1}{2}, \frac{9}{4} \right)$ 
and by $\left( 2, \frac{3}{4}, 3 - \frac{9}{8} \sqrt{2} \right)$ with $\left( 2, \frac{3}{4}, 3 + \frac{9}{8} \sqrt{2} \right).$
  \label{fig:sphere}}
\end{figure}

Solving Eq.~(\ref{cond2}) for $n_i$, we find
\begin{equation}
n_{i} = \frac{3}{2} \left( \alpha_i + r_i \sqrt{\alpha}_i \right), \quad~{\text{for}}~i=1, 2, ..., N \, ,
\end{equation}
where $r_ i \in \{-1, 1\}$ and $\alpha_i > 0$. 
Once again, dS/AdS solutions are found by taking 
the difference between two Minkowski solutions: 
\begin{equation}
W_{dS} = \lambda_1 \cdot \prod_{i=1}^N \xi_i^{n_{i}} - \lambda_2 \cdot \prod_{i=1}^N \xi_i^{\bar{n}_i}  \, , 
\label{mink5}
\end{equation}
where $\bar{n}_i = \frac{3}{2} \left( \alpha_i + \bar{r}_i \sqrt{\alpha}_i \right)$, with $\bar{r}_i = - r_i$. 
Combining the general $N$-field superpotential form~(\ref{mink5}) with the effective scalar potential~(\ref{effpot}), we find the dS/AdS 
vacuum result 
\begin{equation}
   V = 12 \, \lambda_1 \, \lambda_2.
\end{equation}
An illustration of these solutions for the case $N=3$ is shown in Fig. \ref{fig:sphere}. 

Minkowski vacuum solutions are described by any point on the surface of the unit sphere. To obtain the dS/AdS vacuum solutions, one can combine any arbitrary point lying on the surface of the sphere with its antipodal point: $\Vec{r} \rightarrow -\Vec{r}$. 
In the left panel of Fig.~\ref{fig:sphere}, we illustrate four distinct Minkowski vacuum solutions lying on the surface of a sphere which are combined into two unique dS/AdS vacuum solutions. In the right panel of Fig.~\ref{fig:sphere}, 
we show dS/AdS vacuum solutions for possible choices for powers $n_i$ (blue sheet) and $\bar{n}_i$ (yellow sheet) as 
functions of the radial vector component $|r_i|$ and curvature parameter $\alpha_i$. To recover a Minkowski vacuum solution, one can choose any arbitrary point lying on either the yellow or blue sheet, which yields $V = 0$. If we combine the arbitrarily chosen point with a perpendicular point on the opposite sheet (which would correspond to the antipodal point $\vec{r} \rightarrow - \vec{r} \, $ on the surface of a sphere as shown in the left panel of Fig.~\ref{fig:sphere}), it yields the dS/AdS vacuum solution with $V = \, 12 \, \lambda_1 \, \lambda_2$.

Finally, we note that these solutions can be further generalized by adding $M-1$ matter fields per modulus as follows. 
Starting with Eq.~(\ref{kahmult}), we now define
\begin{equation}
\mathcal{V}_i =T_i + T^*_i - \sum_{j = 1}^{M - 1} \frac{|\phi_{ij}|^2}{3} \, .
\end{equation}
As previously, we then fix the VEVs of the imaginary fields to zero, so that $T_i = T^*_i$ and  $\phi_{ij} = \phi^*_{ij}$. 
Using the same notation:
\begin{equation}
\mathcal{V}_i \longrightarrow \xi_i,~\text{when}~T_i = T^*_i~\text{and}~\phi_{ij} = \phi^*_{ij} \, ,
\end{equation}
the argument inside the logarithm in the K\"ahler potential becomes
\begin{equation}
\xi_i = 2T_i - \sum_{j = 1}^{M-1} \frac{|\phi_{ij}|^2}{3} \, .
\label{xiN}
\end{equation}
With this definition of $\xi_i$,  Minkowski  and dS/AdS vacuum solutions are found using 
Eq.~(\ref{mink5}) with the new definition of $\xi_i$ in (\ref{xiN}). 

\section{Generalized No-Scale Starobinsky-Like Inflationary Models}
\label{avatars}

Following this discussion of the connection between no-scale supergravity and higher-order theories of gravity,
we now generalize our previous construction of Starobinsky-like models of inflation in the context of no-scale supergravity.

\begin{figure}[ht!]
\centering
\includegraphics[scale=0.4]{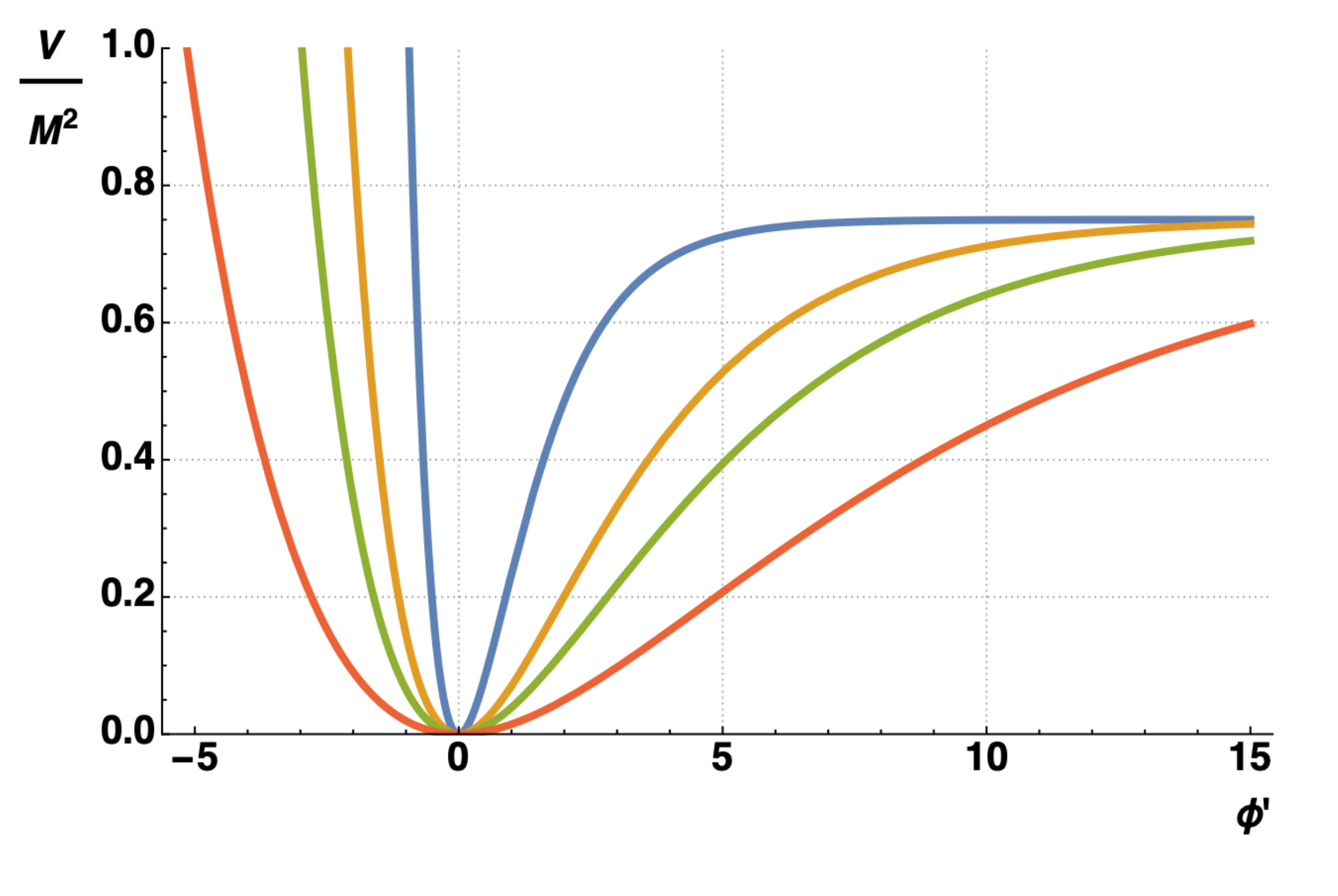}
\caption{\it The $\alpha$-Starobinsky potential~(\ref{alstaro}) for different values of the curvature parameter $\alpha = 1~(\rm{blue}),  \, 5~(\rm{yellow}),  \, 10~(\rm{green}),  \, 30~(\rm{orange})$, where $\alpha = 1$ corresponds to the original Starobinsky model of inflation~(\ref{staropot}).
}
\label{staroplot}
\end{figure}

Assuming an SU(2,1)/SU(2)$\times$U(1)  K\"ahler potential of the form
\beq
K \; = \; - \, 3 \, \alpha \, \ln \left(T + T^*  -\frac{|\phi|^2}{3} \right) \, ,
\label{K2-1a}
\eeq
we consider the following form of superpotential with an arbitrary function $f(\phi)$:~\cite{enov3}
\begin{equation}
W_{I} \; = \; \sqrt{\alpha}  \, f(\phi) \cdot \left(2T - \frac{\phi^2}{3} \right)^{\frac{3}{2} \left(\alpha - \sqrt{\alpha} \right)} \, . 
\label{Wf}
\end{equation}
This reduces to the following relatively simple form in the real direction $ \phi = \phi^*$ and $T = T^*$, for $\langle T \rangle = 1/2$:
\begin{equation}
V =  \left(1-\frac{\phi ^2}{3}\right)^{1-3 \sqrt{\alpha }} \cdot f'(\phi )^2 \, ,
\label{genstaro}
\end{equation}
where $f'(\phi) = df/d\phi$.
Then, if
\beq
 f'(\phi ) = \frac{\sqrt{3} M \,\phi}{\left(\phi +\sqrt{3}\right)}   \left(1-\frac{\phi ^2}{3}\right)^{(1-3 \sqrt{\alpha })/2}  \, ,
 \label{difff}
\eeq
we obtain a Starobinsky-like potential
\begin{equation}
V = \frac{3}{4} M^2 \left(1 - e^{- \sqrt{\frac{2}{3 \alpha}} \phi'} \right)^2
\label{alstaro}
\end{equation}
after a field redefinition is made to a canonically-normalized field, $\phi'$: 
\begin{equation}
\phi = \sqrt{3} \tanh \left(\frac{\phi^\prime}{\sqrt{6 \alpha} } \right).
\label{kinphi}
\end{equation}
This is a concrete realization of the generalization discussed earlier leading to Eq.~(\ref{avatarB}).
Some $\alpha$-Starobinsky potential forms with different values of $\alpha$ are plotted 
in Fig.~\ref{staroplot}. We can see in the Figure that increasing the value of the curvature parameter 
$\alpha$ stretches the Starobinsky potential horizontally, reducing the flatness of the plateau at any fixed value of $\phi^\prime$, and thereby increasing $r$ as was seen in Eq.~(\ref{rstaro}).

The values of the cosmological observables predicted by these $\alpha$-Starobinsky potentials are shown in the $(n_s, r)$ plane in Fig.~\ref{fig:plancklimits},
where they can be compared with the results of the Planck collaboration 
in combination with other CMB data. 
We see that the scalar tilt $n_s$ changes only 
slightly as the curvature parameter $\alpha$ increases,
remaining within the range $\sim 0.96-0.97$, whereas the value of the
tensor-to-scalar ratio $r$~(\ref{rstaro}) increases with the value of $\alpha$.
The combined CMB data have established a 68\% upper bound on $r_{0.002}$ of $\sim 0.041$, 
which is reached with $n_s \sim 0.967$ when $\alpha \sim  27$ and
$N_* \simeq 55$, as indicated by the blue star. 
The prediction of the original Starobinsky model with $N_* \in [50, 60]$, which
corresponds to the case $\alpha = 1$,
is indicated by the green dots and line at small $r$.
Future measurements of $r$ will impose stronger constraints on $\alpha$, 
and more precise measurements of $n_s$
could constrain $N_*$ and hence the history of the Universe after inflation,
in particular the rate of decay of the inflaton into low-mass particles~\cite{reheating} (see Fig.~\ref{fig:Nstar} below).

The superpotential is found by solving Eq.~(\ref{difff}), and has the form of a 
hypergeometric function for arbitrary $\alpha$ \cite{enov3}:
\begin{equation}
f (\phi) \; = \;   M \left[\frac{3-3^{-m} \left(3-\phi ^2\right)^{m+1}}{2 (m+1)}-\frac{\phi ^3 \, _2F_1\left(\frac{3}{2},-m;\frac{5}{2};\frac{\phi ^2}{3}\right)}{3 \sqrt{3}} \right]  \, ,
\label{alphastaro}
\end{equation}
where $m = \frac{3}{2} \left(\sqrt{\alpha} - 1 \right)$. 
Remarkably, the expression in (\ref{alphastaro}) simplifies for particular
values of $\alpha$. Specifically, when $\alpha = 1$, $f(\phi)$ is of the Wess-Zumino (WZ) form given in Eq.~(\ref{wi})~\cite{eno6} with $\lambda = M/\sqrt{3}$,
 as in this case $m=0$ and $_2F_1\left(\frac{3}{2},0;\frac{5}{2};\frac{\phi ^2}{3}\right)=1$.~\footnote{  
 The function $f(\phi)$ is a polynomial whenever $9 \alpha$ is an odd perfect square other than 1.}
With this choice, one recovers the exact form of the Starobinsky potential.

The underlying SU(2,1)/SU(2)$\times$U(1) no-scale symmetry can be used to 
generate additional forms for the superpotential that
lead to the same physical scalar potential, and thereby Starobinsky inflation.
For example, we can start from a more symmetric representation of
the SU(2,1)/SU(2)$\times$U(1) coset space~\cite{EKN}:
\begin{equation}
K \; = \; - 3 \ln \left(1 - \frac{|y_1|^2 + |y_2|^2}{3} \right) \, ,
\label{K21symm}
\end{equation}
where the complex fields $y_{1,2}$ are related to the fields $T, \, \phi$ appearing in (\ref{K2-1a}) for $\alpha = 1$ by
\begin{equation}
y_1 \; = \; \left( \frac{2 \phi}{1 + 2 T} \right) \; ; \qquad \; y_2 \; = \; \sqrt{3} \left( \frac{1 - 2 T}{1 + 2 T} \right) \, ,
\label{Tphiwrite}
\end{equation}
with the inverse relations
\begin{equation}
T \; = \; \frac{1}{2} \left( \frac{1 - {y_2}/{\sqrt{3}}}{1 + {y_2}/{\sqrt{3}}} \right)\, ; \qquad \;
\phi \; = \; \left(\frac{y_1}{1 + {y_2}/{\sqrt{3}}} \right) \, .
\label{Tphirewrite}
\end{equation}
When the coordinates are transformed as in (\ref{Tphiwrite}, \ref{Tphirewrite}), the
effective superpotential becomes:
\begin{equation}
W(T, \phi) \; \to \; {\widetilde W}(y_1, y_2) \; = \; \left( 1 + {y_2}/{\sqrt{3}} \right)^3 W  \, .
\label{Wtilde}
\end{equation}
In the $(y_1, y_2)$ basis,
one has an effective potential
\begin{equation}
V \; = \; \frac{{\hat V}}{(1 - (|y_1|^2 + |y_2|^2)/3)^2} \, ,
\label{Vhatphii}
\end{equation}
where
\begin{eqnarray}
\hat{V} = |W_1|^2 + |W_2|^2 - \frac{1}{3} |3W - W_1 y_1 - W_2 y_2|^2 \, ,
\label{fullVhat}
\end{eqnarray}
with $W_{1,2} = \partial W/\partial y_{1,2}$.
If $\langle y_2 \rangle = 0$, one finds
\begin{eqnarray}
V & = & \frac{{\hat V}}{(1 - |y_1|^2/3)^2} \, , \nonumber \\
{\rm where} \; \; \;{\hat V} & = & (1 - |y_1|^2/3) |W_1|^2 + |W_2|^2 - 3 |W|^2 +
(y_1 W_1 W^* + {\rm h. \, c.}) \, ,
\label{useful}
\end{eqnarray}
and $y_1$ can be converted into a canonically-normalized inflaton field $x$ by the transformation
\begin{equation}
y_1 \; = \; \pm \sqrt{3} \tanh (\chi/\sqrt{3} )\; = \; \pm  \sqrt{3} \tanh (x/\sqrt{6} ) \, ,
\label{arctanh}
\end{equation}
where $\chi = (x + iy)/\sqrt{2}$ and the latter equality holds for $y=0$.
From this perspective, one would expect that inflation could be driven by either the matter-like field $\phi$,
as in Eq.~(\ref{wi}), or the volume modulus, $T$.

The WZ model defined by Eq.~(\ref{wi}), can be rewritten in the symmetric basis as  \cite{eno7}
\begin{equation}
W \; = \; M \left[ \frac{y_1^2}{2} \left(1+\frac{y_2}{\sqrt{3}} \right) - \frac{y_1^3}{3 \sqrt{3}} \right] \, ,
\label{W1}
\end{equation}
which is a WZ model for $y_1$ along with an interaction term $y_1^2 y_2$.
In this model $W$, $W_1$, and $W_2$ are all non-zero, even with
the assumption that $y_2$ is fixed so that $\langle y_2 \rangle = 0$.
Using (\ref{useful})
we obtain 
\begin{equation}
V \; = \; \frac{M^2 |y_1|^2 ~ |1 - y_1/\sqrt{3}|^2}{(1 - |y_1|^2/3)^2} \, ,
\label{V1}
\end{equation}
which is of the form needed to yield the Starobinsky potential. 

Returning again to the $(T, \phi)$ basis, inflation requires one of the two fields to be fixed.
In the WZ-like model (\ref{wi}), we had assumed that $T$ is fixed
with a vacuum expectation value of Re~$\langle T \rangle = \frac{1}{2}$ and Im~$\langle T \rangle = 0$,~\footnote{We discuss the question of stabilization in Section~\ref{sec:stable} below.}  in which case the kinetic term becomes:	
\begin{equation}\label{Lkinphi}
{\cal L}_{\rm kin} = \frac{1}{\left(1 - \frac{\phi \phi^*}{3} \right)^2} \partial_{\mu}\phi \partial^{\mu} \phi^* \, .
\end{equation}
 We also assume that the 
imaginary part of the matter field $\phi$ is fixed to Im~$\langle \phi \rangle = 0$ by the dynamics of the potential. 
The following redefinition leads to a canonically-normalized field:
\begin{equation}\label{phigen}
\phi = \pm \sqrt{3} \tanh \left(\frac{x}{\sqrt{6}}\right) \, ,
\end{equation}
where $x$ is a real scalar field. 

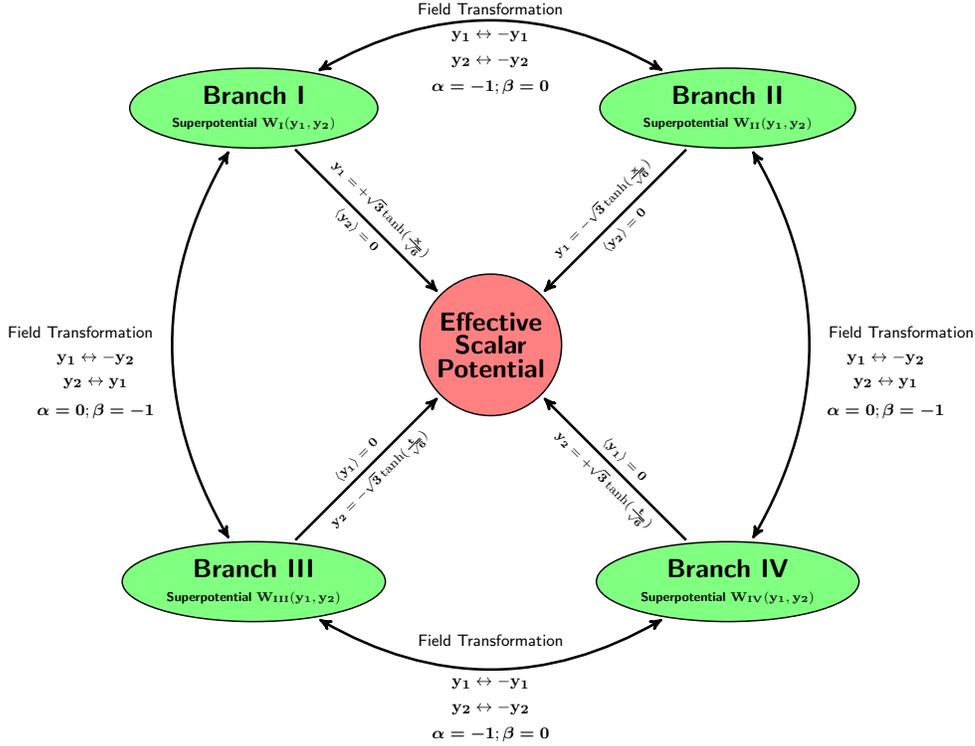
\begin{figure}[!ht]
\begin{center}
\resizebox{\textwidth}{!}{
\begin{tikzpicture}[->,>=stealth',shorten >=1pt,auto,node distance=6.5cm,
                    thick,main node/.style={align=center,circle,draw,font=\linespread{0.8}\sffamily\Large\bfseries},Arrow/.style = {line width=2mm, draw=gray, 
                -{Triangle[length=3mm,width=4mm]},
                shorten >=1mm, shorten <=1mm}],
                    	
\tikzset{
    pil/.style={
           ->,
           line width=0.5mm,
           shorten <=2pt,
           shorten >=2pt,}
},               
  \node[fill=red!50, main node] (1) {Effective \\ Scalar \\ Potential};			
  \node[fill=green!50, main node, ellipse, draw] (I)[above left of=1]{Branch I \\ \scriptsize Superpotential $\mathbf{W_I(y_1, y_2)}$};
  \node[fill=green!50, main node, ellipse, draw] (II)[above right of=1] {Branch II \\ \scriptsize Superpotential $\mathbf{W_{II}(y_1, y_2)}$};
 
    \node[main node, ellipse, draw] (III)[fill=green!50, below left of=1] {Branch III \\ \scriptsize Superpotential $\mathbf{W_{III}(y_1, y_2)}$};
  \node[fill=green!50, main node, ellipse, draw] (IV)[below right of=1]{Branch IV \\ \scriptsize Superpotential $\mathbf{W_{IV}(y_1, y_2)}$};
  \path[every node/.style={font=\sffamily\small}]
    (I) edge [pil, right, above] node[above, rotate=-45] {\scriptsize $\mathbf{~y_1 = +\sqrt{3} \tanh(\frac{x}{\sqrt{6}})}$} node[below, rotate=-45] {\scriptsize $\mathbf{\langle y_2 \rangle = 0}$}(1)
         edge [pil,<->, bend left=30] node[above] {Field Transformation} node[below] {$\mathbf{y_1 \leftrightarrow -y_1}$}node[below, yshift=-5mm] {$\mathbf{y_2 \leftrightarrow -y_2}$}node[below, yshift=-10mm] {$\bm{\alpha=-1};\bm{\beta = 0}$}(II)

edge [pil,<->, bend right=30] node[above, xshift=-15mm] {\hspace {-0.7cm} Field Transformation} node[below,xshift=-15mm] {$\mathbf{y_1 \leftrightarrow -y_2}$}node[below, yshift=-5mm, xshift=-15mm] {$\mathbf{y_2 \leftrightarrow y_1}$} node[below, yshift=-10mm, xshift=-15mm] {$\bm{\alpha = 0};\bm{\beta = -1}$}(III)         
         
   (II) edge [pil, left, above]  node[above, rotate=45] {\scriptsize $\mathbf{y_1 = -\sqrt{3} \tanh(\frac{x}{\sqrt{6}})}$} node[below, rotate=45] {\scriptsize $\mathbf{\langle y_2 \rangle = 0}$}(1)
   
edge [pil,<->, bend left=30] node[above, xshift=15mm] {\hspace{0.5cm} Field Transformation} node[below,xshift=15mm] {$\mathbf{y_1 \leftrightarrow -y_2}$} node[below, yshift=-5mm, xshift=15mm] {$\mathbf{y_2 \leftrightarrow y_1}$}node[below, yshift=-10mm, xshift=15mm] {$\bm{\alpha = 0};\bm{\beta = -1}$}(IV)     
   
   (III) edge [pil, left, below]  node[above, rotate=45] {\scriptsize $\mathbf{\langle y_1 \rangle = 0}$} node[below, rotate=45] {\scriptsize $\mathbf{~y_2 = -\sqrt{3} \tanh(\frac{t}{\sqrt{6}})}$}(1)  
   
edge [pil,<->, bend right=30] node[above, yshift=3mm] {Field Transformation} node[below] {$\mathbf{y_1 \leftrightarrow -y_1}$}node[below, yshift=-5mm] {$\mathbf{y_2 \leftrightarrow -y_2}$} node[below, yshift=-10mm] {$\bm{\alpha=-1};\bm{\beta = 0}$}(IV)
   
   (IV) edge [pil, right, below] node[above, rotate=-45] {\scriptsize $\mathbf{\langle y_1 \rangle = 0}$} node[below, rotate=-45] {\scriptsize $\mathbf{y_2 = +\sqrt{3} \tanh(\frac{t}{\sqrt{6}})}$}(1);
\end{tikzpicture}}
\end{center}
\caption{\it Diagram illustrating the transformation laws between the superpotentials in different branches, together with the field fixings and canonical field redefinitions that yield the same Starobinsky-like effective scalar potential.} \label{fig:branches2}
\end{figure}

Alternatively, we can set $\langle \phi \rangle = 0$. The kinetic term in this case is:
\begin{equation}\label{kinT}
{\cal L}_{\rm kin} = \frac{3}{\left(T + T^* \right)^2} \partial_{\mu}T \partial^{\mu} T^* \, .
\end{equation}
We again assume a vacuum expectation value for $T$ with Im~$\langle T \rangle = 0$, 
so that the real part of the volume modulus $T$ can be redefined as a canonically-normalized field given by:
\begin{equation}\label{Tgen}
T = \frac{k}{2} e^{\pm \sqrt{\frac{2}{3}} t} \, ,
\end{equation}
where the field $t$ is real and the coefficient in front of~(\ref{kinT}) is compatible with the symmetric field redefinitions~(\ref{Tphirewrite}). In this way, by fixing one of the complex scalar fields $(T, \phi)$ and performing 
a canonical field redefinition~(\ref{kinphi}) or~(\ref{kinT}), the SU(2,1)/SU(2)$\times$U(1) symmetry can be broken into 
one of four different branches,~\cite{enov1}
defined as
\begin{small}
\begin{align}\label{symtrans1}
& \textbf{Branch I:}~\left(\phi = +\sqrt{3k} \tanh \left(\frac{x}{\sqrt{6}} \right);~\langle T \rangle = \frac{k}{2} \right) \longrightarrow
\left(y_1 = +\sqrt{3} \tanh \left(\frac{x}{\sqrt{6}} \right);~\langle y_2 \rangle = 0 \right) \, , \\ \label{symtrans2}
& \textbf{Branch II:}~\left(\phi = -\sqrt{3k} \tanh \left(\frac{x}{\sqrt{6}}\right);~\langle T \rangle = \frac{k}{2} \right) \longrightarrow
\left(y_1 =- \sqrt{3} \tanh \left(\frac{x}{\sqrt{6}} \right);~\langle y_2 \rangle = 0 \right) , \\ \label{symtrans3}
& \textbf{Branch III:}~\left( \langle \phi \rangle = 0;~ T = \frac{k}{2} e^{+ \sqrt{\frac{2}{3}} t} \right) \longrightarrow
\left(\langle y_1  \rangle = 0;~ y_2  = -\sqrt{3} \tanh\left(\frac{t}{\sqrt{6}} \right) \right) \, , \\ \label{symtrans4}
& \textbf{Branch IV:}~\left( \langle \phi \rangle = 0;~ T = \frac{k}{2} e^{- \sqrt{\frac{2}{3}} t} \right) \longrightarrow
\left(\langle y_1  \rangle = 0;~ y_2  = +\sqrt{3} \tanh \left(\frac{t}{\sqrt{6}}\right) \right) \, .
\end{align}
\end{small}
\hspace{-2.75mm}
Redefining fields with canonically-normalized kinetic terms in the $(y_1, y_2)$ symmetric basis using
equations~(\ref{symtrans1})-(\ref{symtrans4}), we may consider a general expression $W(y_1, y_2)$ for the superpotential.
The SU(2,1)/SU(2)$\times$U(1) symmetry of the coset space includes the following transformation laws for the fields $y_1$ and $y_2$ \cite{enov1}:
\begin{equation}\label{syminv}
y_1 \rightarrow \alpha y_1 + \beta y_2, \qquad y_2 \rightarrow -\beta^* y_1 + \alpha^* y_2 \, .
\end{equation}
The K\"ahler potential is invariant under these transformations, but the superpotential $W(y_1, y_2)$ transforms non-trivially in general. We find it more convenient to use the symmetric $(y_1, y_2)$ basis when starting the analysis of a general superpotential.
Starting with $W(y_1, y_2)$ in any one of the four different branches, 
the corresponding superpotential in other branches can be obtained by making the transformations~(\ref{syminv}). The relations between the superpotentials 
in different branches obtained by these transformations
are indicated in Fig.~\ref{fig:branches2}.

The general expressions for superpotentials in all the four branches that yield the 
Starobinsky inflationary potential with canonically-normalized kinetic terms
can be obtained from a general superpotential expression for Branch I of the form:
\begin{align}\label{genstaro1}
& \textbf{Branch I:}~&W(y_1, y_2)& &=& &a y_1+b y_1^2+c y_1^3+d y_2+e y_2 y_1+f y_2 y_1^2 + g(y_1, y_2) \, , 
\end{align}
where $g(y_1, y_2)$ has the following properties: $g(y_1, 0)
= 0$, $\partial g/ \partial y_1 (y_1 , 0)$  $= 0$ and $\partial g/ \partial y_2 (y_1, 0) = 0$.
The function $g$ may also include terms containing factors $y_2^n$, 
but these would not contribute to $V$, since we impose 
the condition $\langle y_2 \rangle = 0$. 
In order to avoid supersymmetry breaking close to the inflationary scale,
we do not include a constant term in (\ref{genstaro1}),
nor in the general form of $W$ in the other branches.  

Performing the transformation (\ref{syminv}) with $\alpha = -1$ and $\beta = 0$,  we obtain the superpotential for Branch II. 
If, instead, we apply the transformation with $\alpha = 0$ and $\beta = -1$ to the general expression for Branch I,
we obtain the superpotential for Branch III.
Finally, applying either the same transformation to Branch II or applying the previous transformation  with $\alpha = -1$ and $\beta = 0$  to Branch III,
we obtain the superpotential of Branch IV.

Using the form (\ref{genstaro1}) for the Branch I superpotential, we can derive $\hat{V}$ from (\ref{fullVhat})
and match to a known solution from Ref.~\refcite{eno7}:
\beq
\hat{V} = M^2 |y_1|^2 |1-y_1/\sqrt{3}|^2 \, ,
\eeq
corresponding to the WZ model found in Ref.~\refcite{eno6}. Matching coefficients leads to four sets of solutions:
\begin{small}
\begin{equation}
\begin{cases}\label{cases1}
a = 0, \qquad c = +\frac{b \left(\sqrt{1-4 b^2}-2\right)}{3 \sqrt{3}}, \qquad d = 0, \qquad e = \pm \sqrt{1-4 b^2}, \qquad f =\mp \frac{\sqrt{1-4 b^2}+2 b^2}{\sqrt{3}} \, , \\
a = 0, \qquad c = -\frac{b \left(\sqrt{1-4 b^2}+2\right)}{3 \sqrt{3}}, \qquad d = 0, \qquad e = \pm \sqrt{1-4 b^2},\qquad  f = \mp \frac{\sqrt{1-4 b^2} - 2 b^2}{\sqrt{3}} \, ,
\end{cases}
\end{equation}
\end{small}
\hspace{-2.75mm}
where all the coefficients are expressed in terms of a free parameter $b$. 
There are two additional solutions:
\begin{eqnarray}
\label{cases2}
&b = -\frac{\sqrt{3} a}{2 a^2+3} \, , & \; c = \frac{16 a^6+72 a^4+108 a^2+27}{36 a \left(2 a^2+3\right)^2} \, , \; d = \pm i a \, , \nonumber \\
&  e = \mp \frac{2 i \sqrt{3} a}{2 a^2+3} \, , & \; f = \mp \frac{i \left(4 a^2 \left(2 a^2+3\right)^2+27\right)}{12 a \left(2 a^2+3\right)^2} \, ,
\end{eqnarray}
where now the coefficients are expressed in terms of a free parameter $a$.
Eqs.~(\ref{cases1}) and~(\ref{cases2}) encompass all of the Branch I solutions corresponding to the Starobinsky inflationary potential with canonically-normalized kinetic terms. 

The WZ model in Eq.~(\ref{W1}) is a special case of the Branch I superpotential (\ref{genstaro1})
with 
\beq
 a, \, d, \, e=0; \qquad b = \frac{1}{2}; \qquad c = - \frac{1}{3 \sqrt{3}}; \qquad f = \frac{1}{2 \sqrt{3}} \, ,
 \label{wzgen}
 \eeq 
as in Eq.~(\ref{cases1}) with $b=1/2$. This is just one specific example of the full set of Branch I solutions,
all of which yield a Starobinsky potential. 
We could equally well have chosen a solution with $b = 0$ giving
\beq
 a, \, b, \, c, \, d=0; \qquad e =  -1; \qquad f =  \frac{1}{\sqrt{3}} \, ,
 \label{wzgen0}
 \eeq 
 corresponding to the superpotential
 \beq
 W = M y_1 y_2 \left( -1+ \frac{y_1}{\sqrt{3}}  \right) .
 \eeq
We now consider this last solution and rotate it to Branch III as described in Fig.~\ref{fig:branches2},
i.e., we take $y_1 \to -y_2$ and $y_2 \to y_1$ giving
\beq
\label{cecsym}
W = M y_1 y_2 \left( 1+\frac{y_2}{\sqrt{3}} \right) \, .
\eeq
Finally, we rotate back to the $(T,\phi)$ basis to obtain (dropping an irrelevant overall sign)
\beq
W = \sqrt{3} M \phi \left(T - \frac{1}{2} \right) \, ,
\label{cec}
\eeq
which is equivalent to another well-studied no-scale analogue of the $R+R^2$ theory \cite{Cecotti, FeKR,EGNO2,others,EGNO3}.
Several specific and equivalent examples of these avatars of no-scale Starobinsky models were discussed in Ref.~\refcite{eno7}
(see also Refs.~\refcite{reheating,EGNO4,FKP,Moreothers,king1,dgmo}).

 We conclude this Section by mentioning briefly an alternative, non-oscillatory no-scale supergravity model of inflation (NO-NO inflation) proposed recently.~\cite{NONO} In this scenario there are no inflaton oscillations at the end of the inflationary era, and the expansion of the Universe is instead dominated by the kinetic energy density of the inflaton, a possibility called kination. When the Universe transitions from inflation to kination, it preheats instantly through a coupling to ``Higgs" fields that annihilate rapidly and scatter producing matter particles. These later dominate the energy density, reheating the Universe to a temperature higher than during BBN. 
The NO-NO model makes predictions for
CMB observables that are consistent with Planck~2018 data,
produces gravitational waves with a density that is compatible with BBN, and produces via gravitino decay a density of supersymmetric cold dark matter that is also consistent with cosmological data. 

\section{Stabilizing Moduli Fields in No-Scale Supergravity}
\label{sec:stable}

Up until this Section, we have implicitly assumed that all the
real components of the complex fields that are not driving inflation have been stabilized,
whereas the inflaton field has remained dynamical. 
In order not to spoil the inflationary dynamics, 
it is important to ensure that during inflation the other stabilized real fields remain fixed or at least strongly constrained. In this Section we do not attempt to review the problem of moduli stabilization, but instead we focus on certain specific examples of field stabilization mechanisms. However, the field stabilization mechanisms we discuss here do serve as existence proofs that lead to successful stabilization, whose origin we leave open.

In all the cases discussed here, we consider adding higher-order correction terms into the K\"ahler potential as first proposed in~{\cite{EKN3}}, and discussed more recently in~{\cite{eno7,EGNO4}}:
\begin{equation}
\label{kstab1}
K \; = \; -3 \, \ln\left(T+T^* - \frac{|\phi|^2}{3}+ \frac{(T+T^*)^{2n}}{\Lambda_T^{2n - 2}} \right) \, ,
\end{equation}
where $n > 1$ and $\Lambda_T$ is a mass scale which is smaller than the Planck scale $M_P$. For simplicity, hereafter we take the lowest possible value $n = 2$, corresponding to a quartic term $\left( T + T^*\right)^4$. 

The higher-order term in the K\"ahler potential~(\ref{kstab1}) stabilizes the volume modulus $T$ in the real direction with ${\rm{Re}} \, T = 0$. In order to stabilize the imaginary part of $T$, one may consider the following generalized K\"ahler potential form:
\begin{equation}
\label{kstab2}
K \; = \; -3 \, \ln \left(T+T^{*}-\frac{|\phi|^{2}}{3}+\frac{\left(T+T^{*}-2c\right)^{4}+d\left(T-T^{*}\right)^{4}}{\Lambda_T^{2}}\right) \, ,
\end{equation}
where the term $d\left(T-T^{*}\right)^{4}$ stabilizes the potential in the imaginary direction of $T$ with ${\rm{Im}} \, T = 0$ , and we also introduced a field shift in the term $\left(T+T^{*}-2c\right)^{4}$ that stabilizes the real part and leads to ${\rm{Re}} \, T = c$. 
In the case of the Wess-Zumino model (\ref{wi}), we saw how the Starobinsky potential can be obtained for the illustrative choice $c =1/2$. This modification of the K\"ahler potential fixes the volume modulus $T$ during inflation and generates the following masses
\begin{equation}
m_{{\rm{Re}} \, T}^2 = 144 \, \frac{m_{3/2}^2}{\Lambda_T^2}; \qquad m_{{\rm{Im}} \, T}^2 = 144 \, d \, \frac{m_{3/2}^2}{\Lambda_T^2},
\label{mT}
\end{equation}
which are hierarchically larger than the gravitino mass if $\Lambda_T\ll 1 $.  Since the quartic stabilization term in the K\"ahler potential~(\ref{kstab2}) should be treated as an effective interaction, one must require that $\Lambda_T > m_{{\rm{Re}} \, T}, \, m_{{\rm{Im}} \, T}$, i.e., $\Lambda_T > 12 m_{3/2}/\Lambda_T, \, 12 \sqrt{d} \,m_{3/2}/\Lambda_T$.

In the case of the Wess-Zumino model that is expressed in the symmetric basis $(y_1, y_2)$ and given by Eq.~(\ref{W1}), the inflationary dynamics are driven by the rolling inflaton field, $y_1$. During inflation, the effective potential is stabilized only in the imaginary direction of $y_2$ and we must prevent the real component of $y_2$ 
from acquiring a non-zero VEV, which would spoil the inflationary dynamics. 
To address this problem and stabilize the field $y_2$ in the real direction, one may introduce the following higher-order term in the symmetric K\"ahler potential~(\ref{K21symm}):
\begin{equation}
\label{kahstab3}
K \; = \; -3 \, \ln \left(1-\frac{\left|y_{1}\right|^{2}+\left|y_{2}\right|^{2}}{3}+\frac{\left|y_{2}\right|^{4}}{\Lambda^{2}}\right) \, ,
\end{equation}
where $\Lambda <1$, and the addition of the quartic stabilization term $|y_2|^4/\Lambda^2$ does not affect the inflationary potential $V(y_1)$ in the real $y_1$ direction.

Alternatively, if we consider models of inflation where inflation is driven by the volume modulus, $T$, e.g., the Cecotti superpotential~(\ref{cec}) that also reproduces the Starobinsky model of inflation, one can modify the SU(2,1)/SU(2)$\times$U(1) K\"ahler potential~(\ref{K2-1a}) with $\alpha = 1$ by including a higher-order stabilization term for the matter-like field, $\phi$:
\begin{equation}
\label{kstab4}
K \; = \; -3 \, \ln \left(T+T^{*}-\frac{|\phi|^{2}}{3}+ \frac{|\phi|^4}{\Lambda_{\phi}^2}\right) \, ,
\end{equation}
where the quartic stabilization term $|\phi|^4/\Lambda_{\phi}^2$ stabilizes the effective potential in both the real and imaginary directions of $\phi$, and $\Lambda_{\phi} < 1$.

The moduli stabilization mechanism discussed in this Section can be extended easily to multi-field models of inflation that are characterized by an SU(N, 1)/SU(N) $\times$ U(1) coset space or more complicated coset structures, and the field stabilization of such models was discussed in~\cite{enno, enov3, ennov} .

We mention in passing that quartic stabilization terms
have also been considered\cite{strongpol} in the context of the Polonyi model \cite{pol} discussed earlier. Adding a term
\beq
\delta K = -\frac{|z|^4}{\Lambda_z^2}
\eeq
in either minimal or no-scale supergravity shifts
the Planck-scale minimum. For example, the minimum found in minimal supergravity in Section \ref{sec:susyb} is shifted  to $\langle z \rangle = \Lambda_z^2/2\sqrt{3}$ with $\zeta = 1/\sqrt{3}$. \cite{dlmmo,ADinf,ego,EGNO4,dgmo,kmov,enov4} The mass of the Polonyi field is then $m_z = \sqrt{12} m_{3/2}/\Lambda_z$, which is significantly heavier than in the minimal model when $\Lambda_z \ll 1$. Strong stabilization in this case can lead to a resolution of the cosmological problems associated with the Polonyi field that were discussed earlier. We discuss the strong stabilization of the Polonyi field in more detail in Section~\ref{sec:ttype}. 

We conclude this Section by mentioning an alternative stabilization mechanism, which does not require higher-order terms in the logarithm.~\cite{EGNO2, EGNO3}. In typical orbifold string compactifications with three moduli that are fixed by some unspecified mechanism at a high scale to be proportional, the K\"ahler potential takes the generic form~{\cite{Dixon:1989fj, Casas:1998qx, Antusch:2011ei}
\beq
K \; = \; - 3 \ln \left(T + T^* - \frac{1}{3} \sum_i |\phi_i|^2\right) + \sum_a \frac{|\vphi_a|^2}{(T + T^*)^{n_a}} \, ,
\label{finalK}
\eeq
where $T$ is the volume modulus, the $\phi_i$ are untwisted matter fields, and the $\varphi_a$ are twisted matter fields with modular weights $n_a$. Coupling a twisted matter field with modular weight 3 to $T$ with a Cecotti-like superpotential~(\ref{cec}), and
\begin{align}
\label{stab6}
K \; &= \; -3 \, \ln (T+T^*)+\frac{|\varphi|^{2}}{(T+T^*)^{3}} \, ,
\end{align}
the Starobinsky potential is recovered in the direction of the canonically-normalized ${\rm Re}\, T$, and there is a quadratic potential along the ${\rm Im}\, T$ direction. During inflation the effective scalar potential is proportional to the exponential factor $e^K$, or
\begin{equation}
V \propto e^{|\varphi|^2/(T+T^*)^3} \simeq e^{|\varphi|^2} \, ,
\end{equation}
and the twisted field $\varphi$ is rapidly stabilized at the origin at the beginning of inflation. It is also worth noting that the inflationary dynamics of this model differ crucially from those with only untwisted fields. The real and imaginary parts of $T$ mix through their
kinetic terms, leading to a coupling between curvature and isocurvature perturbations, and therefore an enhancement of the curvature modes at super-horizon scales. A correct discussion of their behavior during inflation requires a more sophisticated analysis than for the single-field models (see Ref.~\refcite{EGNO3}).

\section{No-Scale Inflation and Phenomenology}
\label{pheno}

We turn now to some phenomenological aspects of no-scale inflationary models. 
For this we work in the $(T,\phi)$ basis, 
and discuss the possibilities for supersymmetry breaking,
the incorporation of matter and the generation of soft mass terms in the
separate cases in which either 
$T$ or $\phi$ plays the role of the inflaton.

\subsection{$\phi$-Type Inflation}

As illustrated in (\ref{finalK}), matter fields may be included in the K\"ahler potential as either untwisted or twisted fields
depending whether their kinetic terms originate inside or outside the logarithm. We first consider the WZ model (\ref{wi}),
where the inflaton is one of the untwisted matter fields,
$\phi_1$. The superpotential for the matter fields can be written as 
\beq\label{w_phi}
\begin{aligned}
W&= (T+c)^{\beta}W_2(\phi_i) + (T+c)^{\gamma}W_3(\phi_i) \\
&\qquad +(T+c)^{\sigma}W_2(\vphi_a) +(T+c)^{\rho}W_3(\vphi_a) +  \mu \, ,
\end{aligned}
\eeq
where $c$ is an arbitrary constant, which we take to be 1/2
for illustration, and
$W_{2,3}$ denote bilinear and trilinear terms with modular weights
that are in general non-zero.
Here $\mu$ is a constant that contributes to supersymmetry breaking.
Assuming that $T$ is properly stabilized with $\langle T \rangle = 1/2$, 
the gravitino mass is simply
\beq
m_{3/2} = \mu \, ,
\eeq
and soft supersymmetry-breaking terms
are easily calculated to be \cite{EGNO4}
\begin{equation}
\begin{aligned}[c]
& {\rm Untwisted~Matter~Fields:} \\
& m_0 = 0 , \\
& B_0 = - \beta m_{3/2} , \\
& A_0 = - \gamma m_{3/2},
\end{aligned}
\qquad \qquad
\begin{aligned}[c]
& {\rm Twisted~Matter~Fields:} \\
& m_0^2 = (1-n_a)^{\frac{1}{2}} m_{3/2} , \\
& B_0 =  2\left(1-n_a-\frac{\sigma}{2}\right)m_{3/2} , \\
& A_0 = 3\left(1-n_a-\frac{\rho}{3}\right) m_{3/2} .
\label{sft_mu_1}
\end{aligned}
\end{equation}
The form of Eq.~(\ref{sft_mu_1}) opens up various phenomenological possibilities.

If all matter fields are of the untwisted type, 
we see that there are no supersymmetry-breaking contributions
to scalar masses, as expected in pure no-scale supergravity. 
If in addition, the modular weights $\gamma$ and $\beta$ vanish, then $A_0 = B_0 = 0$,
and we recover the full set of  {\it no-scale} boundary conditions \cite{LN}.  Radiative electroweak symmetry breaking \cite{ewsb} 
can be accommodated if these boundary conditions
are fixed at scales above the GUT scale \cite{eno5,pro,emo2,ENO8,Ellis:2016qra,eenno}. 
In this case the parameter space is more restricted
than in CMSSM-like models, since the ratio
of the Higgs VEVs, $\tan \beta$, is determined by the 
Higgs minimization conditions and is no longer a free parameter~\cite{vcmssm}. 

If matter fields are of the twisted type,
and the kinetic modular weights are 0,
we obtain universal soft scalar masses as in {CMSSM}-like models, which
are determined by the gravitino mass \cite{bfs}. 
When the superpotential weights are equal ($\rho = \sigma$),
we obtain minimal {mSUGRA}-like boundary conditions, with $A_0 = (3-\rho) m_{3/2}$ and $B_0 = (2-\rho) m_{3/2}$,
i.e., $B_0 = A_0 - m_0$~\cite{bfs,vcmssm}. 
These mSUGRA-like models also yield a more restrictive 
parameter space.
In the symmetric ($y_1,y_2$) basis with no superpotential
weights, we would find $\rho = \sigma = 3$, in
which case $A_0 = 0$ and $B_0 = -m_{3/2}$.
If, in addition, there are no tree-level sources
for gaugino masses, the models would
be equivalent to {\it pure gravity mediation} (PGM) with radiative electroweak symmetry breaking \cite{eioy,evno}. 
Finally, we note that if the weights $n_a \ne 0$, we have a source for non-universal scalar masses in the twisted sector. 

One can also consider the effects of a Polonyi sector on $\phi$-type models \cite{EGNO4},
but we defer a discussion of alternatives to the later
Section on unified no-scale attractor models.
We note that in Ref.~\refcite{king1} a term linear in $\phi$ is included, 
which plays the role of the Polonyi field, and Starobinsky-like inflation is possible 
so long as the gravitino mass $m_{3/2} \lesssim 1$ PeV.
The soft supersymmetry-breaking parameters for this model were derived in Ref.~\refcite{king2}.

\subsection{$T$-Type Inflation}
\label{sec:ttype}

As an example of $T$-type inflation, we consider the model
first suggested in Ref.~\refcite{Cecotti}, which is
described by the superpotential  (\ref{cec}),
generalized by allowing 
the K\"ahler curvature to differ from 2/3. 
We take the superpotential to be \cite{klr}:
\begin{equation}
\label{C1}
W_I = \sqrt{3 \alpha} \, M \phi \,  \left(T - \frac{1}{2} \right) \,\left(2T \right)^{\frac{3 \alpha - 3}{2}} \, .
\end{equation}
In this case, supersymmetry cannot be 
broken by a constant term, as the minimum of the scalar potential is now found at 
\beq\label{Tinfmu}
T=\frac{1}{2}-\alpha \frac{\mu^2}{M^2} \ , \quad \phi_1 = \sqrt{3 \alpha}\frac{\mu}{M} \, ,
\eeq
and the cosmological constant $V_0 \simeq  -3 \, \mu^2 < 0$.
We can add in this case a Polonyi field \cite{pol}
with K\"ahler potential given by
\beq\label{K_pol}
K \supset z z^*-\frac{(z z^*)^2}{\Lambda_{z}^2} \, ,
\eeq
where we include the strong stabilization of $z$ \cite{strongpol,dlmmo,ADinf,ego,EGNO4,enov4}.
We consider both possibilities of untwisted and twisted Polonyi fields. 
The superpotential is given by Eq.~(\ref{polonyi}), but strong stabilization 
shifts the minimum to $z\simeq  \Lambda_z^2/3\sqrt{12} \; (\Lambda_z^2/\sqrt{12})$, for the untwisted (twisted) case, 
with the parameter $\zeta\simeq 1/\sqrt{3}$ tuned to yield a 
vanishing cosmological constant \cite{dlmmo,ADinf,ego} when $\alpha = 1$.  
More generally, if we combine the Polonyi sector with the 
inflationary sector there is a shift in the
supersymmetry-breaking minimum \cite{enov4}:
\begin{equation}
\label{shiftone}
\begin{aligned}[c]
& {\rm Untwisted~Case}: \\
& \langle T \rangle \simeq \frac{1}{2} + \left(\frac{2 \alpha - 1}{ 3 \alpha}\right) \Delta^2,\\
& \langle \phi \rangle \simeq \Delta,\\
& \langle z \rangle \simeq \frac{\sqrt{\alpha}}{6 \sqrt{3}} \Lambda_z^2, \\
& \zeta  \simeq \frac{1}{\sqrt{3 \alpha}} - \left( \frac{1 + 3 \alpha(\alpha - 1)}{6 \sqrt{3} \alpha^{3/2}} \right) \Delta^2,
\end{aligned}
\qquad \qquad
\begin{aligned}[c]
& {\rm Twisted~Case}: \\
& \langle T \rangle \simeq \frac{1}{2} + \frac{2 \alpha}{3} \Delta^2,\\
& \langle \phi \rangle \simeq {\sqrt{\alpha}} \Delta,\\
& \langle z \rangle \simeq \frac{1}{2 \sqrt{3}} \Lambda_z^2, \\
& \zeta  \simeq \frac{1}{\sqrt{3}} - \frac{\sqrt{3} \alpha^2}{6} \Delta^2 \, ,
\end{aligned}
\end{equation}
where we define $\Delta \equiv \mu/M$ and assume that $\Delta, \, \Lambda_z \ll 1$. 
In this case, the form of the inflationary potential is unmodified from the Starobinsky form, save for the shift of the position of the minimum from $t_0 = 0$ to $t_0=(\frac{1}{\sqrt{\alpha}}-2\sqrt{\alpha}) \sqrt{2/3}\Delta^2$ and $t_0=-2\alpha^{3/2} \sqrt{2/3}\Delta^2$ for the untwisted and twisted cases respectively, where
\beq
T=\frac{1}{2}\left(e^{-\sqrt{\frac{2}{3\alpha}}t}+i\sqrt{\frac{2}{3}}\sigma\right) \, ,
\eeq
and $t$ and $\sigma$ denote the canonically-normalized real and imaginary part of $T$, respectively, and we associate $t$ with the inflaton. 
The supersymmetry-breaking scale given by the gravitino mass is $m_{3/2}= \mu/\sqrt{3\alpha} \, (\mu/\sqrt{3})$ for untwisted (twisted) $z$.
The mass of  the canonically-normalized Polonyi field is $m_z^2 = 36 m_{3/2}^2/\Lambda_z^2 \; (12 m_{3/2}^2/\Lambda_z^2)$
and, as discussed earlier, is hierarchically larger than $m_{3/2}$, thereby alleviating the cosmological Polonyi problem \cite{ego,enov4}. 

When combined with the superpotential (\ref{w_phi}) that includes matter fields, 
we obtain the following universal soft supersymmetry-breaking parameters \cite{EGNO4,enov4} when the Polonyi field is untwisted:
\begin{equation}
\begin{aligned}[c]
& {\rm Untwisted~Matter~Fields:} \\
& m_0^2 = \left(\alpha - 1 \right)m_{3/2}^2 , \\
& B_0 = -m_{3/2} , \\
& A_0 = 0,
\end{aligned}
\qquad \qquad
\begin{aligned}[c]
& {\rm Twisted~Matter~Fields:} \\
& m_0^2 = m_{3/2}^2 , \\
& B_0 =  -m_{3/2} , \\
& A_0 = 0 .
\label{softpoluntw}
\end{aligned}
\end{equation}
We note that in this case there is no dependence on the modular weights. 
As one can see, the only dependence of the soft supersymmetry-breaking terms on the curvature parameter $\alpha$
that appears in the soft scalar masses for untwisted matter fields. 
When $\alpha = 1$, we have vanishing input scalar masses, as is typical in no-scale models.
When $\alpha = 2$, we obtain $m_0 = m_{3/2}$, $B_0 = -m_{3/2}$, and $A_0 = 0$, which is the same as the pattern of soft terms when matter fields are twisted. These are of the mSUGRA type when the gaugino masses are of order $m_{3/2}$ and of the PGM type if gaugino masses are generated through anomalies. 

When the Polonyi field is twisted we find
\begin{equation}
\begin{aligned}[c]
& {\rm Untwisted~Matter~Fields:} \\
& m_0^2 = \alpha \, m_{3/2}^2 , \\
& B_0 = -m_{3/2} , \\
& A_0 = 0,
\end{aligned}
\qquad \qquad
\begin{aligned}[c]
& {\rm Twisted~Matter~Fields:} \\
& m_0^2 = m_{3/2}^2 , \\
& B_0 =  -m_{3/2} , \\
& A_0 = 0.
\end{aligned}
\end{equation}
The soft terms for twisted matter fields are unchanged from Eq. (\ref{softpoluntw})
and, as before, only the untwisted matter fields have a dependence on $\alpha$. In this case, because $m_0 = \sqrt{\alpha} m_{3/2}$,
the only restriction we have is  $\alpha > 0$. 

\subsection{Unified no-scale attractors}
\label{unified}

As an alternative to breaking supersymmetry with a Polonyi field,
we can use the two fields in the inflationary sector to 
break supersymmetry, leaving a small residual vacuum energy 
that could be identified as dark energy~\cite{enov2,enov3,enov4}.
Such models have been called unified no-scale attractors.   
We consider a simple case with a single modulus, $T$, and a single matter field, $\phi$,
that we associate with the inflaton, 
so that $\xi = 2T - \phi^2/3$.
We specialize to the case $\alpha = 1$,
and refer the interested reader to Ref.~\refcite{enov3}
for a generalization. Such models can also 
be formulated with a twisted inflaton, $\vphi$, see Ref.~\refcite{enov4}. 

In this simple set-up, the inflationary superpotential (\ref{Wf}) 
reduces to (\ref{wi}), and the de Sitter superpotential (\ref{mink5})
reduces to 
\begin{equation}
W_{dS} =  \lambda_1 M^3 - \lambda_2 M^3 \left(2T - \frac{\phi^2}{3} \right)^3 \, ,
\end{equation}
as $n=0$ and ${\bar n} = 3$,
and we have scaled the constants in $W_{dS}$ with the cube of the inflaton mass, $M$.
The superpotential then becomes 
\begin{equation}
W = W_{I} + W_{dS} = M \left(\frac{\phi^2}{2} - \frac{\phi^3}{3 \sqrt{3}} \right) + {\lambda}_1 M^3 - { \lambda}_2 M^3 \left(2T - \frac{\phi^2}{3} \right)^3 \, .
\label{uniwz}
\end{equation}
The unified Wess-Zumino model~(\ref{uniwz}) with the fields fixed at $\langle T \rangle = \frac{1}{2}$ and $\langle {\rm Im}~\phi \rangle = 0$
then yields the following scalar potential:
\begin{equation}
V \; = \; 12  { \lambda}_1 { \lambda}_2 M^6 + 12 { \lambda}_2  M^4 \left(\frac{\phi^2}{2} - \frac{\phi^3}{3 \sqrt{3}} \right) + 3M^2 \left(\frac{\phi}{\sqrt{3} + \phi} \right)^2 \, ,
\label{potphi}
\end{equation}
which becomes
\begin{equation}
V \; = \; 12  { \lambda}_1 { \lambda}_2 M^6 + 6 { \lambda}_2 M^4 \tanh^2 \left(\frac{x}{\sqrt{6}} \right) \left(3 - 2 \tanh \left(\frac{x}	{\sqrt{6}} \right) \right) + \frac{3}{4}M^2 \left(1 - e^{- \sqrt{\frac{2}{3}} x} \right)^2
\label{unistaro1}
\end{equation}
after the canonical field redefinition~(\ref{phigen}).

The first term in (\ref{unistaro1}) is a cosmological constant
with the value $\Lambda = 12 { \lambda}_1 { \lambda}_2  M^6$.  
The vacuum energy density will be modified by contributions from
phase transitions occurring after inflation, which are negative in general. 
If ${ \lambda}_{1,2} \sim \mathcal{O}(1)$,
these contributions should be of order $M^6 \sim 10^{-30}$ 
in order to (almost) cancel the first term in (\ref{unistaro1}) and
yield a net cosmological constant 
of order $10^{-120}$ today, in natural units. We note in this connection that
the GUT phase transition in a flipped SU(5)$\times$U(1) GUT
model is expected to occur after inflation \cite{EGNNO3} and to contribute
$\Delta V \sim - M_{\rm susy}^2 M_{\rm GUT}^2 \sim - ({ \lambda}_1 - { \lambda}_2)^2 M^6 M_{\rm GUT}^2$, suggesting that a hierarchy
${ \lambda}_1/{ \lambda_2}$
or ${ \lambda}_2/{ \lambda_1} \sim  (M_{\rm GUT}/M_P)^2$ should be preferred (see below).

The second term in (\ref{unistaro1}) perturbs the inflaton potential given by the
third term in (\ref{unistaro1}), but
 is unimportant for the inflationary dynamics, since it has a prefactor of $M^4$,
 whereas the inflationary potential is scaled by $M^2$. 
 This term adds a relatively small amount, $6 { \lambda}_2 M^4$,
 to the Starobinsky plateau height $(3/4) M^2$ at large $x$.

At the end of inflation, supersymmetry is broken through an $F$-term for $T$, which is given by~\cite{enov2, enov3}
\begin{equation}
\label{fterm}
\sum_{i = 1}^{2} |F_i|^2 = F_T^2 \simeq \left(\lambda_1 + \lambda_2 \right)^2 M^6 \, ,
\end{equation}
where $F_T$ is derived using Eq.~(\ref{fterm0}),
and the gravitino mass is given by
\begin{equation}
\label{grav}
m_{3/2} = e^{G/2} = e^{K/2} W = \left(\lambda_1 - \lambda_2 \right) \frac{M^3}{M_P^2} \, .
\end{equation}
To obtain a gravitino mass $m_{3/2} \simeq \mathcal{O} (1) \, \text{TeV}$, we choose $\lambda_2 \ll \lambda_1$, in which case 
\begin{equation}
\label{lambda}
m_{3/2} = \left(\lambda_1 - \lambda_2 \right) \frac{M^3}{M_P^2} \simeq \lambda_1 \frac{M^3}{M_P^2} \, ,
\end{equation}
and $F_T \simeq m_{3/2} $.
By scaling $W_{dS}$ with $M^3$,
we are able to obtain a TeV mass scale for supersymmetry breaking 
without fine-tuning. Furthermore, we relate the supersymmetry-breaking
scale to the inflation scale $M$ (see also Refs.~\refcite{dgmo,kmov}).

It is now relatively straightforward to add Standard Model fields to the model, and the full superpotential can be written as 
\begin{equation}
W = W_I + W_{dS} + W_{SM} \, , 
\label{fullW}
\end{equation}
where $W_{SM}$ can be written in a form similar to Eq.~(\ref{w_phi}). 
For $\alpha \ne 1$ and recalling the definitions 
$n_{\pm} = \frac{3}{2} \left( \alpha \pm \sqrt{\alpha} \right)$ (\ref{npm}),
we should multiply the expression in (\ref{w_phi}) by a factor of
$\sqrt{\alpha} \xi^{n_-}$, where
$\xi = 2T - \frac{\phi^2}{3}$ ($\xi = 2T$), assuming here that the inflaton is untwisted (twisted).
Similar results can be obtained using instead $ \sqrt{\alpha} \xi^{n_+}$.
The resulting soft terms are very similar to those in Eq.~(\ref{sft_mu_1}) \cite{enov4}:
\begin{equation}
\label{termsphi}
\begin{aligned}[c]
& {\rm Untwisted~Matter~Fields:} \\
& m_0^2 = \left(\alpha - 1 \right) m_{3/2}^2 \, , \\
& B_0 = \left(2 \sqrt{\alpha} - 2 - \beta \right) m_{3/2} \, , \\
& A_0 = \left(3 \sqrt{\alpha} - 3 - \gamma \right) m_{3/2} \, ,
\end{aligned}
\qquad \qquad
\begin{aligned}[c]
& {\rm Twisted~Matter~Fields:} \\
& m_0^2 = \frac{(\alpha - n_a)}{\alpha} m_{3/2}^2 \, , \\
& B_0 = \left(2 \sqrt{\alpha}  - 2 n_a - \sigma \right) m_{3/2}\, , \\
& A_0 = \left(3 \sqrt{\alpha} - 3 n_a - \rho \right) m_{3/2} \, .
\end{aligned}
\end{equation}
For $\alpha = 1$, these results reduce to those in Eq.~(\ref{sft_mu_1})\cite{EGNO4}.

In the superpotentials in Eqs.~(\ref{uniwz}) and (\ref{fullW}), the volume modulus plays the role of the Polonyi field and is responsible for supersymmetry breaking.  As a result, it is subject to similar cosmological constraints as the Polonyi field
to avoid excessive entropy production or an excessive
dark matter abundance produced by modulus decay.  
However the strong stabilization of $T$ helps resolve these issues in a similar way to a 
strongly-stabilized Polonyi field.\cite{ego}

\subsection{Constraints on the Stabilization Parameter $\Lambda_T$}

As noted in Section~\ref{sec:susyb}, cosmological problems\cite{polprob} arise when the modulus field value after inflation is displaced from its potential minimum. In the absence of strong stabilization,
the displacement is $\mathcal{O}(M_P)$. Stabilization reduces the amplitude to a maximum displacement of $\Lambda_T/4\sqrt{3}$. Thus, after the period of exponential expansion, both the inflaton, $\phi$, and the modulus, $T$, undergo scalar field oscillations 
until they decay.  These oscillations begin when the Hubble parameter, $H$, is roughly $\frac23 m_{\phi,t}$,
and $H = (\rho/3)^{1/2}$ may be determined by the energy density, $\rho$, stored in either inflaton or modulus oscillations, or the radiation energy produced by
inflaton decays. In this model, $m_\phi = M$, and the mass of the canonically-normalized modulus is $m_t$, given by $1/\sqrt{3\alpha}$ times the value in Eq.~(\ref{mT}).
We parametrize the inflaton decay rate here as
\beq
\Gamma_\phi = d_\phi^2 \frac{M^3}{M_P^2} \, ,   
\eeq    
where $d_\phi$ is a model dependent gravitational-strength coupling.

We can distinguish several possible histories for the fields $\phi$ and $T$, depending on the various masses and the inflaton decay coupling, $d_\phi$\cite{enov4}.
(See also Ref.~\refcite{Kane:2015jia} for a review of cosmological moduli.)
Here we focus on one scenario in particular, where $T$ oscillations begin before inflaton decay. We refer to this as Scenario I (as opposed to Scenario II when $T$ oscillations begin after inflaton decay). Furthermore, we can also distinguish if $T$ decays before the inflaton (I~a), after the inflaton but before $T$ oscillations dominate the energy density (I~b), and when they do dominate the energy density (I~c). These regions are shown in the $(m_{3/2}, \Lambda_T)$ plane in Fig.~\ref{LambdaT} for $d_\phi = 10^{-3}$. The derivations of the boundaries of these subregions can be found in Ref.~\refcite{enov4}.

\begin{figure}[ht!]
\centering
\includegraphics[scale=0.42]{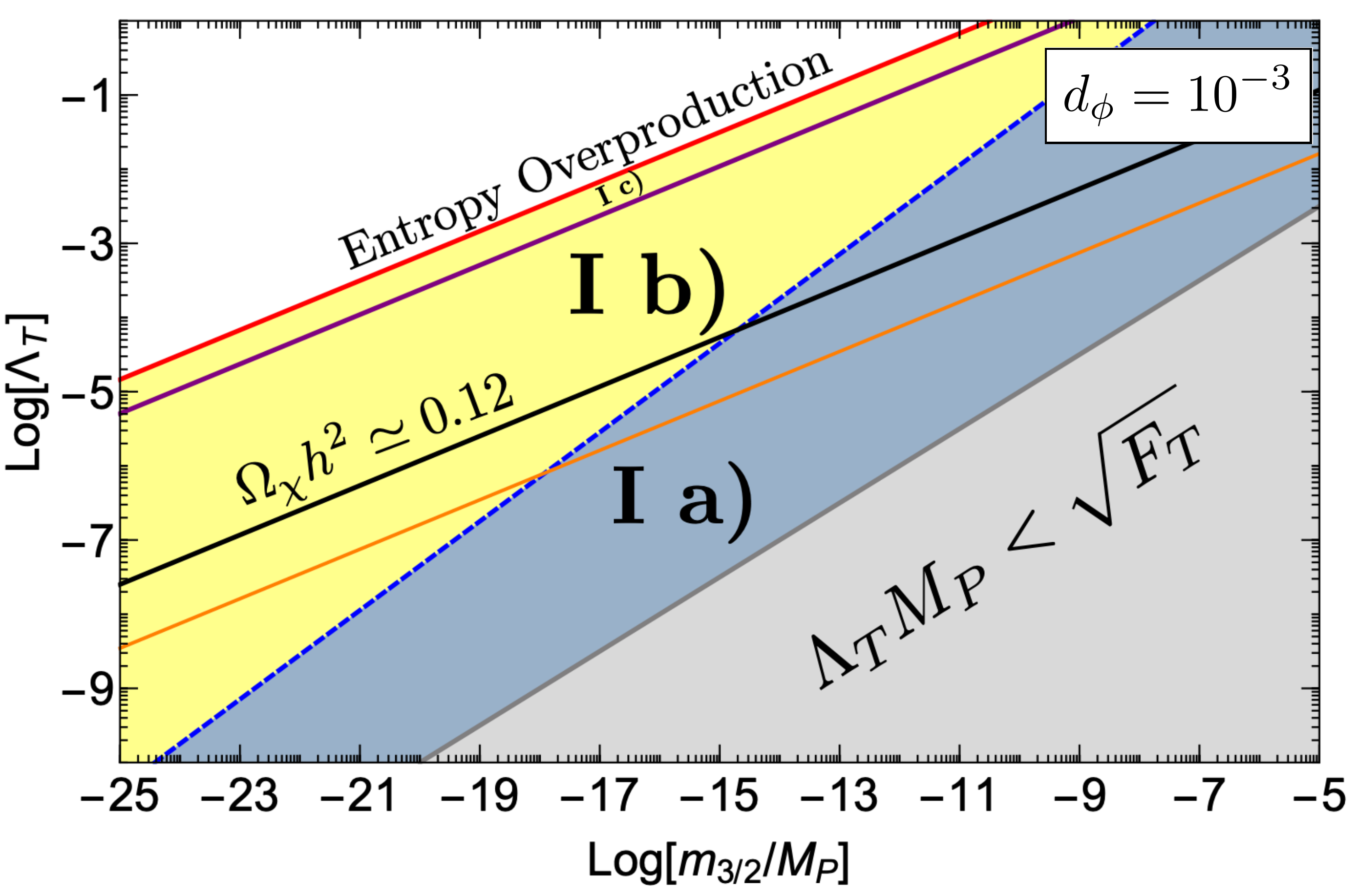}
\caption{\it Plot of the constraints on the modulus stabilization parameter, $\Lambda_T$,
as a function of the gravitino mass, $m_{3/2}$,
for models with $\alpha = 1$ and $d_\phi = 10^{-3}$. 
The regions shaded yellow and blue correspond, respectively, to the
Scenarios  I~b, c), and I~a) described in the text, whereas the grey regions are excluded
by the effective interaction condition (\ref{effint}). Regions between this and the
dark matter density constraint (solid black line) are allowed by all the constraints.
}
\label{LambdaT}
\end{figure}

As long as $T$ decays while the Universe is either dominated by the inflaton oscillations (I~a)
or radiation produced by inflaton decays (I~b),
the entropy produced by $T$ decays is negligible. 
However, if $T$ decays late, when it dominates the total energy budget, a significant amount of entropy production is possible. The amount of allowed entropy production
is model-dependent, and in some cases necessary. For example, in the case of Affleck-Dine
baryogenesis\cite{AD}, the initial baryon-to-entropy ratio may be large, and some dilution due to late inflaton decays\cite{eeno} and/or moduli decays\cite{cgmo}
is welcome.  It is possible to derive an upper limit on $\Lambda_T$:~\cite{enov4}
\begin{equation}
    \Lambda_T \lesssim 2 \alpha^{1/6} \sqrt{3} \left(\frac{256}{\pi} \right)^{1/9} d_{\phi}^{-2/9} \Delta^{2/9} \left( \frac{m_{3/2}}{M} \right)^{1/3} \, ,
    \label{entlim}
\end{equation}
where $\Delta$ is the maximum allowed dilution factor.
This limit is shown in Fig.~\ref{LambdaT} as the 
boundary above region I~c, assuming $\Delta < 100$.

A bound stronger than that due to entropy production
can be derived from the production of cold dark matter,
since $T$ decays into pairs of gravitinos that subsequently decay into LSPs could result in an overabundance of cold dark matter. We can derive the following upper limit on $\Lambda_T$:~\cite{enov4}
\begin{equation}
    \Lambda_T \lesssim 2 \times 10^{-3} \, \alpha^{-1/6} \, d_\phi^{-1/3} \frac{m_{3/2}^{1/3} M_P^{1/6}}{M^{1/2}}
    \left(\frac{m_{\chi}}{100~\text{GeV}} \right)^{-1/3} \, .
    \label{DMdensity}
\end{equation}
This limit is seen as the black line cutting through regions I~a) and I~b) in Fig.~\ref{LambdaT}, as labelled. 
Below the orange line (running parallel and below the black line), thermal production of gravitinos equals that from $T$ decays.

Finally, we note that there is a lower limit on $\Lambda_T$ coming from the postulated form of the stabilization terms in the K\"ahler potential. 
Since these should be treated as effective interactions obtained by integrating out fields with masses ${\cal O}(\Lambda_T)$, we require $\Lambda_T > \sqrt{F_T}$~\cite{strongpol,dgmo,kmov}, and using~(\ref{fterm}) we find that
the limit
\begin{equation}
    \Lambda_T >  \alpha^{-1/4} \left(\frac{m_{3/2}}{M_P}\right)^{1/2}
    \label{effint}
\end{equation}
is imposed by the effective interaction assumption. The area violating this limit is shaded
grey in Fig.~\ref{LambdaT}.

\subsection{Reheating}
\label{sec:reheat}

No model of inflation is complete without a discussion of reheating. On the one hand, it
is essential for making contact with physics at the TeV scale and the Standard Model and, 
on the other hand, the amount of reheating affects the estimate of $N_*$ and hence, in particular, the 
amount of scalar tilt predicted within any given model.
As in the previous Subsections, we treat the cases of $\phi$-type and $T$-type inflation separately.

\subsubsection{Reheating in $\phi$-type models}
\label{sec:reheatingphi}

Reheating is determined by the couplings of the inflaton to SM fields, which depend on the forms of
$K, \, W$, and their derivatives. 
If we assume that all matter fields $\{\phi, \, \vphi\}$ have vanishing VEVs at the end of inflation, we find that
\beq\label{VEV_cond}
\langle W^i\rangle = \langle W^a\rangle = 0 \ , \quad \langle K^i\rangle = \langle K^a\rangle =0 \, ,
\eeq
where $i, a$ are indices for untwisted and twisted fields, respectively, and hence in terms of the K\"ahler function,
\beq
G^{i}=G^a=0.
\eeq
In contrast, $\langle G_T \rangle = -3$, and we have $\langle W \rangle = \mu$.

If, as we have assumed until now, there are no direct superpotential couplings between 
$\phi$ and other fields, so that $W_{1I} = 0$, where the index 1 corresponds to the inflaton,
and the index $I = \{ T, i, a\}$ represents the volume modulus $T$, untwisted matter fields, and twisted matter fields, then there are no decay channels directly to matter scalars \cite{EKOTY,EGNO4}.
Note, however, that by adding a superpotential term such as $ \zeta \phi_1 (T-1/2)^2$, which does not affect the
dynamics of inflation, we obtain a decay to moduli with rate
\beq\label{phitoT}
\Gamma(\phi_1 \rightarrow \delta T\,\delta T) = M \frac{|\zeta|^2}{72\pi}\, ,
\eeq
where $\delta T=\sqrt{3}(T-1/2)$ is the canonically-normalized modulus fluctuation. 
Decays to a pair of gravitinos would proceed at the same rate \cite{EGNO4}. 
However, if these are dominant they could lead to overproduction of 
the LSP that, if it is stable, would exceed the bounds on the cold 
dark matter density. 
 
Alternatively, a direct coupling between $\phi_1$ and the matter sector may be allowed. For example, 
this field may be associated with a heavy singlet sneutrino \cite{snu,ENO8}. In such a case, one can consider  the addition of a Yukawa-like term
\beq\label{w_rhn}
\Delta W = y_{\nu} H_u L\phi_1
\eeq
to the Standard Model superpotential, where $y_{\nu}$ denotes the Yukawa coupling. Such a coupling leads to a scalar mass matrix characteristic of seesaw models:
\beq
\left( \begin{matrix}
\bar{\phi}^1 & \bar{\tilde{\nu}}
\end{matrix}
\right) \left(
\begin{matrix}
M^2 + \tilde{m}^2 & -M\tilde{m} \\
-M\tilde{m} & \tilde{m}^2+\kappa\mu^2
\end{matrix}
\right) \left(
\begin{matrix}
\phi_1\\
\tilde{\nu}
\end{matrix}
\right) \, ,
\eeq
where $\tilde{m}\equiv y_{\nu} \langle H_u \rangle = y_{\nu} v \sin\beta$, $\kappa=(1-n_{\nu})$ for a twisted neutrino, $\kappa=0$ for an untwisted neutrino, and $v\simeq 174$~GeV is the SM Higgs VEV.
Even in the presence of direct couplings, we can consider $\phi_1$ to be
the inflaton mass eigenstate, up to corrections of order $\mu/M,v/M\ll 1$.

In this case, we have a coupling 
$-M y_{\nu}{H^*}_u {\tilde{L}}^*\phi_1$, and
the inflaton decay width is given by
\beq\label{phisneu}
\Gamma(\phi_1 \rightarrow H_u^0\tilde{\nu},H_u^+\tilde{f}_L) = M \frac{|y_{\nu}|^2}{16\pi}\, ,
\eeq
where we have neglected the masses of the final-state particles. This decay rate would be
fast if $|y_\nu | \sim {\cal O}(1)$. However, in order to avoid problems associated with gravitino overproduction 
during reheating, we must set a bound on the Yukawa coupling associated with the inflaton:~\cite{ENO8}
\beq
\label{ylim}
y_\nu \la 10^{-5} \, ,
\eeq
leading to a corresponding constraint on the reheating temperature, $T_{\rm RH}$, whose derivation we discuss below.

As in the scalar case, all couplings to matter fermions vanish for a $\phi_1$-independent matter superpotential. 
However, the decay into a fermion and a higgsino is possible if we identify $\phi_1$ with a singlet neutrino, 
with superpotential (\ref{w_rhn}). In this case, the rate is given by 
\beq\label{phineu}
\Gamma(\phi_1\rightarrow \tilde{H}_u^0\nu,\,\tilde{H}_{u}^+f_L) =  M \frac{|y_{\nu}|^2}{16\pi} \, ,
\eeq
i.e., equal to the rate of decay into scalars. 

If we assume instantaneous inflaton decay and thermalization of decay products, 
we can easily relate the reheating temperature to the inflaton decay rate \cite{dg,nos}. 
After the period of exponential expansion,
the inflaton begins to oscillate about its minimum when the Hubble parameter, $H = \frac23 M$.
If we define $R_\phi$ as the scale factor when inflaton oscillations begin,
we can write the energy density and Hubble parameter as
\begin{equation}
\label{energy1}
\rho_{\phi} \simeq \frac{4}{3} M^2 M_P^2 \left( \frac{R_{\phi}}{R} \right)^3 \, ,
\end{equation}
\begin{equation}
\label{hubble1}
H \simeq \frac{2}{3} M \left(\frac{R_{\phi}}{R} \right)^{3/2} \, .
\end{equation}
We can further define the time of decay by $\Gamma_\phi t = 1$, or $H = \frac23 \Gamma_\phi$. Then, if all of the energy density in oscillations
is converted to radiation we have
\beq\label{trehdef}
T_{\rm RH} = \left(\frac{40}{g_{\rm RH}\pi^2}\right)^{1/4}\left(\Gamma_{\phi}M_P\right)^{1/2}\,, 
\eeq
where $g_{\rm RH} \equiv g \left( T_{\rm RH} \right)$ is the number of effective degrees of freedom in the thermal bath at $T_{\rm RH}$.
For $M=3 \times 10^{13}$ GeV and $g_{\rm RH} = 915/4$,
we have
\beq
T_{\rm RH} = 6.2 \times 10^{14} y_\nu~{\rm GeV} \, .
\eeq
The abundance of gravitinos produced thermally can be expressed as~\cite{egnop}
\beq
\frac{n_{3/2}}{s} \simeq 2.6 \times 10^{-4} \left( 1 + 0.56 \frac{m_{1/2}^2}{m_{3/2}^2} \right) \left(\frac{\Gamma_\phi}{M_P}\right)^{1/2} \, ,
\eeq
and requiring that the abundance of dark matter
produced by gravitino decay is $\Omega_\chi h^2 < 0.12$
we have 
\beq
\frac{n_{3/2}}{s} < 4.4 \times 10^{-12} \left( \frac{100 {\rm GeV}}{m_\chi} \right) \, ,
\eeq
leading to the limit (\ref{ylim}) imposed by the production of gravitinos from the thermal bath \cite{bbb,cefo,Pradler:2006qh,ps2,rs,kkmy,egnop,Garcia:2018wtq}.

If the gauge kinetic term, $f_{\alpha \beta}$, is non-trivial 
and depends on $\phi_1$, $f_{\alpha\beta}=f(\phi_1)\delta_{\alpha\beta}$, a decay channel for
inflaton decay into gauge fields and gauginos is also possible~\cite{EKOTY,klor,EGNO4}. 
So long as supersymmetry is not broken by the inflaton ($F_{\phi_1} = 0$), this term does not contribute to gaugino masses, which are proportional to the $F$-term and derivatives of the gauge kinetic function as seen in Eq.~(\ref{m12}).
The decay widths
to canonically-normalized gauge boson pairs and gauginos are~\cite{EKOTY}
\beq
\Gamma(\phi_1\rightarrow gg) = \Gamma(\phi_1 \rightarrow \tilde{g}\tilde{g}) = \frac{3d_{g,1}^2}{32\pi}\left(\frac{N_G}{12}\right)\frac{M^3}{M_P^2}\,,
\eeq
where $N_G = 12$ in the Standard Model, and $d_{g,1}$ is given by
\beq
d_{g,1} \equiv \langle {\rm Re}\,f\rangle^{-1}\left|\left\langle\frac{\partial f}{\partial \phi_1}\right\rangle\right| \, .
\eeq
This leads to a reheating temperature of 
\beq
T_{\rm RH} = 6.7 \times 10^{9}~d_{g,1}~{\rm GeV} \, .
\eeq
Thus the cold dark matter density is roughly saturated by 
thermally-produced gravitinos during reheating when the coupling $d_{g,1} \sim 1$.

\subsubsection{Reheating in $T$-type models}

We again assume the absence of VEVs for matter fields and that the conditions (\ref{VEV_cond}) hold.
In contrast to the case of a matter-like inflaton,
the $T$ field couples to the matter sector through the supergravity Lagrangian.
Two-body decays are possible with a rate given by \cite{EGNO4}
\beq
\Gamma(T\rightarrow \Phi_I\bar{\Phi}^J) = (n_I + n_L - 3)^2\frac{|W^{IL}\bar{W}_{LJ}|^2}{48\pi M M_P^2}\,,
\eeq
where the $n_{I,L}$ are modular weights, and a sum over the repeated index $L$ is implied. 
This rate is weak-scale suppressed in the case of MSSM scalars.
For example, the rate for decay to two Higgs bosons is
\beq
\Gamma(T\rightarrow H_{u,d}\bar{H}^{u,d}) = (2n_H-3)^2\frac{|\mu_H|^4}{24\pi M M_P^2}\ ,
\eeq
where $\mu_{H}$ denotes the bilinear Higgs coupling in the MSSM. This two-body rate would
lead to an extremely low reheating temperature:
for an inflaton mass $M\sim 10^{-5}M_P$, and $\mu_H \sim 1$ TeV, $T_{\rm RH}\sim 10^{-1}$ eV.
We note, however, that in the case of high-scale supersymmetry \cite{hssusy}, where all superpartners except the gravitino 
are heavier than the inflaton \cite{bcdm,eev,dgmo,dgkmo,egko,kmov,Garcia:2018wtq}, the decay to Higgs pairs may be the dominant decay channel \cite{dgmo,dgkmo,kmo,kmov}. 

Three-body decays to light scalars are  possible and the widths are given by
\begin{align}
\Gamma(T\rightarrow \Phi_I\bar{\Phi}^J\bar{\Phi}^K) &= (n_I+n_L-3)^2 \, \frac{|W^{IL}\bar{W}_{LJK}|^2M}{12(8\pi)^3M_P^2}\ ,\\
\Gamma(T\rightarrow \Phi_I\Phi_J\bar{\Phi}^K) &= (n_I+n_J+n_L-3)^2 \, \frac{|W^{IJL}\bar{W}_{LK}|^2M}{12(8\pi)^3M_P^2}\ .
\end{align}
For example, the decay to the neutral $d$-type Higgs and the left and right stops has the rate
\beq\label{Htt}
\Gamma(T\rightarrow \bar{H}_d^0\bar{\tilde{t}}_R \tilde{t}_L,\, H_d^0\tilde{t}_R \bar{\tilde{t}}_L) = \left((2n_H-3)^2+(2n_t+n_H-3)^2\right)\frac{|\mu_H y_t|^2M}{4(8\pi)^3M_P^2}\,,
\eeq
where $y_t$ denotes the top Yukawa coupling. If this were dominant, the corresponding reheating temperature would also
be low, in the MeV range.

However, we find that the rates for four-body decays are the largest, despite being phase-space suppressed. The decay width
\beq
\Gamma(T\rightarrow \Phi_I\Phi_J\bar{\Phi}^K\bar{\Phi}^M) = (n_I+n_J+n_L-3)^2 \, \frac{|W^{IJL}\bar{W}_{LKM}|^2M^3}{72(8\pi)^5M_P^2} \, ,
\eeq
where we have disregarded the bilinear couplings, implies the following decay rate to four stops
\beq
\Gamma(T\rightarrow \tilde{t}_R\tilde{t}_L\bar{\tilde{t}}_R\bar{\tilde{t}}_L) = (2n_t+n_H-3)^2 \, \frac{|y_t|^4M^3}{8(8\pi)^5M_P^2}\, ,
\label{4stops}
\eeq
which corresponds to
\beq
T_{\rm RH} = |2n_t+n_H-3|(4.3 \times 10^{6}\,{\rm GeV})|y_t|^2 \, .
\label{T4}
\eeq
Thus, as long as the matter fields do not reside in the untwisted sector 
(for which $n_i = 1$ and the rate 
vanishes), we can obtain a reheating temperature well above that required for successful
Big Bang Nucleosynthesis (BBN).  

There are also decays to matter fermions.
The rates for two-body decays to matter fermions take the form
\beq
\Gamma(T\rightarrow\bar{\chi}_I\chi_J)= (n_I+n_J-3)^2\frac{|W^{IJ}|^2M}{192\pi M_P^2} \, ,
\eeq
which are 1/4 of the rate for three-body decays into scalars. 

The dominant rates are for three-body decays involving two fermions and one matter scalar, which are
\beq\label{Tpcbc}
\Gamma(T\rightarrow \bar{\chi}_I\chi_J\Phi_K) = (n_I+n_J+n_K-3)^2\frac{|W^{IJK}|^2M^3}{36(8\pi)^3M_P^2} \, .
\eeq
These are non-vanishing in the MSSM so long the fields are twisted with weights $n_i \ne 1$. In particular, in the case of the top quark one has the decay rate
\beq\label{THtt_f}
\Gamma(T\rightarrow H_u^0t_L\bar{t}_R, \, \tilde{t}_L\tilde{H}_u^0\bar{t}_R,\,\bar{\tilde{t}}_R t_L\tilde{H}_u^0) = (2n_t+n_H-3)^2\frac{|y_t|^2M^3}{12(8\pi)^3M_P^2} \, ,
\eeq
which is somewhat larger than the four-scalar decay rate (\ref{4stops}), because of the three-body phase-space factor.
This decay rate would lead to a reheating temperature 
\beq\label{TR_fer}
T_{\rm RH} = (8.9 \times 10^7\ {\rm GeV})\,|y_t(2n_t+n_H-3)| \, .
\eeq

Finally, if the gauge kinetic function depends on $T$, 
decays to gauge bosons and gauginos are also possible. 
The decay width to the canonically-normalized gauge boson pairs is readily evaluated, resulting in
\beq\label{T_gg_a}
\Gamma(T\rightarrow gg) = \frac{d_{g,T}^2}{32\pi}\left(\frac{N_G}{12}\right)\frac{M^3}{M_P^2}\,,
\eeq
where 
\beq\label{dgT}
d_{g,T} \equiv \langle {\rm Re}\,f\rangle^{-1}\left|\left\langle\frac{\partial f}{\partial T}\right\rangle\right| \, ,
\eeq
and the corresponding reheating temperature is
\beq
T_{\rm RH} =  (3.8 \times10^{9}\ {\rm GeV})\, d_{g,T} \, .
\label{Tgg}
\eeq
The coefficient $d_{g,T}$ might well be $\mathcal{O}(1)$, 
e.g., for a gauge kinetic function linear in $T$ with $\mathcal{O}(1)$ coefficients,
in which case all other decay channels of the volume modulus $T$ would be overwhelmed by the decays to gauge bosons,
and the reheating temperature would be large. In general, the effective reheating temperature generated by
decays into gauge bosons would exceed that due to decays into matter particles, (\ref{TR_fer}), for any $d_{g,T} \gtrsim {\cal O}(1/40)$.

On the other hand, the decays of $T$ to gauginos are subdominant, since
the corresponding decay rate is
\beq
\Gamma(T\rightarrow \tilde{g}\tilde{g}) = \frac{d_{g,T}^2}{16\pi}\left(\frac{N_G}{12}\right)\frac{m_{3/2}^2 M}{M_P^2}\,.
\eeq
We note that a similar suppression for the decay to gauginos was given in Ref.~\refcite{klor}.

\subsection{The Number of e-Folds in Representative No-Scale Inflation Models}
\label{sec:N*}

In the preceding Subsection we have seen the various possibilities for reheating and the 
related model dependence in inflaton decay
due to the reheating process. We now consider the implications for the number of e-folds $N_*$ in some
representative no-scale models of inflation.

In the slow-roll approximation
the number of e-folds to the end of inflation can be expressed as~\cite{LiddleLeach,MRcmb,planck18}
\beq
N_* = 66.9 - \ln\left(\frac{k_*}{R_0H_0}\right) + \frac{1}{4}\ln\left(\frac{V_*^2}{M_P^4\rho_{\rm end}}\right) + \frac{1-3w_{\rm int}}{12(1+w_{\rm int})}\ln\left(\frac{\rho_{\rm RH}}{\rho_{\rm end}}\right) - \frac{1}{12}\ln g_{\rm RH}\ ,
\label{howmany}
\eeq 
where $R_0$ and $H_0$ are the present 
cosmological scale factor and Hubble expansion rate, respectively, 
$V_*$ is the inflationary energy density at the reference scale, 
$\rho_{\rm end}$ and $\rho_{\rm RH}$ are the energy densities at the end of inflation and after reheating, 
respectively, and $w_{\rm int}$ is the {\it e-fold} average of the equation-of-state parameter during reheating.
Entropy conservation after reheating has been assumed, and we refer the interested reader to Ref.~\refcite{reheating} for details.

For Starobinsky-like models, Eq.~(\ref{howmany}) can be written in the form~\cite{reheating}
\begin{align}
N_* \;=\; &68.66 - \ln\left(\frac{k_*}{R_0H_0}\right) + \frac{1}{4}\ln\left(A_{s}\right) - \frac{1}{2} \ln \left( N_*-\sqrt{\frac{3}{8}}\frac{\phi_{\rm end}}{M_P} + \frac{3}{4}e^{\sqrt{\frac{2}{3}}\frac{\phi_{\rm end}}{M_P}} \right) \notag \\  
& - \frac{1}{12}\ln g_{\rm RH} + \frac{1-3w_{\rm int}}{12(1+w_{\rm int})}\Big[2.030+2\ln\left(\Gamma_{\phi}/M\right) -2\ln (1+w_{\rm eff}) \notag\\ 
&- 2\ln(0.655-1.082\ln\delta) \Big] \, , \label{res1}
\end{align}
where $\phi_{\rm end}$
is the value of the inflaton field at the end of inflation, $w_{\rm eff}$ is the {\it time}-average of the 
equation-of-state parameter during the reheating epoch, and $\delta$ parametrizes the degree of completion of reheating:
\beq
\Omega_{\rm rad} \;=\; \frac{\rho_{\rm rad}}{\rho_{\phi}+\rho_{\rm rad}} \;\equiv\; 1-\delta\,.
\eeq
We see from (\ref{res1}) that $N_*$ depends on $\Gamma_\phi$ both explicitly and implicitly via the dependences
in $w_{\rm int}$ and $w_{\rm eff}$.
The e-fold-averaged equation of state parameter, $w_{\rm int}$ in Starobinsky-like models
may be fit by~\cite{reheating}
\beq
\label{w_fit}
w_{\rm int} = \frac{0.782}{\ln(2.096 M/\Gamma_{\phi})}\,,
\eeq
and the time-averaged equation of state
$w_{\rm eff} \simeq 0.27$.
To calculate $N_*$ as a function of $\Gamma_\phi$, we use the Planck pivot point
$k_* = 0.05$/Mpc, corresponding to $k_*/R_0 H_0 = 221$, and take the MSSM value of $g_{\rm RH} = 915/4$.
Fig.~\ref{fig:Nstar} displays the calculated value of $N_*$ over a wide range of $\Gamma_\phi$, parametrized by
\beq\label{rate_1}
\Gamma_{\phi} \; = \; M \frac{|y|^2}{8\pi}\,,
\eeq
with a coupling
ranging from
$y = 1$ 
to $y \sim 10^{-18}$, in which the latter would correspond to a reheating temperature $T_{\rm RH} \sim 1$~MeV,
below which the successful conventional BBN calculations would need to be
modified substantially.  Within this overall range, we indicate in Fig.~\ref{fig:Nstar} the values of $N_*$ corresponding to the decay processes (\ref{THtt_f}) and (\ref{T_gg_a}), which are consistent with the CMB 
and gravitino constraints.

\begin{figure}[!ht]
\vspace{1cm}
\centering
	\scalebox{0.40}{\includegraphics{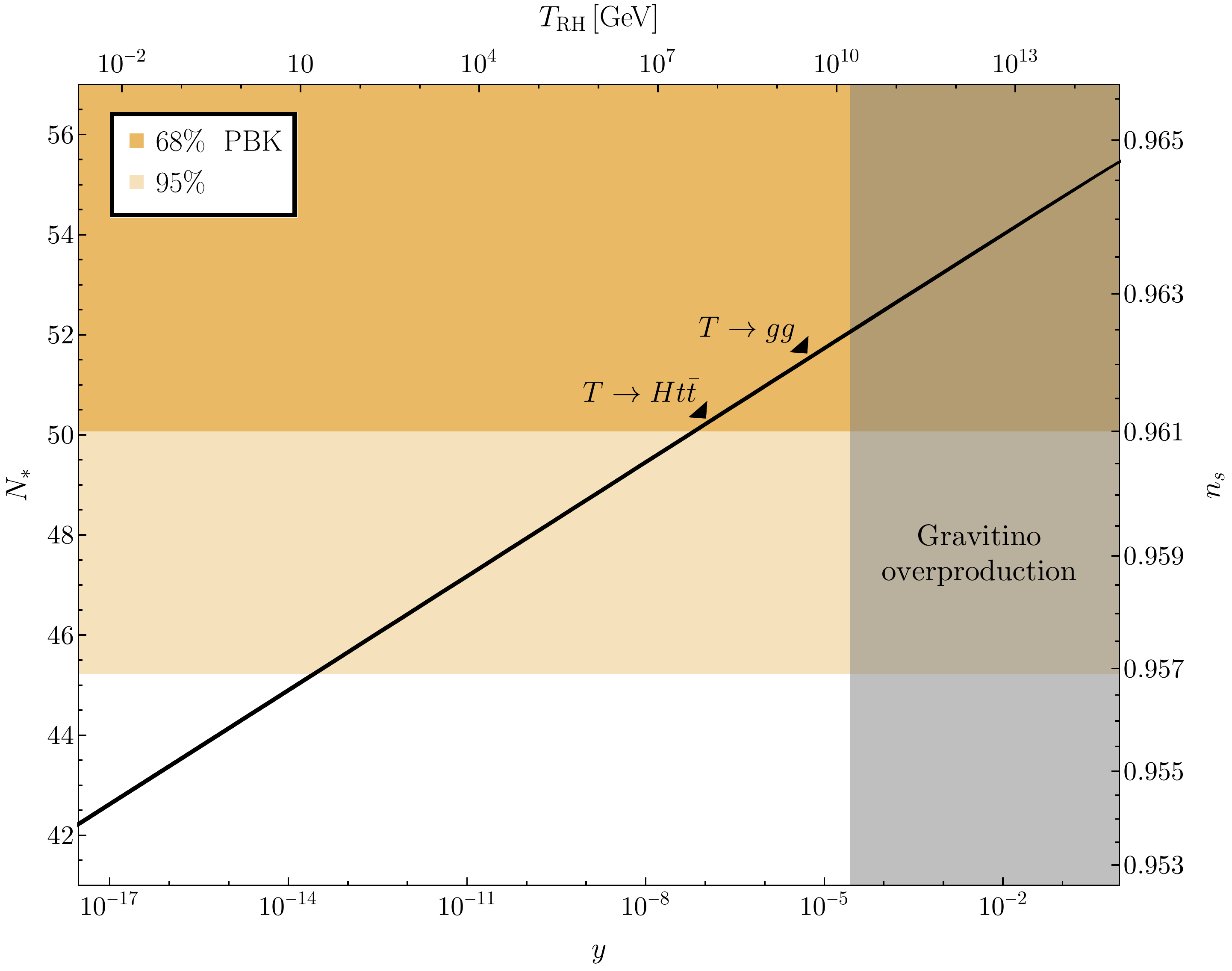}}
	\caption{\it The values of $N_*$ in no-scale Starobinsky-like models as a function of $y$ ($T_{\rm RH}$), for a wide range of decay rates. The diagonal black strip corresponds to the solution of (\ref{res1}) in the range $10^{-3}<\delta<10^{-1}$. The right vertical axis shows the values of $n_s$ in Starobinsky-like no-scale models, for which the tensor-to-scalar ratio varies over the range $0.0034<r<0.0057$ for $N_*$ in the displayed range. The horizontal light beige (orange) shaded region corresponds to the 95\%  (68\%) Planck+BICEP2/Keck (PBK) CL region from Eq.~(\ref{ns}). The vertical shaded region is excluded due to gravitino overproduction. Within it, in-medium and non-perturbative effects may also affect the inflaton decay rate. For illustrative purposes, the results corresponding to the decay processes (\ref{THtt_f}) and (\ref{T_gg_a}) are shown.} 
	\label{fig:Nstar}
\end{figure}

\section{Inflation and UV phenomenology}
\label{sec:UV}

So far we have seen that Starobinsky-like inflation can be constructed  
naturally in the context of 
no-scale supergravity models. The underlying $\mathrm{SU}(2,1)/\mathrm{SU}(2)\times \mathrm{U}(1)$ symmetry leads to a continuous class of phenomenological
models~\cite{eno7,enov1} in which the inflaton may be associated with either a modulus or a matter-like field. 
Ultimately though, we would like to be able to connect these ``low'' energy phenomenologies with
a UV completion of the theory. We anticipate that this should be a string theory incorporating all the gauge interactions
as well as gravity. However, because the appropriate theory is not known, we are more modest in our attempts
here, and examine the consequence of embedding the inflationary theory in the context of a GUT. Specifically, we consider three GUT models: SU(5), which offers a minimal way to realize the no-scale inflation in a GUT, and SO(10)~\cite{EGNNO1}, neither of which can in principle be obtained from perturbative heterotic string theory, and flipped SU(5)$\times$U(1)~\cite{EGNNO2,EGNNO3,EGNNO4,EGNNO5,EGNNO6}, which has been derived within such a string model.

\subsection{SU(5) GUTs and No-Scale Inflation}
\label{sec:unflipped}

We first consider a supersymmetric SU(5) GUT model that incorporates no-scale inflation. As in the minimal supersymmetric SU(5)~\cite{Dimopoulos:1981zb}, three generations of SM quarks and leptons are embedded into ${\bf 10}$ and $\overline{\bf 5}$ representations, $\Psi_i$ and $\Phi_i$, respectively, where $i$ is the generation index, while the MSSM Higgs fields, $H_u$ and $H_d$, reside in ${\bf 5}$ and $\overline{\bf 5}$ representations, $H$ and $\overline{H}$, respectively. The SU(5) GUT symmetry is spontaneously broken by a VEV of a ${\bf 24}$, $\Sigma$, down to the SM gauge group. In addition to these fields, we introduce an SU(5) singlet field, $S$, as the inflaton. We also assume that this model respects $R$-parity, so as to suppress dangerous baryon/lepton-number violating renormalizable operators; $H$, $\overline{H}$, $\Sigma$, and $S$ are $R$-parity even and the rest of the fields are $R$-parity odd.

The renormalizable superpotential for this model is given by 
\begin{align}
  W_5 &=  \mu_\Sigma {\rm Tr}\Sigma^2 + \frac{1}{6} \lambda^\prime {\rm
  Tr} \Sigma^3 + \mu_H \overline{H} H + \lambda \overline{H} \Sigma H
 \nonumber \\
 &+ \left(h_{\bf 10}\right) 
  \Psi \Psi H +
  \left(h_{\overline{\bf 5}}\right) \Psi \Phi
  \overline{H}
  \nonumber \\ 
  & + \frac{M}{2} \phi^2 - \frac{\lambda_\phi}{3} \phi^3 
  + \lambda_{H\phi} \phi \overline{H} H + \lambda_{\phi \Sigma} \phi {\rm Tr}\Sigma^2
  ~,
 \label{W5}
\end{align}
where we have suppressed the tensor structure
and omitted generation indices, for simplicity.
We choose $\lambda_\phi = M/(\sqrt{3} M_P)$, so as to obtain the Starobinsky potential~\cite{eno6}, and $\mu_H = 3\lambda V_{\Sigma} $, where $\langle \Sigma \rangle = V_{\Sigma}\, {\rm diag} (2,2,2,-3,-3)$ is the VEV of $\Sigma$, with $V_{\Sigma} \equiv 4\mu_{\Sigma}/\lambda^\prime$,  to realize doublet-triplet mass splitting for the $H$ and $\overline{H}$ multiplets. The coupling $\lambda_{\phi\Sigma}$ needs to be small in order to realize successful Starobinsky-like inflation, so we assume $\lambda_{\phi\Sigma} \ll 1$ in what follows.  

The K\"{a}hler potential includes the inflaton field as an untwisted field. The rest of the fields can be included in the K\"{a}hler potential as either untwisted or twisted fields; for concreteness, we assume all of the fields are untwisted in the following discussion. 

All of the above fields except $\phi$ and $\Sigma$ have vanishing field values
in the instantaneous potential minimum during inflation. We assume that the adjoint Higgs field is displaced by a small amount from its vacuum value during the inflationary period, i.e., 
\begin{equation}
  \langle \Sigma \rangle = (V_{\Sigma} +\sigma) \, {\rm diag} (2,2,2,-3,-3)  ~,
\end{equation}
with $|\sigma| \ll V_{\Sigma} $. We show below that this condition can be satisfied for sufficiently small $\lambda_{\phi\Sigma}$. In this case, the scalar potential during inflation is given by 
\begin{align}
  V &= \frac{\hat{V}}{(T+T^* -|\phi|^2/3 - 10 |V_\Sigma+\sigma |^2 )^2} 
   ~,
\end{align}
with
\begin{align}
  \hat{V} &= \left|M \phi - \lambda_\phi \phi^2 + 30 \lambda_{\phi\Sigma} (V_\Sigma +\sigma )^2 \right|^2  \nonumber \\
  &+ \frac{15}{2} \left| V_\Sigma +\sigma\right|^2 
  \left|4(\mu_\Sigma + \lambda_{\phi\Sigma}\phi) - \lambda^\prime (V_\Sigma +\sigma )  \right|^2   ~.
\end{align}
The instantaneous value of $\Sigma$ during inflation is determined by the second term in the above equation for $\lambda_{\phi\Sigma} \ll 1$: 
\begin{equation}
  \sigma \simeq \frac{4 \lambda_{\phi\Sigma}\phi}{\lambda^\prime}  ~.
\end{equation}
To assure $|\sigma| \ll V_{\Sigma}$, we thus assume $\lambda_{\phi\Sigma} \ll \mu_\Sigma / M_P $. This condition is generically weaker than the limit obtained from the bounds on the inflation parameters, as we see below. 

With $\langle T \rangle = 1/2$ and the canonically-normalized field $x$ given by Eq.~(\ref{phigen}), the inflaton potential can be approximated by 
\begin{align}
  V \simeq \frac{3}{4} M^2 \left(1-e^{- \sqrt{\frac{2}{3}}x}\right)^2 
  + \Delta V ~,
\end{align}
where 
\begin{align}
  \Delta V &= 15 M^2 V_\Sigma^2 e^{- \sqrt{\frac{2}{3}} x} 
  \sinh^2 \biggl(\sqrt{\frac{2}{3}}x\biggr) \nonumber \\
 & + 60 \sqrt{3} \lambda_{\phi\Sigma}\, M V_\Sigma^2 
  e^{-\frac{1}{\sqrt{6}}x} \cosh^2 \biggl(\frac{x}{\sqrt{6}}\biggr)
  \sinh \biggl(\frac{x}{\sqrt{6}}\biggr) \nonumber \\ 
  &\simeq \frac{3}{4} M^2 V_{\Sigma}^2\biggl(
  5 + \frac{30 \lambda_{\phi\Sigma} }{\sqrt{3}M}   
  \biggr) e^{\sqrt{\frac{2}{3}}x} ~. 
  \label{eq:delvsu5}
\end{align}
The CMB observables are then estimated to take the values 
\begin{align}
  n_s &\simeq 1 - \frac{2}{N_*} + \frac{320}{27} \biggl(\frac{V_\Sigma}{M_P}\biggr)^2 \biggl(
    1 + \frac{2 \sqrt{3}\lambda_{\phi\Sigma} M_P}{M}   
    \biggr) N_* ~, \\ 
    r &\simeq \frac{12}{N_*^2} + \frac{640}{9} \biggl(\frac{V_\Sigma}{M_P}\biggr)^2 \biggl(
      1 + \frac{2\sqrt{3} \lambda_{\phi\Sigma} M_P }{M}   
      \biggr) ~,
\end{align}
where we explicitly exhibit factors of the Planck mass, which were often set to unity in previous expressions. We see that the predicted values of $n_s$ and $r$ deviate from the Starobinsky values. These deviations are constrained by the Planck measurement on $n_s$~\cite{planck18}. For $N_* = 50 \, (60)$, for instance, we have 
\begin{equation}
 \biggl(
    1 + \frac{2 \sqrt{3}\lambda_{\phi\Sigma} M_P }{M}   
    \biggr)^{1/2}   V_\Sigma \, < 1.1 \, (0.72) \times 10^{16}~{\rm GeV}~,
  \label{eq:limonvsigma}
\end{equation}
at 95\% CL. This estimate shows that the CMB measurement can probe GUT physics directly in this scenario and that, in fact, the Planck measurement has already imposed a severe limit on the GUT scale. Note that this bound exists even if the inflaton has no direct coupling to the GUT Higgs in the superpotential; as we can see in Eq.~\eqref{eq:delvsu5}, $\Delta V \neq 0$ for $\lambda_{\phi\Sigma} = 0$. This contribution comes from the overall factor of $e^{2K/3}$ in the scalar potential when there are fields that have non-vanishing VEVs during inflation. This type of contribution exists also in the SO(10) model discussed in Section~\ref{eq:so10}, but is absent in the flipped SU(5) model discussed in Section~\ref{sec:flipped}. 

The bound~\eqref{eq:limonvsigma} has important implications for the allowed region in the MSSM parameter space in the minimal scenario, since the value of $V_\Sigma$ is determined by a renormalization-group analysis. Using threshold corrections at the GUT scale, we can determine a combination of the masses of the GUT-scale particles~\cite{Hisano:1992mh, Hisano:1992jj, Hisano:2013cqa}, namely $(M_X^2 M_\Sigma)^{1/3}$, where $M_X$ and $M_\Sigma$ are the masses of the GUT gauge field and the adjoint Higgs field, respectively. The VEV of the GUT Higgs is then given by~\cite{eemno} 
\begin{equation}
  V_{\Sigma} = \frac{1}{5} \biggl(\frac{2}{\lambda^\prime g_5^2}\biggr)^{1/3} 
  (M_X^2 M_\Sigma)^{1/3} ~,
\end{equation}
where $g_5$ is the SU(5) gauge coupling, which is also determined by GUT threshold conditions. As a result, $V_\Sigma$ is given as a function of a free parameter $\lambda^\prime$ once the MSSM mass spectrum is fixed. The bound~\eqref{eq:limonvsigma} then leads to a lower limit on $\lambda^\prime$, which provides an additional restriction on the parameter space, especially in the constrained scenarios discussed in Refs.~\refcite{eemno,eenno,Ellis:2019fwf}.

In the present scenario, reheating proceeds through the coupling $\lambda_{HS}$ in Eq.~\eqref{W5}, with which the inflaton decays into $H_u$ and $H_d$. The decay rate is 
\begin{equation}
  \Gamma (\phi \to H_u H_d) \simeq 2 \times \frac{|\lambda_{H\phi}|^2}{8 \pi} M ~.
\end{equation}
As discussed in Section~\ref{sec:reheatingphi}, the coupling $|\lambda_{HS}|$ must be $|\lambda_{HS}| \lesssim 10^{-5}$ in order to evade the gravitino overproduction problem. If we also introduce right-handed neutrinos to this model, which are SU(5) singlets with $R$-parity odd, we can couple the inflaton also to these fields without modifying the inflation dynamics. In this case, the inflaton can decay into right-handed neutrinos as well, and the gravitino overproduction bound again restricts the inflaton-right-handed neutrino couplings to be $\lesssim 10^{-5}$.

\subsection{SO(10) GUTs and No-Scale Inflation}
\label{eq:so10}

We saw in Section~\ref{sec:reheatingphi} that efficient reheating is possible
when we identify the inflaton with the right-handed sneutrino~\cite{snu,ENO8}.
One might think that this identification could carry over to an SO(10) completion, in
which the right-handed neutrino is included with other Standard Model fields in  the {\bf 16} representation. However, this is not possible as
there are no gauge-invariant {\bf 16}$^2$ or {\bf 16}$^3$
couplings in SO(10). The {\bf
54} or {\bf 210} representations  do allow both quadratic and cubic
couplings in the superpotential, though these are typically associated with Higgs fields that break GUT
symmetry, and would require $M$ to be GUT scale $\sim 10^{16}$ GeV, rather than $\sim 10^{13}$ GeV as expected for the inflaton.
Also, they acquire GUT-scale VEVs, which is also not the case for the Starobinsky model.

The model we consider\cite{EGNNO1} includes an SO(10) singlet as the inflaton. 
The field content is similar to the SO(10) GUT in Ref.~\refcite{GN}, which includes a {\bf 210}
representation, $\Sigma$, to break SO(10) to an intermediate gauge group, 
a pair of $\mathbf{16}$ and $\mathbf{\overline{16}}$ representations, $\Phi$ and ${\bar \Phi}$, to break the intermediate gauge group 
to the SM, a {\bf 10} representation, $H$, which includes the SM Higgs fields that break the Standard Model. There are also three matter multiplets, $\psi_i$, in {\bf 16} representations, and we include
one singlet per generation, $\phi_i$, one of which is identified as the inflaton. 

The superpotential of the theory takes the following generic form:
\begin{align} \notag 
W &= M \left( \frac12 \phi^2 - \frac{1}{3\sqrt{3}} \phi^3 \right) + y H\psi\psi + (M'+b\phi)\bar{\Phi}\psi 
\nonumber \\[3pt]
&+ m_{\Phi}\bar{\Phi}\Phi + \frac{\eta}{4!}\bar{\Phi}\Phi\Sigma + \frac{m_{\Sigma}}{4!}\Sigma^2 + \frac{\Lambda}{4!}\Sigma^3 + m_{H}H^2 + \lambda_{\phi H} \phi H^2
\nonumber \\[3pt]
&+H(\alpha\Phi\Phi+\bar{\alpha}\bar{\Phi}\bar{\Phi}
+ \alpha^\prime \Phi \psi
) + c\phi\bar{\Phi}\Phi + \frac{b^\prime}{4!}\bar{\Phi}\psi\Sigma + \frac{\gamma}{4!}\phi\Sigma^2 + \kappa \,  \label{Wgen},
\end{align}
where we have again suppressed the tensor structure
and omitted generation indices, in the interest of simplicity, and do not discuss here the possibility of mixing
between the singlet superfields $\phi_i$.

The first two terms in (\ref{Wgen}) are a Wess--Zumino superpotential that reproduces Starobinsky inflation in
no-scale supergravity~\cite{eno6,eno7},
and the third term represents the SM
Yukawa couplings. The fourth term includes a coupling between SM fields and the
inflaton $\phi$, as does the tenth term: these couplings fix
the neutrino masses and determine the inflaton decay rate.
The couplings in the fifth through
eighth terms enable the SM singlet
components of $\Phi$, $\bar{\Phi}$, and $\Sigma$ to acquire
non-vanishing VEVs. When these VEVs develop, the terms
$\alpha H\Phi \Phi$
and $\bar{\alpha} \bar{H} \bar{\Phi} \bar{\Phi}$ mix the
SU(2)$_L$ doublet components of the $H$, $\Phi$, and $\bar{\Phi}$ multiplets. Suitable choices of these couplings realize
doublet-triplet splitting by enabling two linear
combinations of these fields, denoted by $H_u$ and $H_d$,
to have masses far below the GUT and intermediate scales \cite{Sarkar:2004ww}. The VEVs of $H_u$ and $H_d$ break electroweak symmetry at the TeV scale, as in the MSSM. 
A weak-scale gravitino mass is obtained for a suitable value of the constant $\kappa$,
through the relation $m_{3/2}=\langle e^{K/2}W\rangle$.
This supersymmetry breaking may be
generated by a separate Polonyi sector \cite{pol}, as discussed previously. 
Finally, the no-scale K\"ahler
potential of the SO(10) model is taken to be
\beq\label{Kgen}
K=-3\ln\left[T+T^*-\frac{1}{3}\left(\phi^*\phi + H^{\dagger}H +
\psi^{\dagger}\psi + \Phi^{\dagger}\Phi + \bar{\Phi}^{\dagger}\bar{\Phi}
+ \frac{1}{4!}\Sigma^{\dagger}\Sigma \right)\right]\,.
\eeq
A full treatment of this model is given in Ref.~\refcite{EGNNO1},
and here we simply review the highlights.

We parametrize the VEVs of the SM singlets in the Higgs representation by
\begin{align}
 p &= \langle \Sigma ({\bf 1}, {\bf 1}, {\bf 1}) \rangle ~, ~~~~~~
 a = \langle \Sigma ({\bf 15}, {\bf 1}, {\bf 1}) \rangle ~, ~~~~~~
 \omega= \langle \Sigma ({\bf 15}, {\bf 1}, {\bf 3}) \rangle ~,
 \nonumber \\
 f_R &= \langle \Phi (\overline{\bf 4}, {\bf 1}, {\bf 2}) \rangle ~,
~~~~~~
 \bar{f}_R = \langle \bar{\Phi} ({\bf 4}, {\bf 1}, {\bf 2}) \rangle ~,
~~~~~~
 \widetilde{\nu}_R = \langle \psi (\overline{\bf 4}, {\bf 1}, {\bf 2})
 \rangle ~,
\label{eq:VEVs}
\end{align}
where we show explicitly their ${\rm SU}(4)_C
\otimes {\rm SU}(2)_L \otimes {\rm SU}(2)_R$ quantum numbers. 
We assume $\widetilde{\nu}_R = 0$ at the minimum. This is stable with a positive
mass-squared if either $b$ or $b^\prime$ is non-zero. 
To study better the scalar potential, we write the superpotential \eqref{Wgen}
 in terms of the SM singlet fields, with the rest of the fields set to
zero:  
\begin{align}\notag
W &=  M \left( \frac12 \phi^2 - \frac{1}{3\sqrt{3}} \phi^3 \right) 
 - (M'+ b \phi)\bar{f}_R\nu_R 
+ (\eta f_R+ b^\prime\nu_R) \bar{f}_R(p+3a+6\omega) \\
&-(m_{\Phi} + c\phi) \bar{f}_R f_R
+ (m_{\Sigma}+\gamma \phi)(p^2+3a^2+6\omega^2) +2\Lambda(a^3+3p\omega^2+6a\omega^2) + \kappa ~.
\label{superpot}
\end{align}
In order to realize inflation, 
the couplings $b,
c$, and $\gamma$ must be small, as all of them break the scale symmetry associated with the
potential.  Here, we take $c = \gamma = 0$. We shall see that $b$ enters into the neutrino mass matrix,
and we show the effects of $b \ne 0$ on the inflaton potential. 
 The Higgs fields are displaced a negligible amount from their vacuum values during inflation,
 and the scalar potential during inflation takes the simple form 
\beq \label{Vsimple}
V\simeq \frac{\hat{V}}{\left[1-\frac{1}{3}(|\phi|^2 + \Delta K )\right]^2} \, ,
\eeq
where $f \equiv |f_R| = |\bar{f}_R|$ and 
\begin{align}
\hat{V} &= M^2 | \phi -  \phi^2/\sqrt{3}|^2+|\phi|^2 |b f |^2 ~,
    \label{Vbcg}  \\[2pt]
    \Delta K &\equiv |p|^2 + 3|a|^2 + 6|\omega|^2 +2|f|^2 ~.
\end{align}
In terms of the canonically-normalized field $x$, the scalar potential takes the form
\begin{align}
V 
&\simeq \frac{3}{4}M^2\left(1-e^{-\sqrt{2/3}\,x}\right)^2 +\Delta V\,,
\label{Vnotfull}
\end{align}
where
\beq\label{deltaV}
\Delta V = \left[\frac{3}{4} |b f|^2 + \frac{1}{2}M^2 e^{-\sqrt{2/3}\,x} \Delta K  \right]  \sinh ^2 (\sqrt{2/3}\, x )\,.
\eeq
We show in Fig. \ref{fig:planck1} the effects of the coupling $b$ in $\Delta V$,
plotting the tilt of the scalar perturbation spectrum, $n_s$, and the tensor-to-scalar ratio, $r$.
The red (pink) shaded regions correspond to the 68 (95) \% CL limits from 
Planck \cite{planck18}. 
In the limit where $|b f|, \Delta K \ll 1 $, the inflationary parameters can be approximated analytically by:~\cite{EGNNO1}
\begin{align}\label{nsan}
n_s &\;\simeq\; 1 -\frac{2}{N_*} + \frac{8}{3}\left(\frac{b f}{M}\right)^2N_*^2 + \frac{32}{81} N_* \Delta K  \,,\\ \label{ran}
r &\;\simeq\; \frac{12}{N_*^2}  + \frac{32}{3}\left(\frac{b f}{M}\right)^2N_* +\frac{64}{27} \,\Delta K\,.
\end{align}
Figs. 2 and 4 in Ref.~\refcite{EGNNO1} show the effects of the term proportional to $\Delta K$, and set an upper limit on this combination of $10^{-3.1}$. Notice that, similarly to \eqref{eq:limonvsigma}, this limit is present even if the inflaton does not couple to the GUT Higgs fields in the superpotential. If the values of $p, a, \omega, f$ are $\simeq V_{\rm GUT}$, $\Delta K \simeq 12 V^2_{\rm GUT}$, and this limit gives $V_{\rm GUT} \lesssim 2 \times 10^{16}$~GeV.

\begin{figure}[ht]
\centerline{\psfig{file=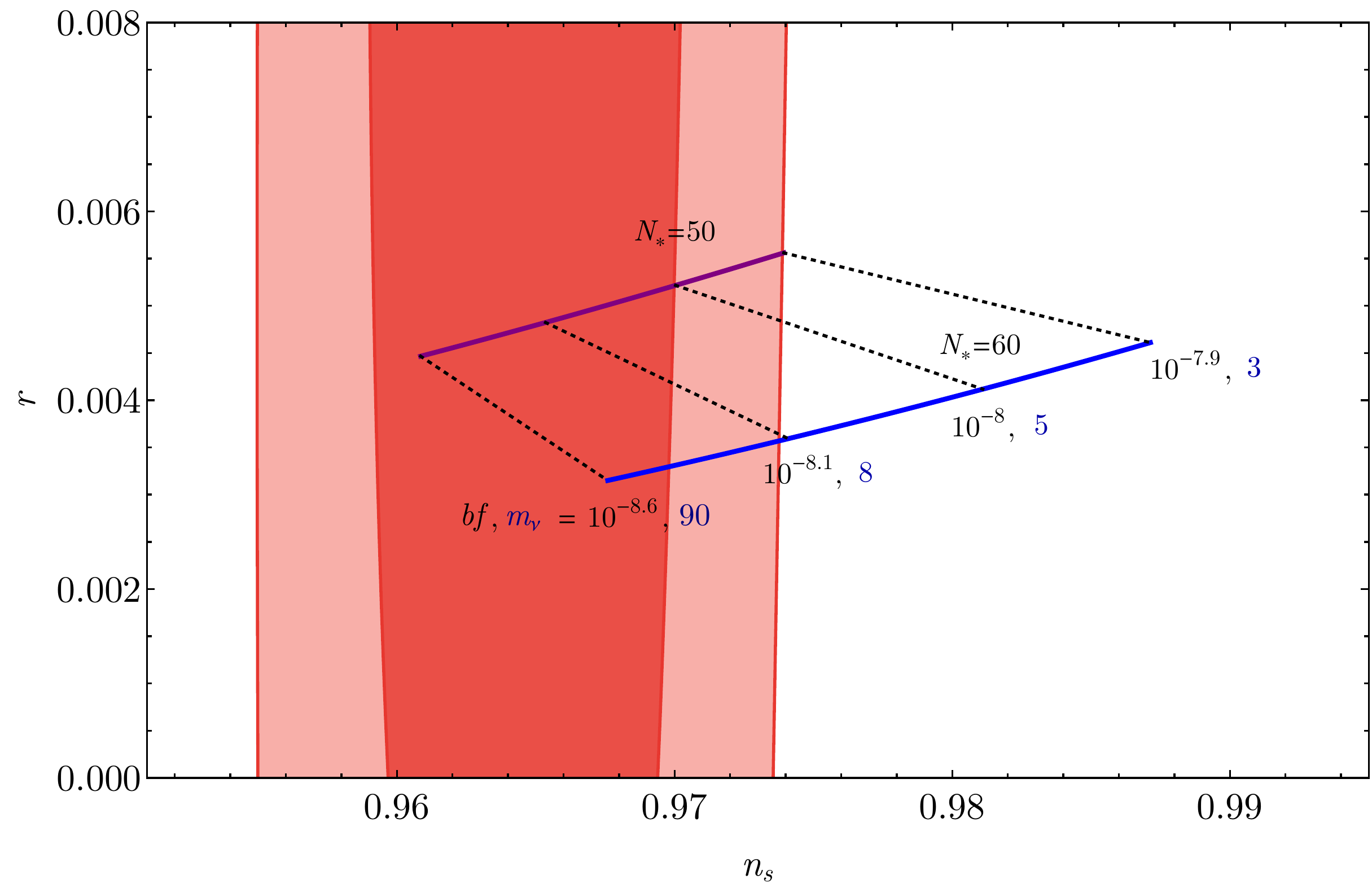,width=10cm}}
 \caption{\it Parametric $(n_s,r)$ curves as functions of $b f$ for $N_*=50,60$, with
    the 68 and 95\% CL Planck constraints shown in the background. The solid curves illustrate the parametric dependence using the analytical approximation (\ref{Vnotfull}) and (\ref{deltaV}) with $\Delta K = 0$. 
The dotted curves illustrate particular values of $b f$, quantified in units of $M_P$, 
and we indicate the corresponding left-handed neutrino masses in units of $10^{-4}$~eV, assuming 
$y_\nu \sin \beta = 10^{-5}$ and $M = 10^{-5} M_P$. 
        }
    \label{fig:planck1}
\end{figure}

We see in Fig.~\ref{fig:planck1} the effect of a non-zero value of $b$,
recalling that $b f = 0$ corresponds to the exact Starobinsky
result. In order to obtain values of $(n_s, r)$ consistent with Planck, we must
require that the product $b f < 10^{-7.9}$ $(10^{-8.1})$ for $N_*\simeq
50$ $(60)$ $e$-folds of inflation. Since the VEV of $\Phi$ is less
than the GUT scale, $f \la 10^{-2.1}$, and the constraint we
have on $b$ is $b < 10^{-5.8}$ $(10^{-6})$. The scalar potential for
several choices of $b f$ is shown in Fig. \ref{fig:Vcorr1}. As one can
see, so long as $b f \la 10^{-2.5} M \sim 10^{-7.5}$, the potential is
indistinguishable from the Starobinsky potential out to the value $x
\sim 5.5$ needed for 60 $e$-folds of inflation. 

Reheating in this model is largely controlled by the coupling $b$. 
The inflaton decay rate is
\beq
\Gamma(\phi\rightarrow H_u\tilde{L}) + \Gamma(\phi\rightarrow \tilde{H}_u L) \simeq
\frac{M}{4\pi}\left| b\right|^2\,,
\eeq
which leads to a reheating temperature 
\beq
T_{\rm RH} \simeq 10^{15} ~{\rm GeV}\times
\left| b \right| \lesssim 10^{9} ~{\rm GeV}
~,
\eeq
compatible with the success of conventional BBN.

We note also that
CP violation in the inflaton decay process may generate a lepton asymmetry non-thermally~\cite{nos, dg}, which is then converted to a baryon asymmetry~\cite{fy} through sphaleron processes~\cite{Manton:1983nd, Kuzmin:1985mm}.
For related work see Refs. \refcite{kmov,cdo,my,CPviol,giudicereh,leptinf1,leptinf2}.

\begin{figure}[ht]
\centerline{\psfig{file=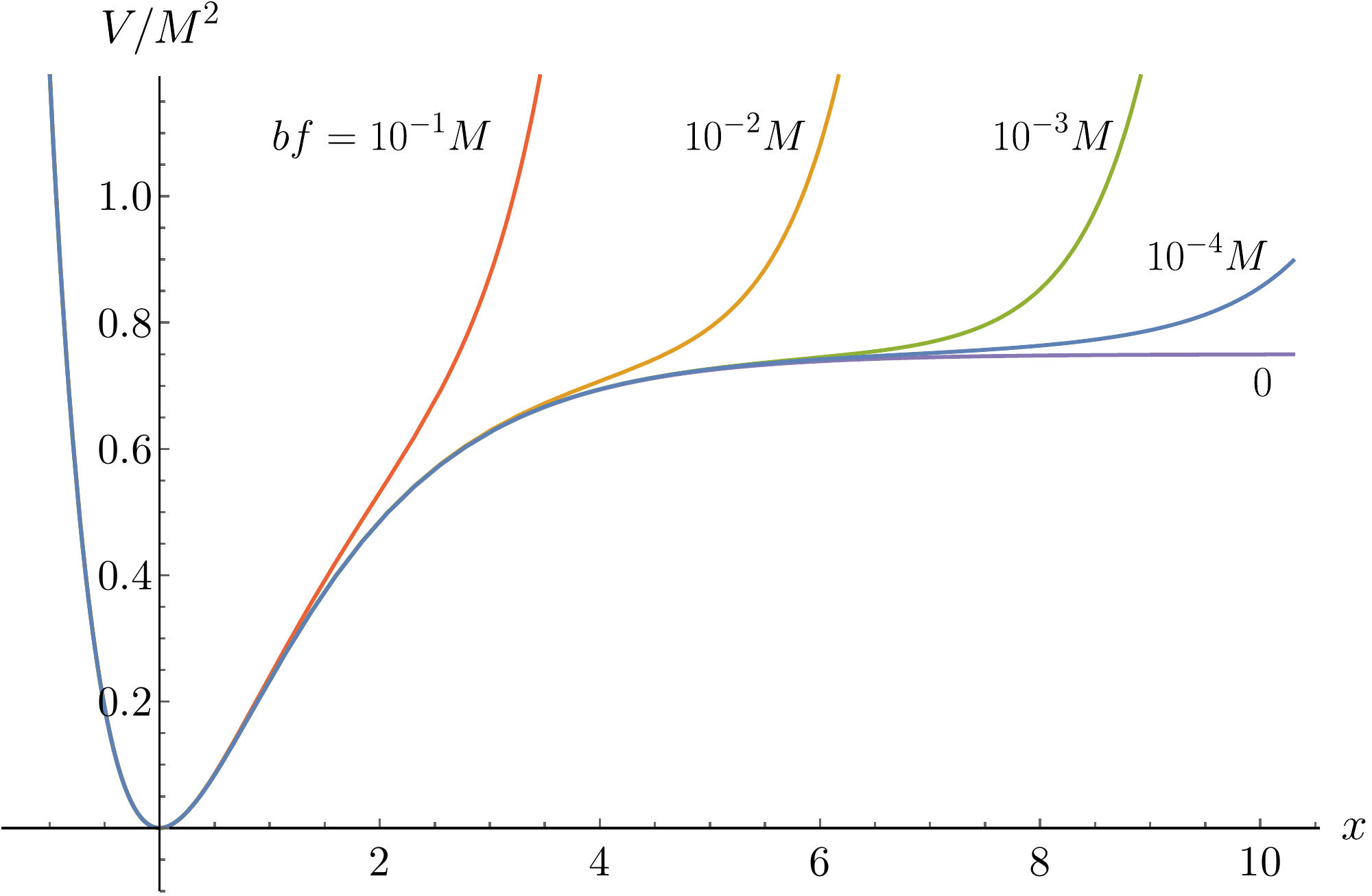,width=9cm}}
\caption{\it The inflationary potential for different values of $b f$, in units of the inflaton mass $M\simeq 10^{-5}M_P$. 
The curve labeled $b f = 0$ is the Starobinsky potential.  \label{fig:Vcorr1}}
\end{figure}

Before we conclude this Subsection, we comment on the generation of neutrino masses in this model. 
A non-zero value of the
coupling $b$ induces mixing between right-handed neutrinos and the
``singlinos" $\tilde{\phi}_i$, which are the fermionic components of the
singlet superfields $\phi_i$. Disregarding Planck-suppressed
factors, the neutrino-singlino fermion mass matrix is given by
\cite{GN}  
\beq \label{seeM}
{\cal L}_{\rm mass} = -
\begin{pmatrix}
\overline{\nu}_L & \overline{\nu}^c_R & \overline{\tilde{\phi}} 
\end{pmatrix}
\begin{pmatrix}
0 & -y_\nu\,v\sin\beta & 0 \\
-y_\nu\,v\sin\beta & 0 & -b f\\
0 & -b f & M
\end{pmatrix}
\begin{pmatrix}
\nu_L \\[3pt]
\nu_R^c\\[3pt]
\tilde{\phi}
\end{pmatrix}
 \, .
\eeq
A similar form for the
mass matrix is found in flipped SU(5)~\cite{flipped2,ENO3} (see the following Section). 
For the first-generation neutrinos,  sufficient
inflation restricts the coupling $b$ as we have seen previously. 
In this case, the couplings satisfy the hierarchy
\beq
y_\nu\,v\sin\beta \ll b f \ll M ~,
\eeq
and the fermion mass eigenstates are 
\begin{align} \label{nuM1}
\nu_L^M \;&\simeq\;  \nu_L \,-\, \frac{M\,y_\nu v\sin\beta}{(b
 f)^2}\,\nu_R^c \,-\, \frac{y_\nu v\sin\beta}{b  f}\,\tilde{\phi} \, , \\
\nu_R^M \;&\simeq\;  \nu_R^c \,+\, \frac{b f}{M}\,\tilde{\phi} \,+\,
 \frac{M\,y_\nu v\sin\beta}{(b f)^2}\,\nu_L \, , \\ \label{nuM3}
\tilde{\phi}^M \;&\simeq\;  \tilde{\phi} \,-\, \frac{b f}{M}\,\nu_R^c \,+\,
 \frac{b f \,y_\nu v\sin\beta}{M^2}\,\nu_L \, .
\end{align}
For the second and third generations, the coupling
$b$ (recall we have suppressed all generation indices) 
can be arbitrary. For all generations, the masses of the light neutrinos are given by
\begin{equation}
 m_\nu \simeq M\left(\frac{y_\nu\,v\sin\beta}{b f}\right)^2 
 \simeq \biggl(\frac{M}{10^{-5} M_P}\biggr)
 \biggl(\frac{y_\nu \sin\beta}{10^{-5}}\biggr)^2 
 \biggl(\frac{b f}{10^{-8} M_P}\biggr)^{-2} \times 10^{-4} ~ {\rm eV}
 ~.
 \label{eq:mnuso10}
\end{equation}
As we see from this equation, we cannot take $b f$ to be arbitrary small; to obtain a sufficiently small neutrino mass, $m_\nu \lesssim 0.1$~eV, we need $|b f| \gtrsim 3.5 \times 10^{-10} M_P$ for $M = 10^{-5} M_P$ and $y_\nu \sin \beta = 10^{-5}$, which is approximately equal to the size of the up-quark Yukawa coupling. If we instead take $f_\nu \sin \beta \simeq 10^{-2}$, corresponding to the charm-quark Yukawa coupling, the condition $m_\nu \lesssim 0.1$~eV leads to $|b f| \gtrsim 3.5 \times 10^{-7}$, which conflicts with the bound set by the Planck observation, as shown in Fig.~\ref{fig:planck1}. We thus conclude that in this model the inflaton field $\phi$ must predominantly couple to the first-generation neutrino.

This coupling yields a strong correlation between the first-generation neutrino mass and the inflation parameters $n_s$ and $r$ as shown in Fig.~\ref{fig:planck1}. It becomes apparent if we write Eqs.~\eqref{nsan} and \eqref{ran} using Eq.~\eqref{eq:mnuso10} as 
\begin{align}
n_s &\;\simeq\; 1 -\frac{2}{N_*} + \frac{8}{3}\frac{(y_\nu v \sin\beta)^2}{M m_\nu}  N_*^2 + \frac{32}{81} N_* \Delta K  \,,\\ 
r &\;\simeq\; \frac{12}{N_*^2}  + \frac{32}{3}\frac{(y_\nu v \sin\beta)^2}{M m_\nu}N_* +\frac{64}{27} \,\Delta K\,.
\end{align}
As we see, both $n_s$ and $r$ get larger for a smaller neutrino mass. 

\subsection{Flipped SU(5)$\times$U(1) and No-Scale Inflation}
\label{sec:flipped}

Another possible UV completion of the no-scale inflationary scenario 
is a flipped  SU(5)$\times$U(1) GUT~\cite{Barr,DKN,flipped2,AEHN}.
Unlike the previous SU(5) and SO(10) GUT models, it can be derived in the fermionic
formulation of weakly-coupled heterotic string theory~\cite{AEHN}. This is
because, unlike the two previous models, it does not require an adjoint Higgs or
larger Higgs representation for gauge symmetry breaking.

The flipped  SU(5)$\times$U(1) GUT contains three generations of SM
matter fields, each with a right-handed neutrino, which are each placed in
$\mathbf{10}$, $\bar{\mathbf{5}}$, and $\mathbf{1}$ representations of SU(5).
The representation assignments of the
right-handed leptons and the right-handed up- and down-type
quarks, are ``flipped'' with respect to standard SU(5). Moreover, the $I = \pm 1/2$ partners
in the left-handed lepton and quark doublets are also flipped.
The SU(5)$\times$U(1) GUT group is
broken to the SM via
$\mathbf{10}+\overline{\mathbf{10}}$ Higgs representations of SU(5), and
subsequently to the unbroken SU(3)$\times$U(1) symmetry via electroweak doublets
in $\mathbf{5}+\bar{\mathbf{5}}$ representations. Our notations for the
fields and their gauge representations are as follows:
\begin{alignat}{3}
&F_i && = ({\bf 10},1)_i  && \ni \; \left\{d^c,Q,\nu^c\right\}_i~,\nonumber\\
&\bar{f}_i &&=(\bar{\bf 5},-3)_i && \ni \; \{u^c,L\}_i~,\nonumber\\
&\ell^c_i &&=({\bf 1},5)_i && \ni \; \{e^c\}_i~,\nonumber\\
&H &&=({\bf 10},1)~, && \; \; \bar{H} \; = \; (\overline{\bf 10},-1)~,\nonumber\\
&h &&=({\bf 5},-2)~, && \; \; \bar{h} \; = \; (\bar{\bf 5},2) \, ,
\label{eq:charges}
\end{alignat}
where the subscripts $i = 1, 2, 3$ are generation indices that we
suppress unless they are necessary. 
The model
contains four singlet fields, which have no U(1) charges and are
denoted by $\phi_a=({\bf 1},0)$, $a=0,\ldots,3$.  

The superpotential up to third order in the chiral superfields is:
\begin{align} \notag
W &=  \lambda_1^{ij} F_iF_jh + \lambda_2^{ij} F_i\bar{f}_j\bar{h} +
 \lambda_3^{ij}\bar{f}_i\ell^c_j h + \lambda_4 HHh + \lambda_5
 \bar{H}\bar{H}\bar{h}\\ 
&\quad   + \lambda_6^{ia} F_i\bar{H}\phi_a + \lambda_7^a h\bar{h}\phi_a
 + \lambda_8^{abc}\phi_a\phi_b\phi_c + \mu^{ab}\phi_a\phi_b\,, 
\label{Wgen2} 
\end{align}
where the indices $i,j$ run over the three fermion families, and we have suppressed gauge group indices. We impose
a $\mathbb{Z}_2$ symmetry  
$H\rightarrow -H$
that prevents the mixing of SM matter fields with Higgs colour triplets and
elements of the Higgs decuplets. This symmetry also suppresses the
supersymmetric mass term for $H$ and $\bar{H}$, and thus suppresses dimension-five proton decay operators.

The first 3 terms of the superpotential (\ref{Wgen2}) provide the Standard Model Yukawa couplings. 
The fourth and fifth terms provide for the doublet-triplet separation.
The sixth term accounts for neutrino masses. The seventh term plays the role of the MSSM
$\mu$-term. The last two terms determine the inflationary potential and 
also play roles in neutrino masses.
Without loss
of generality, we take $\lambda_2^{ij}$ and $\mu^{ab}$ to be
real and diagonal in what follows. 

There is one linear combination of the Standard Model singlet components
$\nu_H^c$ and $\nu_{\bar H}^c$ in $H$ and $\bar{H}$, respectively,
corresponding to a $D$- and $F$-flat direction in the potential,
that is massless in the
supersymmetric limit.
This combination is denoted by $\Phi$, and referred to as the flaton,
and SU(5)$\times$U(1) GUT symmetry is broken
along this direction. A soft supersymmetry-breaking mass term
destabilizes the symmetric value $\Phi =0$, and the degeneracy
along this flat direction is also
lifted by a non-renormalizable superpotential term of the form
\begin{equation}
    W_{\text{NR}} = \frac{\lambda}{n! M^{2n-3}_{P}} (H \bar{H})^n ~,
\end{equation}
so that the effective potential for the flaton 
field takes the form 
\begin{equation}
    V_{\text{non-th}} (\Phi) = 
    V_0 -\frac{1}{2}m_{\Phi}^2\Phi^2 + \frac{|\lambda|^2}{[(n-1)!]^2 M_P^{4n-6}} \Phi^{4n-2} \, ,
    \label{eq:treeV}
\end{equation}
where $m_\Phi$ denotes the soft mass of $\Phi$. Minimizing the potential, 
we find
\begin{equation}
    \langle \Phi \rangle 
= 
\biggl[
\frac{\{(n-1)!\}^2m_\Phi^2 M_P^{4n-6}}{(4n-2) |\lambda|^2}
\biggr]^{\frac{1}{4(n-1)}} ~.
\end{equation}
Therefore, to obtain a GUT-scale VEV with a $\lambda = {\cal O}(1)$, 
we should have $n\geq 4$, and here we take $n = 4$. The absence of unwanted terms, e.g., $(H\bar{H})^2$, may be attributed to an additional symmetry, such as $R$ symmetry~\cite{Hamaguchi:2020tet}. With the flat direction lifted, the flaton (and flatino) mass is 
of the order of the supersymmetry-breaking scale. For further details, see Ref.~\refcite{EGNNO2}.

We concentrate here on the {strong reheating} scenario discussed
in Ref.~\refcite{EGNNO3}, in which
GUT symmetry is unbroken at the end of inflation, and also assume
that the GUT symmetry remains unbroken during reheating.
In this scenario the GUT phase transition arises because the number of
light degrees of freedom, $g$, differs between the broken and unbroken phases
\cite{supercosm,NOT,Campbell:1987eb, EGNNO2, EGNNO3}, and massless
superfields make a temperature-dependent correction $-
g\pi^2 T^4/90$ to the effective potential. Because $g = 103$
in the unbroken phase, compared to $g = 62$ in the broken phase, at high temperatures
$\Phi$ is stabilized at the origin. However, when the temperature drops below
the SU(5) confinement scale, $\Lambda_c$,
the number of light degrees of freedom decreases significantly to $g \leq
25$, favoring the broken phase~\cite{EGNNO2}. 
In this strong reheating scenario the phase transition is driven by
the incoherent component of the
flaton if $\Lambda_c \gtrsim 2.3 (m_\Phi
M_{\rm GUT})^{1/2}$~\cite{EGNNO3}.
This condition becomes $\Lambda_c
\gtrsim 2.3 \times 10^{10}$~GeV for $m_\Phi = 10^4$~GeV and
$M_{\rm GUT} =10^{16}$~GeV.

Figure~\ref{fig:veff} shows the shape of the effective potential 
as a function of $\Phi$  for $0.03\leq T/\Lambda_c \leq 1.2$~\cite{EGNNO3}. 
We note that the minimum near the origin is metastable for $1\gtrsim T/\Lambda_c\gtrsim 0.03$, 
separated from the true vacuum by a shrinking barrier that finally disappears for $T\lesssim 0.03\,\Lambda_c$.

\begin{figure}[ht]
\centerline{\psfig{file=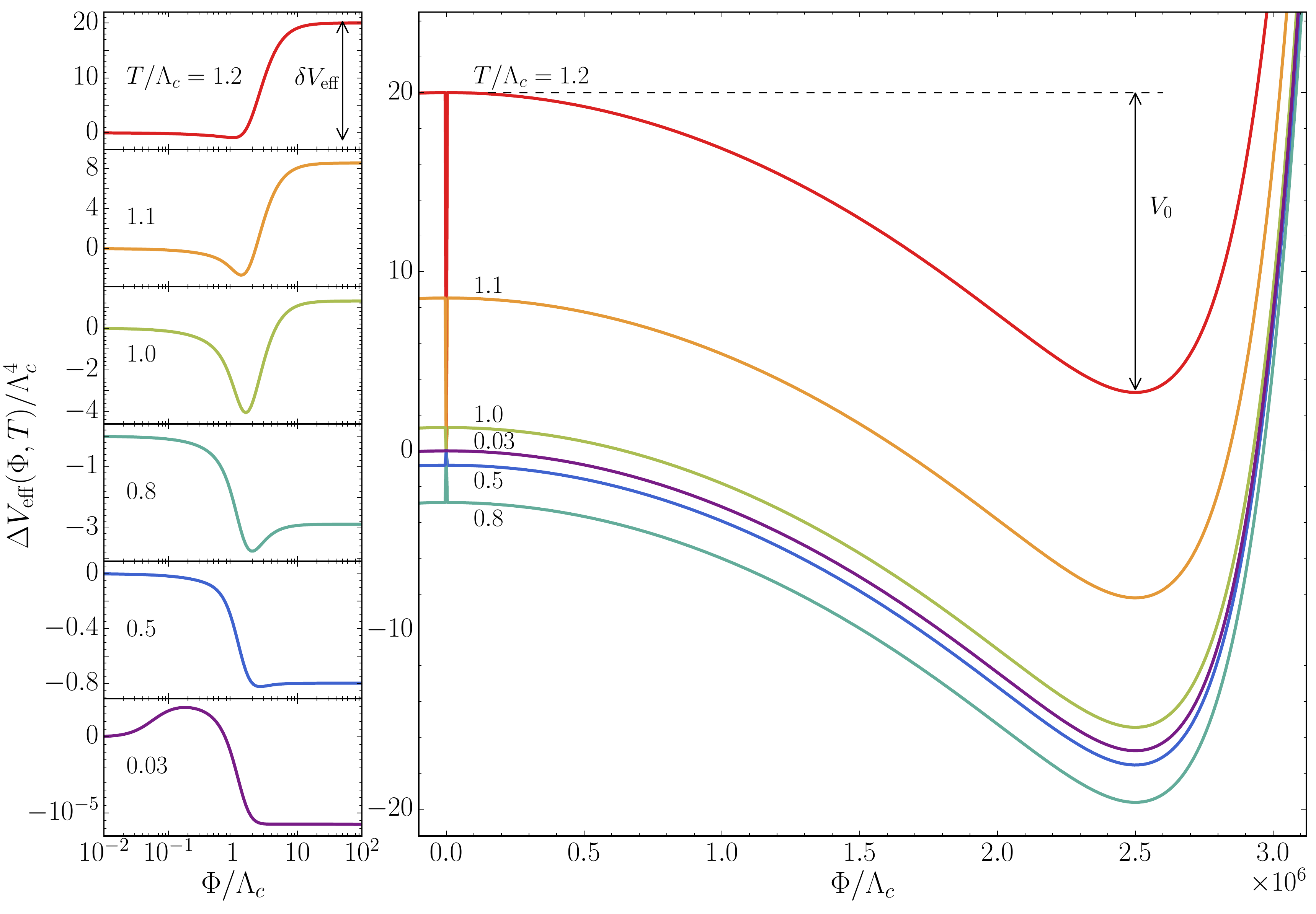,width=12cm}}
\caption{\it The evolution with temperature of the effective potential in strongly-coupled SU(5)$\times$U(1)~\cite{EGNNO3}. Here $\Delta V_{\rm eff} = V_{\rm eff}(\Phi,T)-V_{\rm eff}(0,T)$, where $V_{\rm eff}$ includes the non-thermal contribution (\ref{eq:treeV}) with $n=4$, $m_{\rm \Phi}=10\,{\rm TeV}$, $\Lambda_c = 4\times 10^9\,{\rm GeV}$ (for which $\delta V_{\rm eff}\sim V_0$) and $\langle \Phi\rangle= 2.5\times 10^6 \Lambda_c$ at low temperature. The heights of the left and right sides of the barrier $\delta V_{\rm eff}$ and $V_0$ are labelled for $T/\Lambda_c = 1.2$.   \label{fig:veff}}
\end{figure}

In the case of strong reheating, 
the flaton decouples
from the thermal bath. When $T \lesssim m_\Phi$ it becomes
non-relativistic and eventually dominates the energy density of the Universe until
it decays. Flatons decay when the Hubble expansion rate is comparable to the flaton decay rate, $\Gamma_\Phi$, releasing, in general, a large amount of entropy, which is estimated to be \cite{EGNNO3,EGNNO4}
\begin{equation}
 \Delta \simeq 1.6 \times 10^4\, \lambda^{-2}_{1,2,3,7} \,
\biggl(\frac{M_{\rm GUT}}{10^{16}~{\rm GeV}}\biggr)
\biggl(\frac{10~{\rm TeV}}{m^2_{\rm soft}/m_\Phi}\biggr)^{1/2} ~,
\label{flatonDelta}
\end{equation} 
where $m_{\rm soft}$ represents a typical sfermion
mass. The Universe is reheated again, this time to 
a temperature $T^\prime_{\rm RH} \propto (\Gamma_\Phi M_P)^{1/2}$.

As in the case of SO(10), the inflaton is identified with one (the lightest) of the singlets, $\phi_a$,
which we denote simply by $\phi$. As noted above, we take $\mu^{ab}$ to be diagonal,
\beq
\mu^{ab} = {\rm
diag}\left(M/2,\mu^{11},\mu^{22},\mu^{33}\right)\,,\qquad
\mu^{ab}\leq M_{\rm GUT}\, .
\label{muab}
\eeq
For Starobinsky inflation, the coupling $\lambda_8$ must satisfy 
\beq\label{StaroCond1}
-3\sqrt{3}\, \lambda_{8}^{000} = M\, ,
\eeq
in which case the potential for the SM singlet fields is 
\begin{align} \notag
 V_F &\;\simeq\; \frac{3}{4}M^2\left(1- e^{-\sqrt{2/3}\,x}\right)^2 +
 \frac{3}{4}\sinh^2(\sqrt{2/3}\,x) \sum_i |\lambda_6^{i0}|^2 \left(
 |\tilde{\nu}_{\bar{H}}^c |^2 + |\tilde{\nu}_{i}^c |^2 \right)\\
 & \qquad + \frac{1}{8}M^2e^{\sqrt{2/3}x}\left(|\tilde{\nu}^c_{\bar{H}}|^2 + \sum_i |\tilde{\nu}^c_i|^2\right) + \cdots \, ,
\end{align}
where $x$ is the canonically-normalized inflaton once again. 
As one can see, the large VEV for $x$ during inflation produces a mass term for 
 $\tilde{\nu}^c$ and the GUT-breaking field $\tilde{\nu}_H^c$ that are then driven to the origin as
 assumed above. 

It was shown in Ref.~\refcite{EGNNO3} that reheating is completed in the symmetric phase if $|\lambda_{6}^{i0}| \gtrsim
{\cal O}(10^{-4})$.
When $\langle \Phi\rangle < M$, the inflaton decays to $F$ and $\bar{H}$, with a rate given by
\beq\label{eq:decvan}
\Gamma(x\rightarrow F_i\bar{H}) \;\simeq\; 10\times \frac{|\lambda_6^{i0}|^2}{8\pi} \left( 1- \frac{\langle \Phi\rangle^2}{M^2}\right) M\,.
\eeq
The reheating
temperature in this case is given by
\begin{equation}
 T_{\rm RH} \simeq 1.7  \times 10^{15} ~\mathrm{GeV}
\times \sqrt{\sum_{i} |\lambda_6^{i0}|^2} ~,
\label{eq:treh}
\end{equation}
indicating a direct relation between
$T_{\rm RH}$ and $\lambda_6$.
This coupling is similar in nature to $b$ in the SO(10) GUT model discussed in the previous Section. 

During reheating, gravitinos are produced via the
scattering and decay of particles in the thermal bath~\cite{weinberg,elinn,nos,ehnos,kl,ekn,Juszkiewicz:gg,mmy,Kawasaki:1994af,
Moroi:1995fs,enor,Giudice:1999am,bbb,kmy,stef,Pradler:2006qh,ps2,rs,kkmy,
egnop,Garcia:2018wtq}. For
the calculation of the gravitino production rate, we use the formalism 
outlined in Ref.~\refcite{rs}, but with the 
group-theoretical factors and couplings
appropriate to flipped SU(5)$\times$U(1).
As discussed in Section \ref{sec:reheat}, these gravitinos eventually decay into
LSPs, making a non-thermal contribution to the LSP abundance:
\begin{align}
 \Omega_{\rm DM} h^2 & \simeq  0.12  \,
\biggl(\frac{1.6 \times 10^4}{\Delta}\biggr)
\biggl(\frac{m_{\rm LSP}}{1~{\rm TeV}}\biggr)
\biggl(\frac{\sqrt{\sum_{i} |\lambda^{i0}_6|^2}}{0.0097}\biggr) \nonumber \\
 &=  0.12  \,
\biggl(\frac{1.6 \times 10^4}{\Delta}\biggr)
\biggl(\frac{m_{\rm LSP}}{1~{\rm TeV}}\biggr)
\biggl(\frac{T_{\rm RH}}{1.6 \times 10^{13} ~{\rm GeV}}\biggr)
~.
\label{eq:odm}
\end{align}
This non-thermal
component of the LSP abundance should be combined with the component generated thermally
to obtain the total LSP density.
This is, however, reduced by the flaton entropy dilution factor
$\Delta$ as discussed below, providing a direct relation between the LSP relic density
and $\lambda_6$.

As we have just seen, the coupling $\lambda_6^{i 0}$ plays a
crucial role in both reheating and the generation of gravitinos, and we 
discuss next a third role of this coupling---the generation of
light neutrino masses~\cite{EGNNO4}. As noted earlier, we adopt the basis where $\lambda_2^{ij}$ and $\mu^{ab}$ are real and diagonal.
In this case, the diagonal components of $\lambda_2^{ij}$ is given by
\begin{equation}
 \lambda_2 \simeq \frac{1}{\langle \bar{h}_0 \rangle}
{\rm diag}(m_u, m_c, m_t) ~, 
 \label{eq:lam2andmu}
\end{equation}
and $\mu^{ab}$ is given in Eq.~(\ref{muab}).
Only three of the four singlets will contribute to the 
neutrino mass matrix and we take these to be $\phi_a$, $a = 0,1,2$, including the inflaton.
In what follows we 
express these matrices as  $\lambda_2^{ij} = \lambda_2^i \delta^{ij}$ and $\mu^{ab} = \mu^a \delta^{ab}/2$. 

The relevant superpotential terms  are 
\begin{equation}
 W = \sum_{i=1}^{3} \lambda_{2}^i \nu_i^c L_i H_d 
+\frac{1}{2} \sum_{a=0}^{2} \mu^a {\phi}_a^2 
+ \sum_{i,a} \lambda_6^{ia} \nu_i^c \nu_{\bar{H}}^c {\phi}_a ~,
\end{equation}
where $\lambda_6^{ia}$ is a $3\times 3$ complex
matrix.
The neutrino/singlet-fermion mass matrix can be written as 
\begin{equation}
 {\cal L}_{\rm mass}=
-\frac{1}{2}
\left(
\begin{matrix}
{\nu}_{i} & {\nu}_{j}^c & \tilde{\phi_a}
\end{matrix}
\right) 
\begin{pmatrix}
 0 & \lambda_2^{i j}\langle \bar{h}_0\rangle & 0 \\
\lambda_2^{i j}\langle \bar{h}_0\rangle & 0 &
 \lambda_6^{j a}\langle \tilde{\nu}_{\bar{H}}^c\rangle \\
0 & \lambda_6^{j a}\langle \tilde{\nu}_{\bar{H}}^c\rangle & \mu^a
\end{pmatrix}
\begin{pmatrix}
\nu_{i} \\[2pt] \nu_{j}^c \\[2pt] \tilde{\phi}_a
\end{pmatrix}
+ {\rm h.c.}~,
\end{equation}
where $i,j = 1,2,3$, $a= 0,1,2$,
and $\tilde{\phi}$ corresponds to the fermionic superpartner of the inflaton field $\phi$. 
(For more
generic expressions, see Ref.~\refcite{Ellis:1993ks}).
The mass matrix of the right-handed
neutrinos is obtained from a first seesaw mechanism~\cite{Minkowski:1977sc,Georgi:1979dq}:
\begin{equation}
 (m_{\nu^c})_{ij} = \sum_{a=0,1,2} \frac{\lambda_6^{ia} \lambda_6^{ja}}{\mu^a}
  \langle \tilde{\nu}_{\bar{H}}^c \rangle^2 ~.
\label{eq:mnuc}
\end{equation}
We can diagonalize the mass matrix (\ref{eq:mnuc})
using a unitary matrix $U_{\nu^c}$:
\begin{equation}
     m_{\nu^c}^D = U_{\nu^c}^T m_{\nu^c}
U_{\nu^c}  ~.
\label{eq:mnucdiagonalization}
\end{equation}
Assuming $\langle \tilde{\nu}_{\bar{H}}^c \rangle = 10^{16}$~GeV, 
the light neutrino mass matrix is then obtained through a second
seesaw mechanism:
\begin{equation}
 (m_\nu)_{ij} = \sum_{k} \frac{\lambda_2^i \lambda_2^j (U_{\nu^c})_{ik}
  (U_{\nu^c})_{jk} \langle \bar{h}_0 \rangle^2 }{(m_{\nu^c}^D)_k} ~.
\label{eq:mnu}
\end{equation}
This mass matrix is diagonalized by a unitary matrix $U_\nu$, 
\begin{equation}
    m_\nu^D
= U_\nu^* m_\nu  U_\nu^\dagger ~.
\label{eq:mnudiagonalization}
\end{equation}
The Pontecorvo-Maki-Nakagawa-Sakata (PMNS) matrix in this model is given by~\cite{Ellis:1993ks}
\begin{equation}
U_{\rm PMNS} =   U_\ell^* U_\nu^{\dagger} ~,
\label{eq:pmnsfl}
\end{equation}
where $U_\ell$ is a unitary matrix that is used to diagonalize $\lambda_3$. We can always reproduce the measured PMNS matrix for arbitrary $U_\nu$ by choosing $U_\ell$ appropriately.

Given a matrix
$\lambda^{ia}_6$, the eigenvalues of the $m_\nu$ and $m_{\nu^c}$ matrices, as well as the mixing matrices $U_{\nu^c}$ and $U_\nu$, are uniquely
determined as functions of $\mu^1$ and $\mu^2$ 
via Eqs.~(\ref{eq:mnuc}--\ref{eq:mnu}). If we further assume that the field $\tilde{\phi}_a$ couples
predominantly to the $i$-th generation neutrino, then we can write its mass approximately as 
\begin{equation}
 m_{\nu_{i}} \simeq \frac{\mu^a\left(\lambda_2^{i}\langle 
 \bar{h}_0\rangle\right)^2}{\left(\lambda_6^{ia} \langle 
 \tilde{\nu}^c_{\bar{H}} \rangle \right)^2} 
\simeq 
\frac{\mu^a m_{u_i}^2}{\left(\lambda_6^{ia} 
\langle \tilde{\nu}^c_{\bar{H}} \rangle \right)^2}
~.
\label{eq:linumassesb}
\end{equation}
This expression indicates that due to the hierarchical structure in $\lambda_2$ as shown in Eq.~\eqref{eq:lam2andmu}, the mass eigenvalues of $m_\nu$ also become hierarchical unless there is a large hierarchy in $\mu^a$ or $\lambda_6$. In particular, the lightest mass eigenvalue tends to be much smaller than the other two masses, $m_{\nu_1} \ll m_{\nu_{2,3}}$. This means that for normal neutrino mass ordering (NO), $m_{\nu_2} \simeq \sqrt{\Delta m_{21}^2} = 8.6 \times 10^{-3}$~{eV} and $m_{\nu_3} \simeq \sqrt{\Delta m_{31}^2} = 5.0 \times 10^{-2}$~{eV}, and for inverted neutrino mass ordering (IO), $m_{\nu_2} \simeq m_{\nu_3} \simeq \sqrt{|\Delta m_{32}|^2} = 5.0 \times 10^{-2}$~{eV}, where the values of the squared mass differences, $\Delta m_{21}^2 \equiv m_2^2 - m_1^2$ and $\Delta m_{3\ell }^2 \equiv m_3^2 - m_\ell^2$ ($\ell = 1,2$) are given in the Table below, which we take from Ref.~\refcite{nufit}. 

\begin{table}[ht!]
\tbl{\it Input values for the squared mass differences of active neutrinos \cite{nufit}.}
 { \begin{tabular}{l |c c|cc}
  \hline \hline
  & \multicolumn{2}{c|}{Normal Ordering} & \multicolumn{2}{c}{Inverted Ordering}\\
   & Best fit & 3$\sigma$ range &  Best fit & 3$\sigma$ range \\
  \hline
  $\Delta m_{21}^2$ [$10^{-5}~\text{eV}^2$] & 7.39 & 6.79--8.01& 7.39& 6.79--8.01\\
  $\Delta m_{3\ell }^2$ [$10^{-3}~\text{eV}^2$] & $2.525$& $ 2.431$--$2.622$& $-2.512$& $-(2.413$--$2.606)$\\
   \hline
  \hline
  \end{tabular}  \label{tab:input}}
  \end{table}

The approximate formula \eqref{eq:linumassesb} also shows that the couplings $\lambda^{ia}_6$ cannot be arbitrary small. For example, if we set $\mu^0 = M = 3 \times 10^{13}$~GeV and $\langle \tilde{\nu}^c_{\bar{H}} \rangle = 10^{16}$\,GeV, $m_{\nu_i} \lesssim 0.1$~eV gives $|\lambda^{i0}_6| \gtrsim 10^{-7}, \, 10^{-4},\, 10^{-2}$ for $i = 1,2,3$.

We recall that $\lambda_6^{i 0}$ also controls the
reheating temperature and hence the thermal gravitino abundance, 
yielding the relic LSP abundance produced
by gravitino decays shown in Eq.~(\ref{eq:odm}). In order to avoid overproduction of dark matter,
we obtain the following upper limit on $\lambda_6^{i 0}$ from Eqs.~(\ref{eq:treh}) and (\ref{eq:odm})
\beq
 \sum_{i} |\lambda_6^{i0}|^2 < 10^{-4}  \biggl(\frac{ \Delta}{1.6 \times 10^{4}} \biggr)^2 \left( \frac{1\, {\rm TeV}}{m_{\rm LSP}} \right)^2
 \, .
\eeq
We note
that, if the thermal relic abundance of the LSP is negligibly small, the non-thermal LSP abundance from gravitino decay saturates the cosmological dark matter density for $|\lambda_6^{i 0}| \sim  10^{-2}$,
at the border of the gravitino bound for $\Delta \sim 10^4$ and $m_{\rm LSP} \sim 1$~TeV.

The right-handed neutrinos are
massive only when $\langle \bar{H}  \rangle \ne 0$. In the strong reheating
scenario considered here, the masses of the right-handed neutrinos vanish and they are in thermal
equilibrium immediately after reheating. 
They subsequently acquire masses at the GUT phase transition,
and rapidly drop out of equilibrium.  
They decay non-thermally~\cite{NOT, EGNNO3,EGNNO4,EGNNO5}, generating a lepton asymmetry
as proposed in Ref.~\refcite{fy}, which then generates a baryon
asymmetry via the sphaleron process~\cite{Kuzmin:1985mm}, yielding
a net baryon number density 
\begin{equation}
 \frac{n_B}{s} = -\frac{28}{79} \cdot \frac{135 \zeta (3)}{4\pi^4 g_{\rm
  reh}
  \Delta } \sum_{i=1,2,3} \, \epsilon_i ~,
  \label{nbs}
\end{equation}
where \cite{Ellis:1993ks, EGNNO3}
\begin{equation}
 \epsilon_i 
= 
\frac{1}{2\pi}\frac{ \sum_{j\neq i} {\rm Im} 
\left[\left(
U_{\nu^c}^\dagger (\lambda_2^D)^2 U_{\nu^c}
\right)_{ji}^2\right] }{\left[
U_{\nu^c}^\dagger (\lambda_2^D)^2 U_{\nu^c}
\right]_{ii}}
g\biggl(\frac{m^2_{\nu^c_{j}}}{m^2_{\nu^c_{i}}}\biggr) ~,
\end{equation}
with \cite{Covi:1996wh}
\begin{equation}
 g(x) \equiv -\sqrt{x} \biggl[
\frac{2}{x-1} + \ln \biggl(\frac{1+x}{x}\biggr)
\biggr] ~.
\label{eq:gx}
\end{equation}
 Thus we see that $\lambda_6$ in fact determines not only 
 the reheating temperature, the non-thermal component of cold dark matter,
 and neutrino masses, but also the baryon asymmetry of the Universe.
 
The estimate based on Eq.~\eqref{eq:linumassesb} is inaccurate if the inflaton field couples to several generations or if the mixing matrices $U_{\nu^c}$ and $U_\nu$ have sizable off-diagonal components. In order to consider such cases, we have investigated numerically the effect of the $\lambda_6$ coupling matrix on neutrino masses, the reheating temperature, the non-thermal component of the LSP density, and the baryon asymmetry of the Universe, by performing a scan over the $\lambda_6$ parameter space~\cite{EGNNO3}. 
We write $\lambda_6$ in the form \cite{EGNNO4,EGNNO5}
\begin{equation}
  \lambda_6 = r_6 M_6 ~,  
\end{equation}
where $r_6$ is a real constant, which plays a role of a scale factor, and $M_6$ is a generic complex $3\times 3$ matrix. We scan
$r_6$ with a logarithmic distribution over the range $(10^{-4},
1)$ choosing a total of 2000 values. For each value of $r_6$, we generate a random complex $3\times 3$
matrix $M_6$ with each component taking a value of ${\cal O}(1)$. 

We obtain the eigenvalues of the $m_\nu$ and $m_{\nu^c}$ matrices 
as functions of $\mu^1$ and $\mu^2$ for each choice of the $3\times 3$ matrix $\lambda_6$, 
The two $\mu$ parameters are then fixed by reproducing within experimental uncertainties
the measured values of the squared mass differences, $\Delta m_{21}^2 \equiv m_2^2 - m_1^2$ and 
$\Delta m_{3\ell }^2 \equiv m_3^2 - m_\ell^2$, 
where $\ell = 1 (2)$ in the case of NO (IO). 
We use the results of Ref.~\refcite{nufit} as experimental input, which we summarize in Table~\ref{tab:input}. 
Out of the $4 \times 10^6$ model $\lambda_6$ matrices we generate for each ordering of the neutrino masses,
we find 9839 and 730 acceptable matrix choices in
the NO and IO cases, respectively. This difference may suggest a mild preference for the NO case in our model. 

Fig.~\ref{fig:tr} displays histograms of the reheating temperature $T_{\rm RH}$ in the NO and IO scenarios, 
using orange shading and a dashed blue line, respectively. 
In both cases, all parameter points have $T_{\rm RH} \gg (m_\Phi M_{\rm GUT})^{1/2} \simeq 10^{10}$~GeV, 
so the strong reheating condition is satisfied for $\Lambda_c \gtrsim 2.3 \times 10^{10}$ GeV \cite{EGNNO3}.
We also see that values of $T_{\rm RH} \sim 10^{12}$~GeV are favoured in both the NO and IO
cases, with much larger values 
$\lesssim 10^{15}$~GeV also being possible in the NO scenario.

\begin{figure}[!ht]
\centerline{\psfig{file=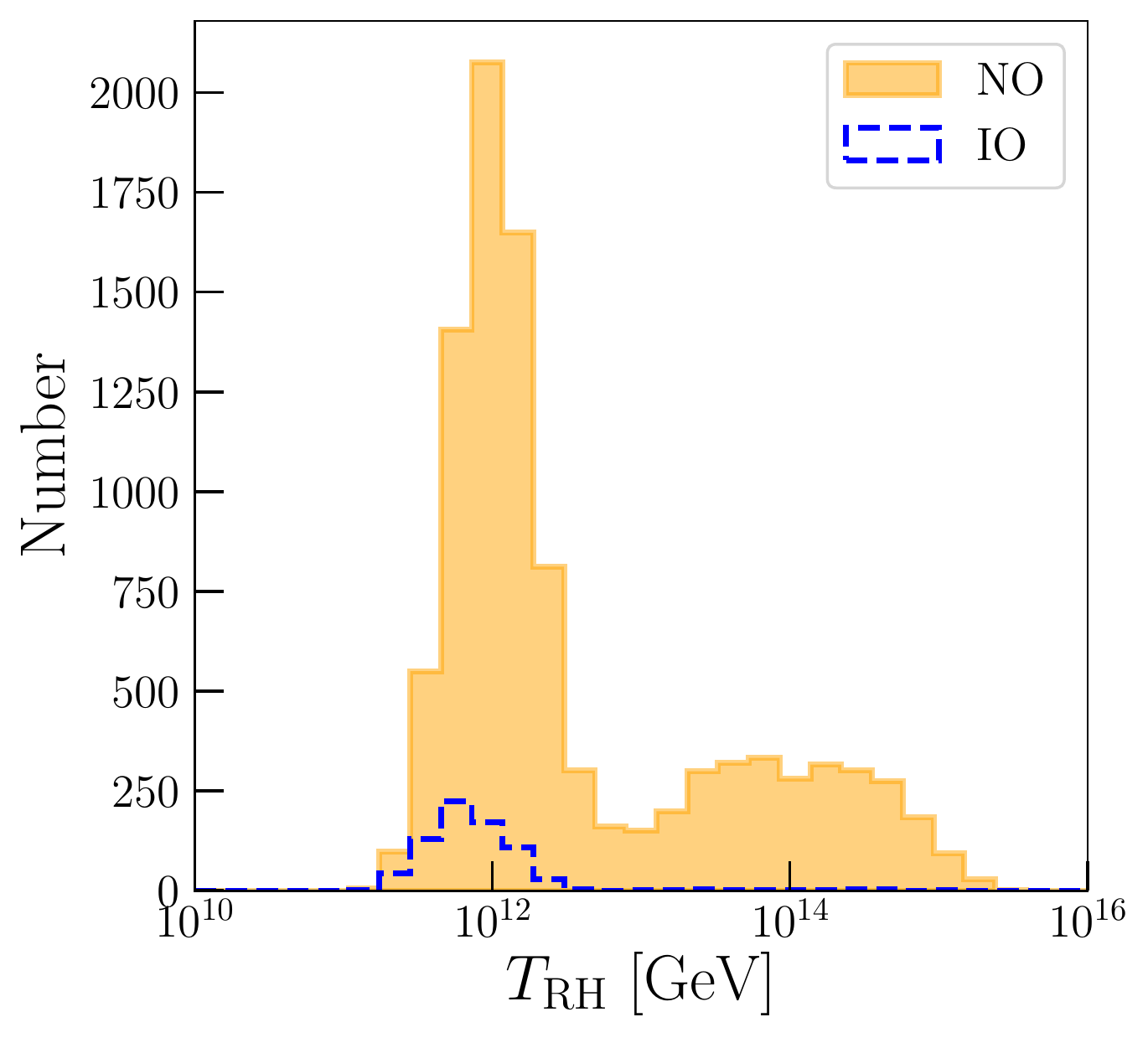,width=8cm}}
\caption{\it Histograms of the values of $T_{\rm RH}$ resulting from the numerical scan of $\lambda_6$ for the NO and IO cases (orange shading and dashed blue line, respectively).}
\label{fig:tr}
\end{figure}

The distributions of the non-thermal dark matter density produced by 
gravitino decays in the NO and IO solutions for $\lambda_6$ is shown in Fig.~\ref{fig:omdm},
assuming that the flaton entropy dilution factor estimated in (\ref{flatonDelta}) has the value $\Delta = 10^4$. 
Many parameter sets yield $\Omega_{\rm DM} h^2 \simeq 10^{-1}$, consistent with the 
density of dark matter measured by the Planck collaboration, $\Omega_{\rm DM}h^2 = 0.12$~\cite{planck18}, 
which is shown as the black line in Fig.~\ref{fig:omdm} for $m_{\rm LSP} = 10$~TeV, 
corresponding to $T_{\rm RH} \simeq 10^{12}$~GeV, as seen in Eq.~\eqref{eq:odm}. 
Some solutions overproduce dark matter for $m_{\rm LSP} = 10$~TeV, 
yielding $\Omega_{\rm DM} h^2 \lesssim 10$, corresponding to
$T_{\rm RH} \simeq 10^{14}$~GeV. In such cases, 
either smaller values of $m_{\rm LSP}$ and/or larger entropy factors 
$\Delta$ would be required for consistency with observation.

\begin{figure}[!ht]
\centerline{\psfig{file=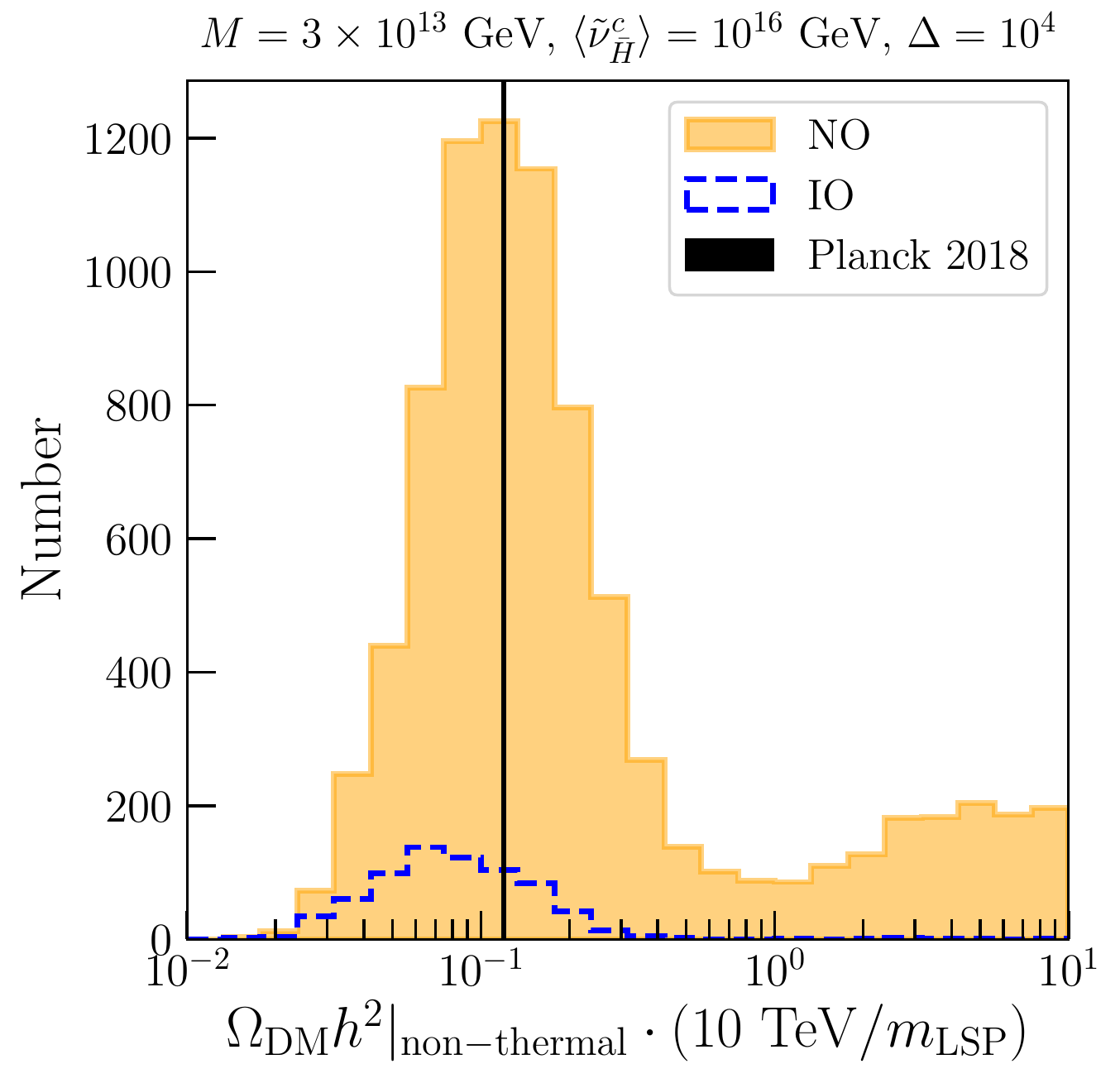,width=8cm}}
\caption{\it Histograms of the values of $\Omega_{\rm DM} h^2$ found in the numerical scan of $\lambda_6$ for the NO and IO cases (orange shading and blue dashed line, respectively) with $\Delta = 10^4$. The vertical black line shows the Planck 2018 value of the dark matter density: $\Omega_{\rm DM}h^2 = 0.12$ \cite{planck18}.
}
\label{fig:omdm}
\end{figure}

In Fig.~\ref{fig:bauhist}, we show histograms of $n_B/s$ assuming an entropy factor $\Delta = 10^4$. 
The observed value of the baryon asymmetry, 
$n_B/s = 0.87 \times 10^{-10}$ \cite{planck18}, is shown as
the vertical solid line in Fig.~\ref{fig:bauhist}, and can easily be explained in our scenario. A value of $\Delta$ much more than two orders of magnitude larger would be unlikely to yield an acceptable value of $n_B/s$.

\begin{figure}[!ht]
\centerline{\psfig{file=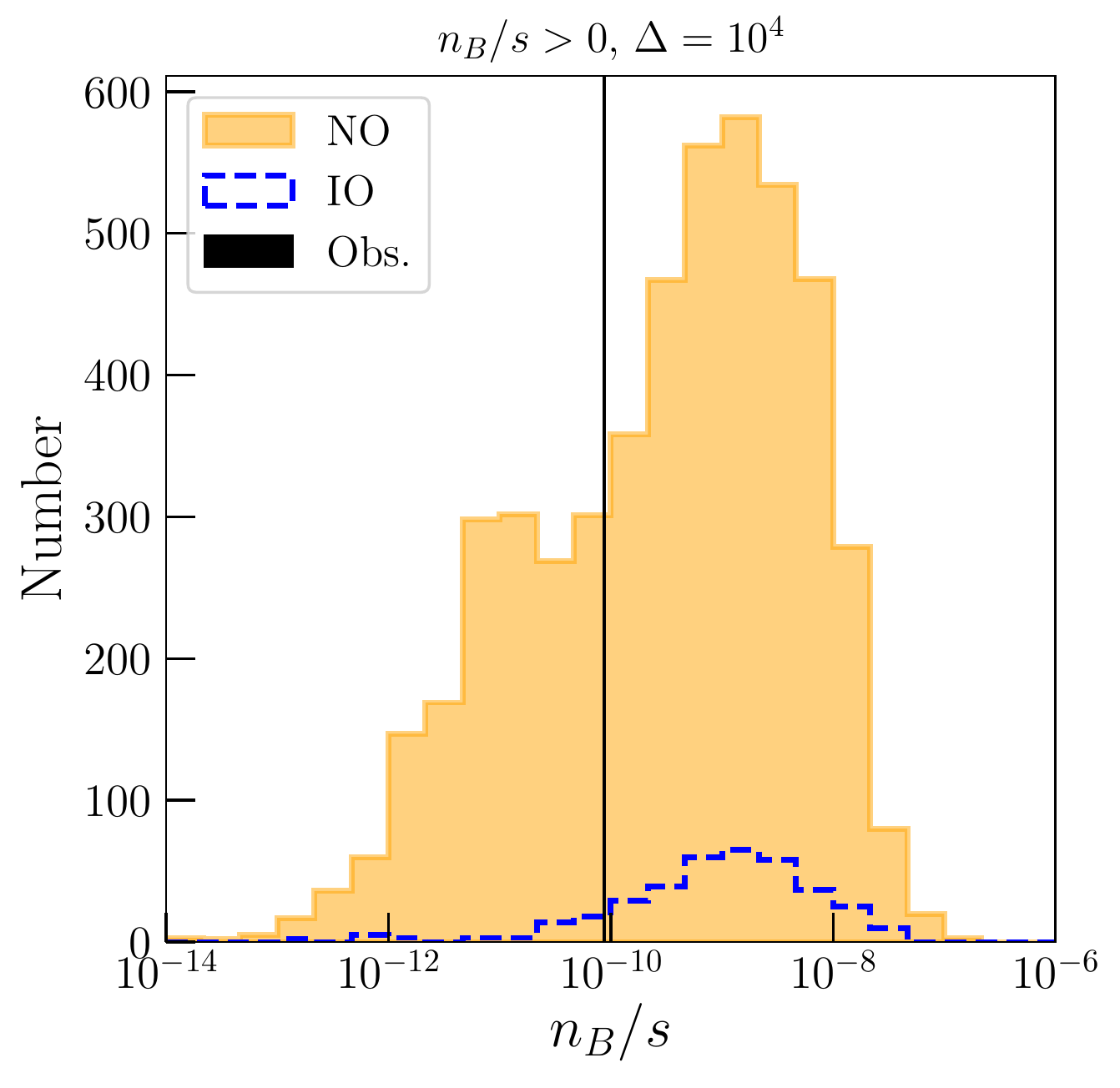,width=8cm}}
\caption{Histograms of values of $n_B/s$ in the NO and IO scenarios (orange shading and blue dashed, respectively) assuming an entropy factor $\Delta = 10^4$. The vertical black solid line shows the observed value. 
}
\label{fig:bauhist}
\end{figure}

The non-thermal contribution to the LSP abundance due to gravitino
decay assuming $m_{\rm LSP} = 10$~TeV is plotted in Fig.~\ref{fig:bau_omdm}
against the value of $n_B/s$ for the same parameter point,
assuming that a factor $\Delta = 10^4$ of entropy is generated. The
horizontal green and vertical black lines show the measured values of
the dark matter abundance $\Omega_{\rm DM} h^2 = 0.12$ and the baryon asymmetry~\cite{planck18}, respectively.
We see that for most of the points
$\Omega_{\rm DM} h^2 \gtrsim
{\cal O}(10^{-2})$ and $n_B/s \lesssim {\cal O}(10^{-7})$. 
There are many points where the non-thermal component
of the LSP abundance from gravitino decays provides all the dark matter density, $\Omega_{\rm DM} h^2 \simeq 0.12$, and $n_B/s \simeq 0.87 \times 10^{-10}$.

\begin{figure}[!ht]
\centerline{\psfig{file=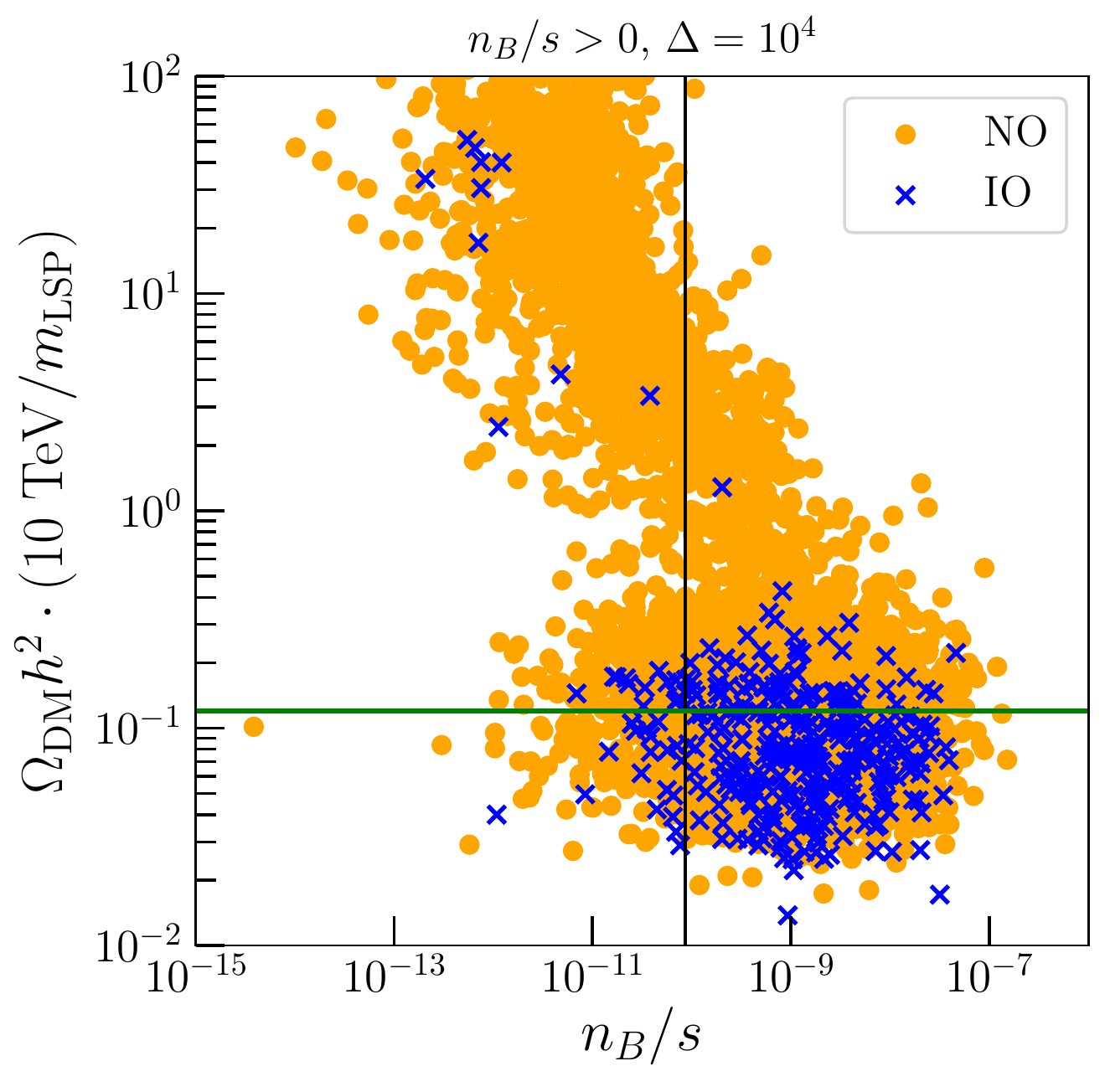,width=8cm}}
\caption{\it Scatter plot of $n_B/s$ vs the non-thermal contribution to the LSP abundance, assuming $m_{\rm LSP} = 10$~TeV and $\Delta = 10^4$, with the
 observed values shown as the horizontal green and vertical black lines,
 respectively. 
}
\label{fig:bau_omdm}
\end{figure}

The amount of entropy production discussed above has important consequences for 
the low-energy supersymmetric parameter space.
In many models~\cite{cmssm,elos,eelnos,eemno,eeloz,ehow++}, 
all the soft scalar masses are assumed to have the same value 
$m_0$ at some high-energy input
scale, $M_{in}$, which may be the same as $M_{\rm GUT}$, as in the CMSSM,
and similarly for the gaugino masses, $m_{1/2}$, and the trilinear 
soft supersymmetry-breaking terms, $A_0$.
The non-observation of supersymmetry
at the LHC~\cite{nosusy} indicates that sparticle masses must be at or
above the TeV scale, in which case one would normally expect the 
relic LSP density following thermal freeze-out to exceed
generically the cold dark matter
density measured by Planck, $\Omega h^2 \simeq 0.12$ \cite{planck18}.
However, this problem can be avoided in the presence of
particular relations between the masses of the LSP
and some other sparticles.  

Consider, for example, a bino LSP $\chi$
that annihilates to SM fermions via $t$-channel sfermion exchange.
In the limit $m_{\tilde f} > m_\chi \gg m_f$, where $m_{\tilde f}$
is a common sfermion mass and $m_\chi$ is the bino mass,
the $p$-wave annihilation cross section may be approximated by~\cite{OS}
\beq
\langle \sigma v \rangle \simeq \frac{g_1^4}{32 \pi} \sum_f ({Y_L}_f^4 + {Y_R}_f^4) \frac{m_\chi^2}{m_{\tilde f}^4} x ,
\label{sv}
\eeq
where $g_1$ is the U(1)$_Y$ gauge coupling, $Y_{L,R}$ are the hypercharges
of the left- and right-handed fermions, respectively,  
and $x = T_f/m_\chi \approx 1/20$ is the ratio of the
annihilation freeze-out temperature to the bino mass. 
In such a case, the relic density is given approximately by~\cite{ehnos,eo}
\beq
\Omega_\chi h^2 \approx 1.9 \times 10^{-11} \left(\frac{T_\chi}{T_\gamma}\right)^3 \sqrt{g_f} \left(\frac{{\rm GeV}^{-2}}{{\frac12 \langle \sigma v \rangle x}}\right) \, ,
\label{oh2}
\eeq
where $g_f$ is the number of relativistic degrees of freedom at freeze-out
and the factor $({T_\chi}/{T_\gamma})^3$ is due to the dilution
of neutralinos between freeze-out and today~\cite{oss,ehnos}. If
$m_\chi \sim 100$~GeV and $m_{\tilde f}\sim 350$~GeV, 
one finds $\Omega_\chi h^2 \sim 0.1$.
However, (\ref{oh2}) shows that $\Omega_\chi h^2 \propto m_{\tilde f}^4/m_\chi^2$,
so that values of sparticle masses
a factor of 100 larger,
i.e., $m_\chi \sim 10$ TeV and $m_{\tilde f}\sim 35$ TeV,
would lead to $\Omega_\chi h^2 = {\cal O}(10^3)$.

This argument may be evaded if the LSP and the next-to-lightest supersymmetric particle (NLSP)
are very similar in mass, in which case coannihilations between them can reduce the
relic density~\cite{gs}. Examples of possible nearly-degenerate NLSPs include the
lighter stop~\cite{stopco,eds,eoz,interplay,raza,eeloz,ehow++} or stau~\cite{stau,celmov,delm}. 
Another density-reduction mechanism comes into play when
the LSP mass is very close to half the mass of the heavy MSSM Higgs scalar and/or pseudoscalar, 
whose $s$-channel exchanges lead to rapid annihilations~\cite{funnel}.
Yet another possibility is that $m_0 \gg m_{1/2}$ and $A_0$ is small. In this case, the
Higgs mixing parameter $\mu$ is also small, the LSP resembles a
Higgsino, and annihilations into $W^\pm$ and $Z^0$ bosons
become important~\cite{fp}. This again requires
quite a finely-tuned relationship between $m_{1/2}$ and $m_0$
for any given values of $A_0$ and $\tan \beta$.

However, the landscape of allowed models changes greatly in the presence
of late-time entropy production. Specifically, a factor $\Delta = 10^4$ of entropy production
in flaton decay would imply an increase in the preferred relic density at freeze-out by $10^4$, 
corresponding to generic sparticle masses of order 10~TeV,
as can be deduced from Eqs. (\ref{sv}) and (\ref{oh2}). Bearing in mind that the 
strong reheating scenario prefers $\Delta = 10^4$
for other reasons, one may regard a supersymmetry-breaking scale of ${\mathcal{O}}(10)$ TeV
as a prediction of the scenario. We can estimate the relic density
in Eq. (\ref{oh2}) as
\beq
\Omega_\chi h^2 \simeq  10^{-7} {\rm \, GeV}^{-2} \Delta^{-1} \frac{m_{\tilde f}^4}{m_\chi^2}
\sim 10^3 \Delta^{-1} \left(\frac{m_{\tilde f}}{30 {\rm \, TeV}}\right)^4 \left(\frac{10 {\rm \, TeV}}{m_\chi} \right)^2 \, ,
\label{oh22}
\eeq
where the entropy release is approximately~\cite{EGNNO3}
\beq
\Delta \sim 10^4 \left( \frac{30 {\rm \, TeV}}{m_{\tilde f}} \right)^{1/2} \, ,
\eeq
yielding
\beq
\Omega_\chi h^2 \sim 10^{-1} \left(\frac{m_{\tilde f}}{30 \, {\rm TeV}}\right)^{9/2} \left(\frac{10 \, {\rm TeV}}{m_\chi} \right)^2\, ,
\label{oh23}
\eeq
where all the relevant couplings are assumed to be ${\cal O}(1)$.

One should also take into account the condition for successful BBN that the  
reheating temperature after flaton decay:~\cite{EGNNO3}
\beq
T_{\rm RH}^\prime \sim 10^{-3} \left(\frac{m_{\tilde f}^3 M_P}{M_{\rm GUT}^2} \right)^{1/2} \sim 1\, {\rm MeV} \left( \frac{m_{\tilde f}}{30 {\rm \, TeV}} \right)^{3/2}\, ,
\label{tr'}
\eeq
be $\gtrsim 1$~MeV, ensuring that the universe is radiation-dominated during BBN. 
Combining Eqs.~\eqref{oh23} and \eqref{tr'}, we find
\beq
 \Omega_\chi h^2 \sim 0.1 \left( \frac{T_{\rm RH}^\prime}{1\, {\rm MeV}} \right)^3 \left(\frac{10\, {\rm TeV}}{m_\chi} \right)^2 \, .
\eeq
We find that the reheating temperature in Eq.~(\ref{tr'}) is
$\gtrsim 1$~MeV, as needed for
BBN, for $m_{\tilde f} \gtrsim {\mathcal O}$(10) TeV.
On the other hand, imposing the cosmological limit
$\Omega_\chi h^2 \le 0.12$ in Eq.~(\ref{oh22}) requires
$m_{\tilde f} \lesssim {\mathcal O}$(10) TeV.
Thus, in this model consistency between these constraints determines the 
supersymmetry breaking scale to be ${\mathcal O}$(10) TeV.
{\it This prediction is consistent with the 
non-observation of SUSY signals 
at LHC so far, while offering hope for detection at a future 100-TeV
proton-proton collider.}

To conclude this discussion, we show illustrative results from one example of a 
super-GUT \cite{superGUT,emo,emo2,emo3} CMSSM based on flipped SU(5) as originally considered in Ref.~\refcite{emo3}.
 In super-GUT models, the universality of the soft supersymmetry-breaking parameters occurs at some scale
 $M_{in}$ above the GUT scale. 
 In  the flipped super-GUT plane shown in Fig.~\ref{fig:superGUT1}, we take
$A_0/m_0 = 0$,  $\tan \beta = 10$,  $M_{in} = M_P$ and $\mu > 0$, and the flipped SU(5) couplings 
are chosen as {\boldmath${ \lambda}$} $= (\lambda_4, \lambda_5) = (0.3,0.3)$. Across the plane, 
in the absence of entropy generation the relic density is significantly larger than its 
observationally-determined value.   
Indeed, $\Omega_\chi h^2$ easily reaches $\mathcal{O}$(1000) when the measured value of the lightest Higgs boson mass
$m_h = 125$ GeV. We see from this Figure that an entropy factor $\Delta = {\cal O}(10^4)$,
as suggested above, would reduce the relic LSP density to an allowed value over a large
range of the soft supersymmetry-breaking gaugino mass $m_
{1/2} \sim  \text{a few} \times 10^3$~GeV (corresponding to strongly-interacting sparticle masses $\sim 10$~TeV, beyond the reach of the LHC) that are compatible with the measured value of $m_h$, within the
expected theoretical error of $\pm 3$~GeV, for values of the soft supersymmetry-breaking scalar mass $m_0 \lesssim 1$~TeV.

\begin{figure}[ht]
\centerline{\psfig{file=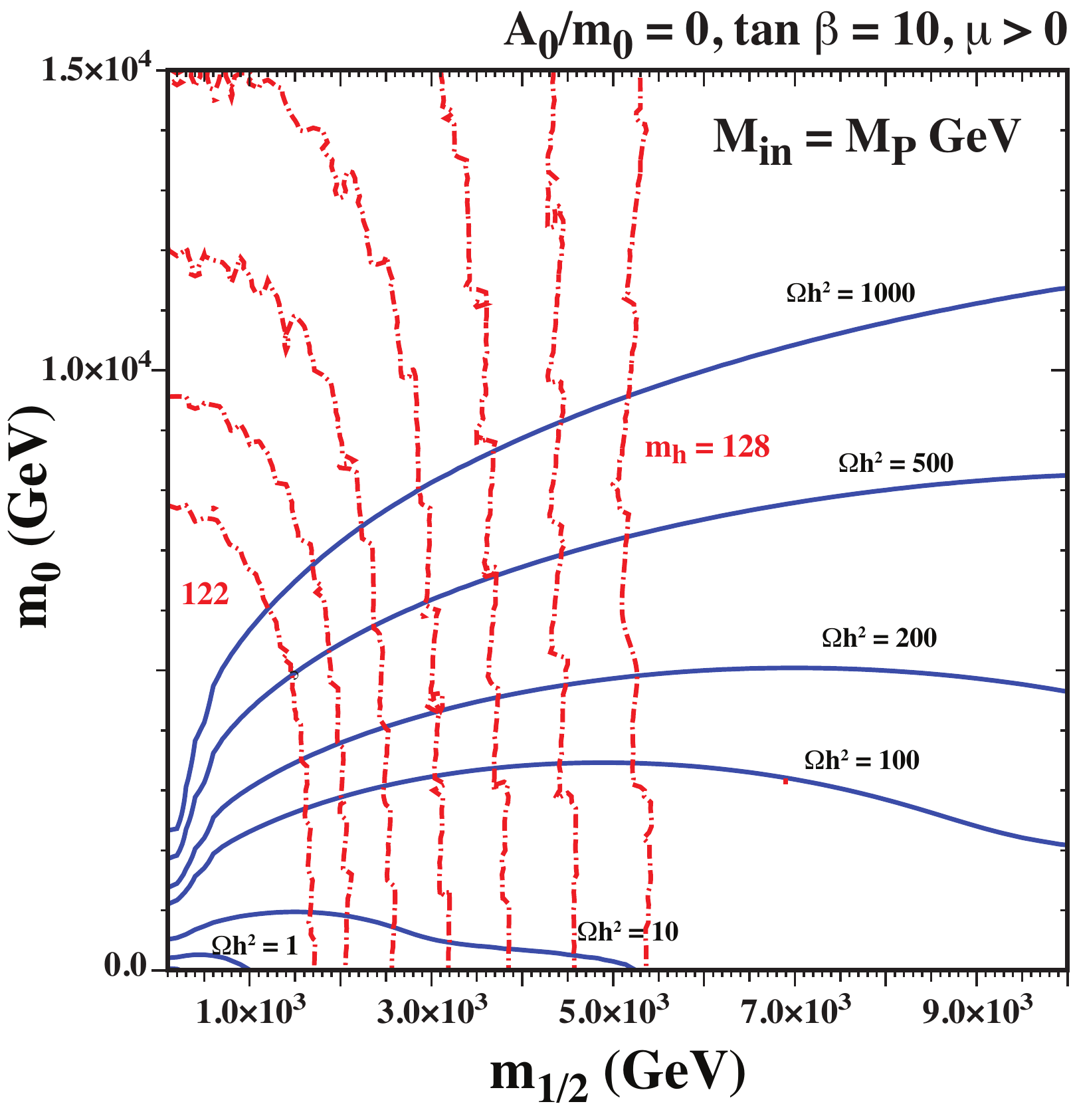,width=8cm}}
\caption{\it A $(m_{1/2}, m_0)$ plane in the flipped super-GUT model with $M_{in} = M_P$, $\tan \beta = 10$, $\mu > 0$, $A_0 = 0$, {\boldmath${ \lambda}$} $= (0.3, 0.3)$.  The red dot-dashed lines are contours of $m_h$ calculated using {\tt FeynHiggs}~\protect\cite{FeynHiggs}, and the solid blue lines are contours of $\Omega_\chi h^2$ in the absence of subsequent entropy generation. This figure is not sensitive to the choice of $\lambda_6$.
  \label{fig:superGUT1}}
\end{figure}

\subsection{Nucleon decay}
\label{sec:pdecay}

In the models discussed in the previous Sections, nucleon decay occurs through the exchange of GUT-scale particles. In particular, the standard and flipped SU(5) GUT models described in Sections~\ref{sec:unflipped} and \ref{sec:flipped}, respectively, have characteristic predictions for nucleon decay branching ratios because of the minimality of these models~\cite{EGNNO6}. In this Section, we briefly review these features and discuss the possibility of distinguishing these models in future proton decay experiments.

In the minimal standard SU(5) model in Section~\ref{sec:unflipped}, $p \to K^+ \bar{\nu}$ induced by the exchange of color-triplet Higgs fields is the dominant decay mode~\cite{Knu}. In fact, the lifetime of this decay mode turns out to be too short if supersymmetry lies around the TeV scale~\cite{Goto:1998qg, mp}, though this problem is alleviated if sparticle have larger masses~\cite{Hisano:2013exa, evno, eelnos, eemno, Ellis:2019fwf}. The $p \to K^+ \bar{\nu}$ mode in constrained supersymmetric models has recently been investigated in detail in Ref.~\refcite{Ellis:2019fwf}; it is found that the predicted rate of this decay mode can exceed the current limit imposed by Super-Kamiokande, $\tau(p \to K^+ \bar{\nu}) > 6.6 \times 10^{33}$~years~\cite{Abe:2014mwa}, in a range of the parameter space that is consistent with the measured value of the Higgs boson mass, and that it is within the reach of future proton decay experiments, such as JUNO~\cite{JUNO}, DUNE~\cite{DUNECDR}, and Hyper-Kamiokande~\cite{HKTDR}. If sparticle masses are as large as $\gtrsim 100$~TeV, as in PGM~\cite{eioy,evno}, the dominant decay mode becomes $p \to e^+ \pi^0$ induced by the exchange of the SU(5) gauge bosons~\cite{evno}, which may also be probed in Hyper-Kamiokande~\cite{HKTDR}. The rates of other decay modes, such as $p \to \mu^+ \pi^0$,\footnote{The ratio $\Gamma(p \to \mu^+ \pi^0)/\Gamma(p \to e^+ \pi^0)$ is predicted to be $\simeq 0.008$ in minimal standard SU(5)~\cite{EGNNO6}.} are quite suppressed unless there is sizable flavor violation in sfermion mass matrices~\cite{Hisano:2013exa} and likely to be beyond the reach of the future proton decay experiments. 

In summary, the prediction for proton decay in the minimal standard SU(5) model in Section~\ref{sec:unflipped} is as follows: i) $p \to K^+ \bar{\nu}$ is the dominant decay mode for relatively light supersymmetric mass spectra; ii) $p \to e^+ \pi^0$ is the dominant decay mode for heavier sparticle masses; iii) only these two decay modes are within the reach of future proton decay experiments. 

Nucleon decay predictions in the flipped SU(5) model discussed in Section~\ref{sec:flipped} are quite different. A detailed comparison of the nucleon decay branching fractions in the flipped and unflipped SU(5) GUTs is given in Ref.~\refcite{EGNNO6}. The decay modes that are most promising for detection in future experiments are: i) $p \to K^+ \bar{\nu}$ and ii) $\Gamma(p \to \mu^+ \pi^0)/\Gamma(p \to e^+ \pi^0)$. 

We note first that in the flipped SU(5) model the color-triplet Higgs exchange process is suppressed due to the absence of a supersymmetric mass term for $H$ and $\bar{H}$. In addition, it is found that the GUT gauge boson exchange process does not induce $p \to K^+ \bar{\nu}$ either~\cite{Ellis:1993ks}. As a result, the branching ratio for decay into $p \to K^+ \bar{\nu}$ is negligible in flipped SU(5), in contrast to standard SU(5), where it is expected to be the dominant decay mode, as discussed above. 

\begin{figure}[t]
  \begin{center}
  \includegraphics[height=75mm]{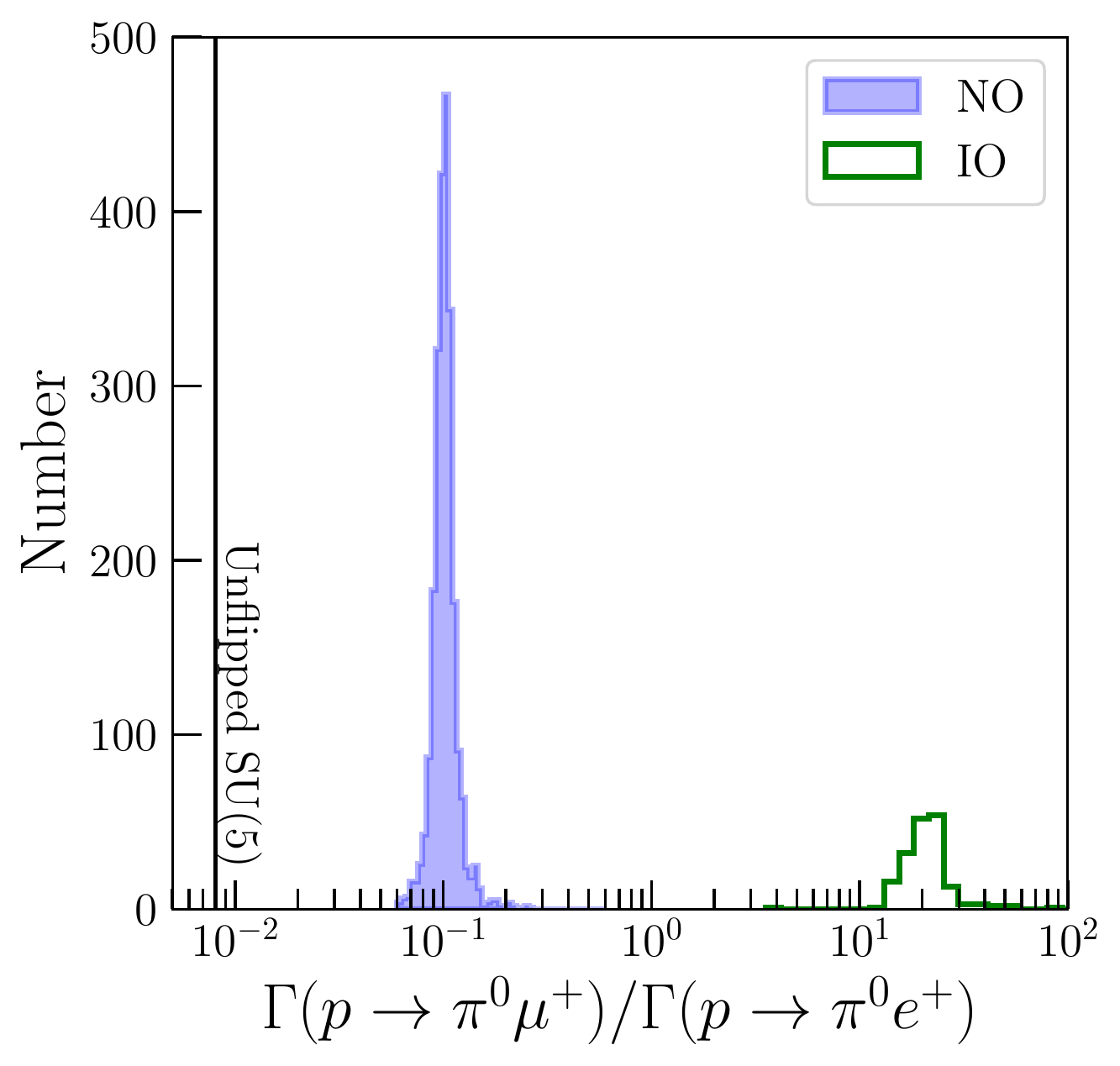}
  \caption{\it Histograms of $\Gamma (p\to \pi^0 \mu^+)/\Gamma (p\to
   \pi^0 e^+)$ in the flipped SU(5) GUT model for the cases of normal-ordered (NO) and inverse-ordered (IO) neutrino mass spectra shown in blue and
   green, respectively. The vertical line corresponds to the unflipped
   SU(5) prediction. }
  \label{fig:rpimuove}
  \end{center}
  \end{figure}

Secondly, the ratio of the decay rates of $p \to e^+ \pi^0$ and $p \to \mu^+ \pi^0$ in flipped SU(5) is found to be~\cite{Ellis:1993ks} 
\begin{equation}
  \frac{\Gamma(p \to \mu^+ \pi^0)}{\Gamma(p \to e^+ \pi^0)} \simeq \frac{|(U_\ell)_{21}|^2}{|(U_\ell)_{11}|^2} ~,
\end{equation}
where $U_\ell$ is a unitary matrix that is used to diagonalize $\lambda_3$, which can be obtained from $U_\nu$ and the PMNS matrix using Eq.~\eqref{eq:pmnsfl}. We show in Fig.~\ref{fig:rpimuove} histograms of the ratio $\Gamma (p\to \pi^0 \mu^+)/\Gamma (p\to \pi^0 e^+)$ in the normal-ordered (NO) and inverse-ordered (IO) neutrino mass scenarios in blue and green, respectively. The vertical black solid line represents the
value predicted in standard SU(5). The flipped SU(5) model
predicts this ratio to be $\sim 0.10$ and $\sim 23$ for the NO and IO
cases, respectively, which is much larger than the standard SU(5) prediction. Hyper-Kamiokande~\cite{HKTDR} is expected to improve by an order of magnitude the present experimental sensitivity to both of these decay modes, which may provide an opportunity to test these predictions.

\section{Summary}

The road to building a no-scale model of inflation has been a long and winding one. 
We have argued that inflation must involve Planck-scale physics~\cite{primordial}, and that inflation cries out for
supersymmetry~\cite{Cries}. As a result, inflationary models should be built from
supergravity, in particular no-scale supergravity~\cite{no-scale,EKN1,EKN,LN}, which
yields naturally a semi-positive potential with flat directions and no AdS holes. 
Moreover, no-scale supergravity emerges naturally in the low-energy limit of string theory~\cite{Witten}.
However, supergravity, even no-scale supergravity, is not sufficient as a framework
to specify the model of inflation. As is usually the case, theory must
rely on experiment to make progress. Measurements of the CMB anisotropy spectrum, (relatively) quickly 
ruled out large numbers of inflationary models, including some
very simple models described in Section~\ref{simple}, based
on minimal and no-scale supergravity (from measurements of $n_s$), as well as models based on single polynomial potentials (from upper limits to $r$). 

Curiously, the models suggesting a singularity-free Universe proposed by 
Starobinsky~\cite{Staro} are equivalent up to a conformal transformation~\cite{WhittStelle} to 
single-field models of inflation with a plateau-like potential - see Eq.~(\ref{staropot}) and Fig.~\ref{fig:staro} - which are in excellent agreement with current CMB measurements. 
A significant theoretical advance was made when it was realized that,
starting with a very simple Wess-Zumino form of superpotential~(\ref{wi}), the Starobinsky potential
could be derived from no-scale supergravity~\cite{eno6}, as we have discussed in Section~\ref{wzeno6}. It was further realized
that the superpotential yielding the Starobinsky potential was not unique \cite{eno7}. 
Indeed,
it had been shown previously~\cite{Cecotti} that $R+R^2$ gravity was equivalent to a different no-scale model with the superpotential given in Eq.~(\ref{cec}). As described in Section~\ref{avatars},
these models are all related by the underlying SU(2,1)/SU(2)$\times$U(1) no-scale symmetry\cite{enov1}. 

The parallels between the conformal transformation in supergravity, and higher-derivative
gravity were reviewed~\cite{eno9} in Sections~\ref{structure} and~\ref{r+r2}.
There we saw that the scale invariance of an $R^2$ theory of gravity
is directly related to the choice of the conformal function, $\Phi$, made in Eq.~\eqref{confPhi}, leading to a K\"ahler potential with a no-scale symmetry. 
Since $R^2$ gravity is equivalent to a de Sitter space-time~\cite{old}, once the connection between $R^2$ gravity and no-scale supergravity is made, 
it is natural to consider the construction of Minkowski, de Sitter, and anti-de Sitter
solutions in no-scale supergravity~\cite{enno,enno}. This connection was reviewed in Section~\ref{pairs}. 

Of course, the inflationary sector cannot be totally isolated from the matter sector,
as reheating and a graceful exit from inflation are necessary.
As a consequence, inflation can not be divorced from the mechanism of supersymmetry breaking\cite{EGNO4,enov4}. No-scale inflation and phenomenology were discussed in Section~\ref{pheno}. In particular, in Section~\ref{unified} we focused on a set of unified no-scale attractor models~\cite{enov3,enov4} in which inflation, supersymmetry breaking, and dark energy
can all be explained in a simple and compact framework given by the superpotential in Eq.~(\ref{uniwz}). In these models, the inflaton mass, $M$, determines the scale of supersymmetry breaking, with $m_{3/2} \sim M^3/M_P^2$. We discussed the mechanisms for reheating and the relation to the number of e-folds of inflation in Sections~\ref{sec:reheat} and~\ref{sec:N*} respectively. 

Finally, in Section~\ref{sec:UV}, we showed how to embed no-scale models of inflation in various GUTs. We considered in turn an SU(5) GUT model, an SO(10) GUT model\cite{EGNNO1},
and a flipped SU(5)$\times$U(1) model\cite{EGNNO2,EGNNO3,EGNNO4,EGNNO5,EGNNO6}. 

We summarize our approach to building a no-scale inflationary cosmology in Fig.~\ref{fig:concept}. The theory is based on
no-scale supergravity (left orange box) which is derived from the superstring
(green box)~\cite{Witten}.
The no-scale Starobinsky-like model of inflation (grey box)
makes successful predictions for the tensor-to-scalar perturbation ratio and the tilt
in the scalar perturbation spectrum (upper left pink boxes).

\begin{figure}[ht]
\centering
    \includegraphics[width=0.85\textwidth]{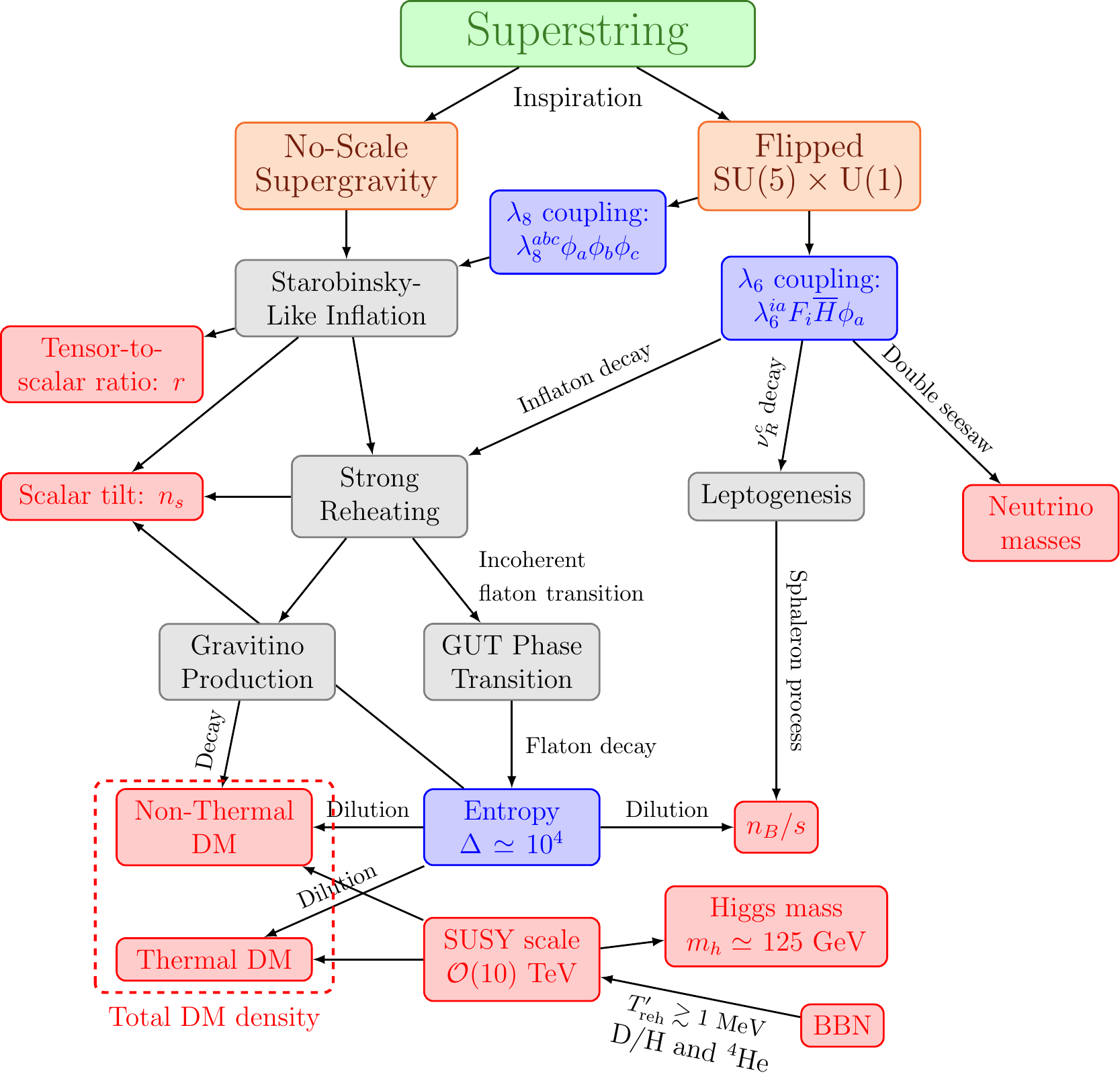}
    \caption{\it The general structure of our preferred scenario for no-scale inflation and particle cosmology~\cite{EGNNO5}.}
    \label{fig:concept}
\end{figure}

As we have emphasized in Section~\ref{sec:flipped}, our preferred field-theoretical framework is
the flipped SU(5)$\times$U(1) GUT (right orange box), which may be derived from weakly-coupled 
heterotic string theory~\cite{AEHN}. In this scenario there is one particular Yukawa coupling, $\lambda_8$,~\cite{EGNNO2,EGNNO3}
which plays key roles in the generation of the inflationary potential. Another coupling, $\lambda_6$, plays a crucial role~\cite{EGNNO4,EGNNO5}
in inflaton decay and the reheating process, in
leptogenesis, and in generating via a double-seesaw mechanism neutrino masses (right pink box) that are compatible with oscillation measurements and cosmological limits. As we have discussed in Section~\ref{sec:pdecay}, this model's predictions for nucleon decay are different from those of conventional SU(5) (which cannot be derived from weakly-coupled string theory), and may be accessible in the next round of neutrino oscillation experiments~\cite{EGNNO6}.

In this scenario, the decay of the inflaton is thought to have caused
strong reheating of the Universe followed by the GUT phase transition (central grey box) associated with a flaton field, which generated a factor $\Delta = {\cal O}(10^4)$ of entropy (lowest blue box). This diluted the baryon asymmetry $n_B/s$ generated by leptogenesis to a value compatible with cosmological measurements (right pink box), and also diluted the gravitino abundance (left grey box) so that the density of cold dark matter produced non-thermally by gravitino decay is also compatible with cosmological measurements as well that produced thermally (left pink boxes) with a supersymmetry-breaking scale that is ${\cal O}(10)$~TeV (lower pink box)~\cite{EGNNO5}. Although placing sparticles beyond the reach of the LHC, this scale is comfortably compatible with the measured value of the Higgs mass $m_h \simeq 125$~GeV, as well as the baryon-to-entropy ratio indicated by the success of conventional Big Bang Nucleosynthesis calculations. 

The scenario depicted in Fig.~\ref{fig:concept} illustrates how no-scale supergravity inflation provides a possible bridge between string theory in the ultraviolet limit and the Standard Model at TeV energies, and in particular a framework for a flipped model of almost everything below the Planck scale.

\newpage
\section*{Acknowledgments}

The work of J.E.~was supported partly by the United Kingdom STFC Grant ST/P000258/1 
and partly by the Estonian Research Council via a Mobilitas Pluss grant. 
The work of D.V.N.~was supported partly by the DOE grant DE-FG02-13ER42020 
and partly by the Alexander S. Onassis Public Benefit Foundation. 
The work of K.A.O. and S.V.~was supported partly
by the DOE grant DE-SC0011842 at the University of Minnesota and
 acknowledges support by the Director, Office of Science, Office of High Energy Physics of the U.S. Department of Energy under the Contract No. DE-AC02-05CH11231.
The work of N.N. was supported by the Grant-in-Aid for Scientific Research B (No.20H01897), Young Scientists B (No.17K14270), and Innovative Areas (No.18H05542). The work of M.A.G.G.~was supported by the Spanish Agencia
Estatal de Investigaci\'on through the grants FPA2015-65929-P (MINECO/FEDER, UE), PGC2018095161-B-I00, IFT Centro de Excelencia Severo
Ochoa SEV-2016-0597, and Red Consolider MultiDark
FPA2017-90566-REDC.


\end{document}